\documentclass[aps,prl,twocolumn,superscriptaddress,10pt,showpacs,longbibliography]{revtex4-2}

\usepackage{blindtext}
\usepackage{lipsum}
\usepackage{graphics}
\usepackage{amsmath}
\usepackage{graphicx}
\usepackage{graphics}
\usepackage{amssymb}
\usepackage{pifont}
\usepackage{verbatim}
\usepackage{physics}
\usepackage{algorithm}
\usepackage{algpseudocode}
\usepackage[normalem]{ulem}
\usepackage[dvipsnames]{xcolor}
\usepackage{bbm}
\usepackage{enumitem} 
\usepackage{multirow}
\usepackage[caption=false]{subfig}
\usepackage{graphicx}
\usepackage{bm}
\usepackage{amsmath}
\usepackage{physics}
\usepackage{amssymb}
\usepackage{amsfonts}
\usepackage{amsthm}
\usepackage{bbm}
\usepackage{mathtools}
\usepackage{braket}
\usepackage[normalem]{ulem}
\usepackage{wrapfig}
\usepackage{tikz}
\usepackage{dsfont}
\usepackage{comment}
\usepackage{thmtools,thm-restate}
\usepackage{placeins}
\usepackage[export]{adjustbox}

\definecolor{blueviolet}{rgb}{0.2, 0.2, 0.6}
\definecolor{webgreen}{rgb}{0,.5,0}
\definecolor{webbrown}{rgb}{.6,0,0}
\usepackage[bookmarks=false,
colorlinks=true,
urlcolor=webbrown, 
linkcolor=blueviolet, 
citecolor=webgreen,
pdfstartpage=1,
pdfstartview={FitH},
bookmarksopen=false
]{hyperref}
\usepackage[nameinlink,capitalize]{cleveref}

\newtheorem{theorem}{Theorem}
\newtheorem*{hypothesis*}{Scrooge Hypothesis}

\newtheorem{definition}{Definition}
\newtheorem{corollary}{Corollary}
\newtheorem{lemma}{Lemma}
\newtheorem{claim}{Claim}

\newtheorem{proposition}{Proposition}
\newtheorem*{theorem*}{Theorem}

\newtheorem*{task*}{Task}
\newtheorem*{proposition*}{Proposition}

\newcommand{\dif}{{\rm d}}

\graphicspath{{.}{./tikz/}}

\newcommand{\be}{\begin{equation}}
	\newcommand{\ee}{\end{equation}}

\newcommand{\Exp}{\mathop\mathbb{E}}

\newcommand{\dist}{\mathrm{dist}}

\DeclareMathOperator*{\E}{{\mathbb{E}}}

\begin{document}
	\title{Conditional dependence and Scrooge ensembles in shallow random quantum circuits}
	
	\author{Yinchen Liu}
	\thanks{These authors contributed equally to this work}
	\affiliation{Department of Combinatorics and Optimization and Institute for Quantum Computing, University of Waterloo}
	
	\author{Max McGinley}
	\thanks{These authors contributed equally to this work}
	\affiliation{TCM Group, Cavendish Laboratory, University of Cambridge}
	
	\author{Thomas Schuster}
	\thanks{These authors contributed equally to this work}
	\affiliation{Walter Burke Institute for Theoretical Physics and Institute for Quantum Information and Matter, California Institute of Technology}
    \affiliation{Google Quantum AI}
	
	\author{David Gosset}
	\affiliation{Department of Combinatorics and Optimization and Institute for Quantum Computing, University of Waterloo}
	\affiliation{Perimeter Institute for Theoretical Physics}
	\begin{abstract}
		The output state of a 2D geometrically local shallow random quantum circuit does not have long range correlations due to its lightcone structure. But this changes if one measures a subset of the qubits: long-range entanglement can be induced by the measurement process, leading to conditional correlations between distant qubits. In this paper we investigate the structure of conditional dependence in these circuits and its consequences for quantum advantage. For a tripartition $ABC$ of the qubits, we consider the ensemble of post-measurement states on $A$ that is conditioned on a specific measurement outcome on $B$ and ranges over all possible measurement outcomes on $C$. For circuit depths exceeding a constant critical value $d^*$, we conjecture that this ensemble is well approximated by a certain generalization of the Haar ensemble, called the Scrooge ensemble~[Jozsa \textit{et al.}, \href{https://doi.org/10.1103/PhysRevA.49.668}{Phys. Rev. A \textbf{49}, 668 (1994)}]; we also provide supporting numerical and analytical evidence. Our conjecture describes a precise sense in which the state retains its lightcone structure on the remaining unmeasured qubits, but also develops some globally random features arising from the measurement. A consequence is that $n$-qubit shallow random quantum circuits in two dimensions are classically efficiently simulable in the presence of a tiny depolarizing noise rate $\Omega(\log(n)/n)$.
	\end{abstract}
	
	\maketitle

	\begin{figure}[t]
		\centering
		\includegraphics[width=\columnwidth]{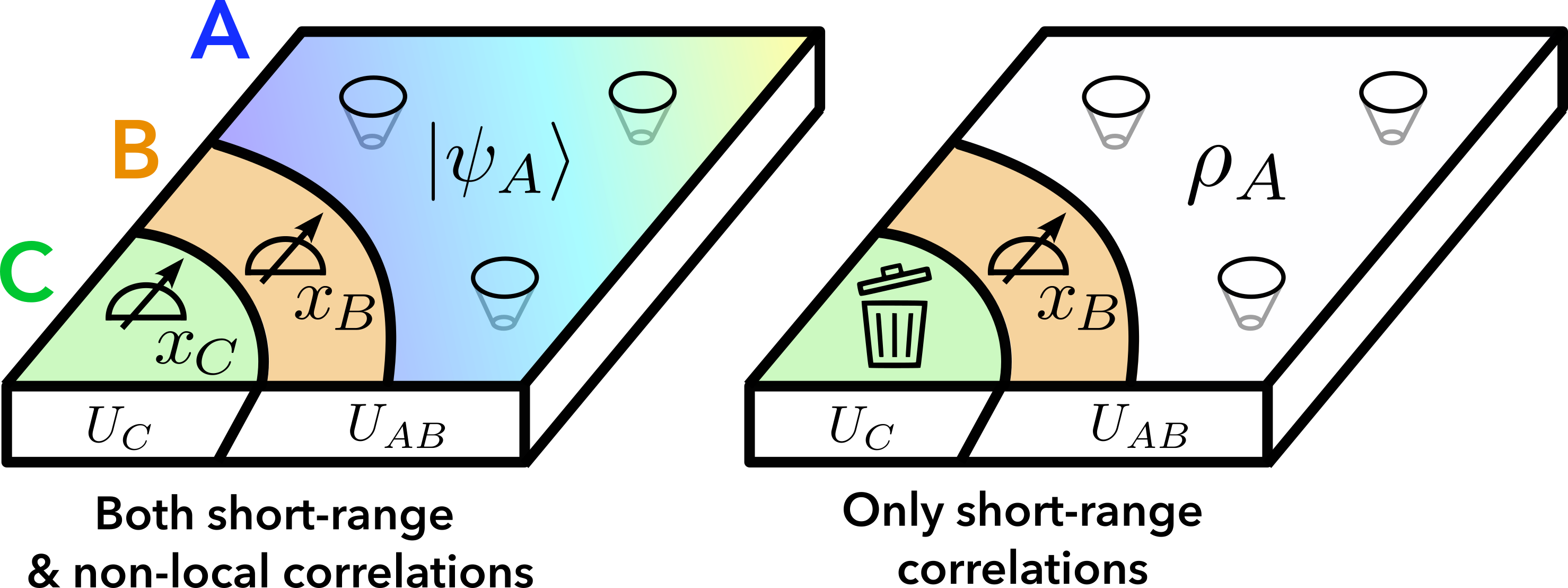}
		\caption{Illustration of the Scrooge hypothesis for constant-depth random quantum circuits. Here $\ket{\psi_A}$ is the post-measurement state with  outcomes $x_B$, $x_C$;  while $\rho_A$ is obtained by measuring $x_B$ and tracing out $C$. We find that moments of $\ket{\psi_A}$ match those of the Scrooge ensemble defined by $\rho_A$. When measured in any fixed basis, $\rho_A$ features only short-range correlations (black light-cones), while $\ket{\psi_A}$ may have non-local correlations (rainbow gradient).}
		\label{fig:1}
	\end{figure}
	Quantum computations composed of random gates are a testbed for understanding generic features of quantum many-body dynamics including entanglement, complexity, and quantum advantage \cite{nahum2018operator, arute2019quantum, google2025observation}. As the circuit depth grows, the quantum state becomes globally Haar random, and its output probabilities converge to approximately independent random variables drawn from the Porter-Thomas distribution. Quantifying the approach to this limit---as measured by signatures of pseudorandomness such as anti-concentration and the $k$-design property---has been the subject of a rich vein of research \cite{brandao2016local, dalzell2022random, harrow2023tdesign, heinrich2025anti, schuster2025random, cui2025unitary,schuster2025strong}. 
    
    Less understood is an opposite limit: \textit{shallow} random quantum circuits (SRQCs), in which the circuit depth is a fixed constant independent of the number of qubits. SRQCs straddle the boundary between classical simulability and quantum advantage. While local features of SRQCs are easy to compute, sampling from the global output distribution is conjectured to be classically hard due to long-range conditional correlations induced by measurement ~\cite{napp2022efficient,bao2024finite,mcginley2025measurement,bene2025quantum}. This dichotomy makes SRQCs a compelling setting for investigating quantum pseudorandomness and advantage outside the $k$-design regime.
	Despite this, our current understanding  is limited to specific features of SRQCs, such as their classical hardness~\cite{napp2022efficient} and measurement-induced entanglement (MIE) ~\cite{bao2024finite,mcginley2025measurement}, and lacks any  broader, universal description.

	Here we propose a bold conjecture concerning the structure of SRQCs. It centers on \emph{projected ensembles}: the set of pure states obtained on a subsystem $A$ after projectively measuring a complementary subsystem (Fig.~\ref{fig:1}). Such ensembles naturally capture the conditional dependence of observables on $A$ on the complementary measurement outcomes. We conjecture that the projected ensembles of SRQCs are described by the \emph{Scrooge ensemble} \cite{jozsa94}, a versatile generalization of the Haar ensemble. Whereas recent works have pointed to the emergence of Scrooge-randomness in dynamics with conservation laws at late times~\cite{goldstein2006distribution,goldstein2016universal,cotler2021emergent, ippoliti2022solvable,  mark2024maximum, chang2025deep, mcginley2025scrooge, mok2026nature,schuster2026fast}, we argue this occurs at constant depth, well before thermalization. We provide numerical evidence for this conjecture in the setting of random Clifford circuits, as well as analytic evidence based on the known mapping between moments of random circuits and statistical mechanical models. 
    
	We also explore the consequences of our conjecture for quantum advantage in SRQCs. We find that the lion's share of information in the output distribution of SRQCs can be computed locally and efficiently on a classical computer, using a simple algorithm based on short-range conditional dependencies. The classical hardness of SRQCs stems entirely from  $O(1)$ additional bits of information, which are stored fully non-locally and give rise to all non-trivial features. A dramatic consequence of this fact is that quantum advantage in SRQCs is much more fragile to noise than previously known~\cite{aharonov2023polynomial,nelson2024polynomial,lee2025classical,nelson2025limitations,cheng2023efficient,wei2025measurement,zhang2025classically}: assuming the Scrooge hypothesis, we provide an efficient classical  algorithm to simulate SRQCs in the presence of a tiny depolarizing noise rate $\gamma=\Omega(\log(n)/n)$. This contrasts with prior work~\cite{wei2025measurement}, which conjectured an efficient classical simulation only when $\gamma=\Omega(1/\log(n))$.
	
	\emph{Background.}---We consider $n$-qubit  two-dimensional (2D) SRQCs of depth $d$ with nearest-neighbor gates arranged in the standard ``brickwork" architecture (Fig.~\ref{fig:canonical_2D_brickwork_circuit}).
    
	The output states $U\ket{0^n}$ of SRQCs $U$ have a local \textit{lightcone} structure that sets them apart from globally random states. Given a  circuit $U$, the lightcone of a subset of (output) qubits $A\subseteq [n]$ is the set $\mathcal{L}(A)\subset [n]$ of (input) qubits that are causally connected to $A$. That is, for any operator $O$ supported completely on $A$, the support of $U^{\dagger} O U$ is contained in $\mathcal{L}(A)$. In particular, we have $U^{\dagger} O U= U_{A}^{\dagger} O U_{A}$,
	where $U_A$ is obtained by removing all gates in $U$ that can be commuted past $O$ and multiplied with their inverses in $U^{\dagger}$. Accordingly, the reduced state of $U\ket{0^n}$ on $A$ coincides with that of $U_{A}|0^{|\mathcal{L}(A)|}\rangle$.
	
\begin{figure}[t]
    \centering
		\subfloat[]{
            \begin{tikzpicture}[scale=0.33]
                \foreach \x in {0,...,10}
                \foreach \y in {0,...,10}
                {
                    \fill (\x,\y) circle (2pt);
                }
                \draw [draw=black] (-0.5,9.5) rectangle (10.5,10.5);
                \draw [draw=black] (-0.5,0.5) rectangle (10.5,9.5);
                \draw [draw=black] (-0.5,-0.5) rectangle (10.5,0.5);
                \fill[red, fill opacity=0.2](-0.5,9.5) rectangle (10.5,10.5);
                \fill[yellow, fill opacity=0.2] (-0.5,0.5) rectangle (10.5,9.5);
                \fill[blue, fill opacity=0.2](-0.5,-0.5) rectangle (10.5,0.5);
                
                \node at (5,10) {\large{$A$}};
                \node at (5,5) {\large{$B$}};
                \node at (5,0) {\large{$C$}};
                \useasboundingbox (-0.5,-0.5) rectangle (10.5,10.5);
            \end{tikzpicture}
		}
        \vspace{0.1cm}
        
		\subfloat[]{
        \begin{minipage}[b]{0.9\columnwidth}
			\begin{ruledtabular}
				\begin{tabular}{ lccc }
					& \multicolumn{3}{c}{$\mathrm{Pr}[I(A;C|B)=r]$} \\ \hline 
					&   2D  Clifford & 2D Clifford  & random \\
					$r$  & depth-$5$& depth-$7$ & stab. state\\
					\hline
					0 & 0.99999 & 0.41815 & 0.41942 \\ 
					\hline
					1 & 0.00001 & 0.42077 & 0.41942 \\
					\hline
					2 & 0 & 0.13952 & 0.13981 \\
					\hline
					3 & 0 & 0.02022 & 0.01997 \\
					\hline
					4 & 0 & 0.00131 & 0.00133 \\
					\hline
					5 & 0 & 0.00003 & 0.00004
				\end{tabular}
			\end{ruledtabular}
		\end{minipage}
		}
		
		\caption{(a) Tripartition ABC of the 2D grid. (b) Numerical estimation of the distribution of the integer $r=I(A;C|B)$ for depth-$5$ and depth-$7$ 2D  random Clifford circuits acting on a $100\times 100$ grid of qubits using $10^5$ random circuit realizations. }
		\label{fig:distribution_of_r}
	\end{figure}
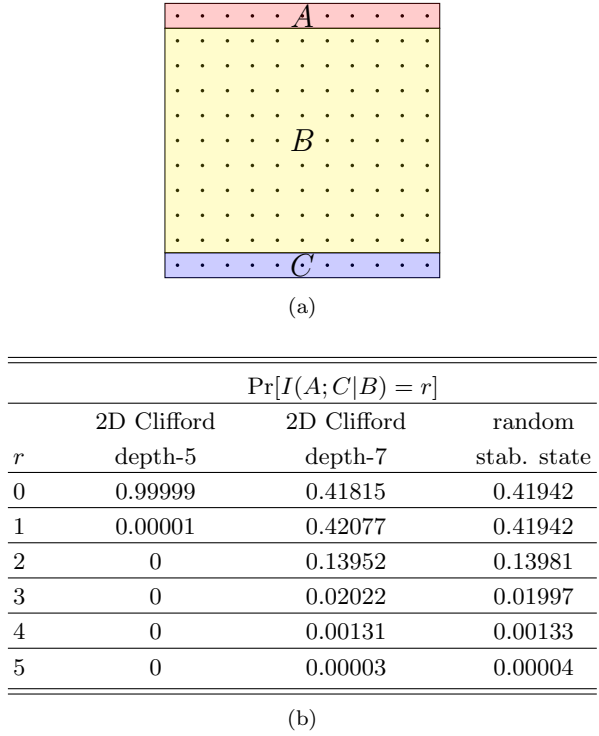

	An important consequence is that the output state $U|0^n\rangle$ of a SRQC can only possess short-range entanglement. Consider any two subsets of qubits $A,C$ that are far enough apart that their lightcones don't intersect---e.g., the top and bottom rows of the 2D grid shown in Fig.~\ref{fig:distribution_of_r}(a). These subsystems are completely independent in the sense that the reduced density matrix factorizes: $\rho_{AC}=\rho_A\otimes \rho_C$. Despite this dramatic restriction, SRQCs can nevertheless exhibit nontrivial \textit{conditional dependencies} in their output distribution, which can render them hard to classically simulate \footnote{Indeed, if we start with the output state $|\psi\rangle$ and then classically simulate measurement of one qubit at a time, then we'll need to sample from the distribution of the current qubit's measurement outcome, conditioned on the outcomes measured so far.}. For example, the post-measurement state on $A,C$ can become entangled as a result of measuring all other qubits (region $B$ in Fig.~\ref{fig:distribution_of_r}(a)), and this may lead to non-local correlations between the outcomes of subsequent measurements on $A$ and $C$.
	
	For SRQCs in the 2D brickwork architecture, Ref.~\cite{napp2022efficient} posited a remarkable phase transition phenomenon: long-range measurement-induced entanglement (MIE) is present above a certain constant critical depth $d^{\star}=\Theta(1)$ and absent below it. The prevalence of long-range MIE for 2D SRQCs with a sufficiently high (constant) depth was recently established rigorously in Ref.~\cite{mcginley2025measurement}. Ref.~\cite{napp2022efficient} also argued that this mechanism underlies quantum advantage in sampling from the output distribution of SRQCs (see below), and that such advantage is only possible when $d>d^{\star}$ (see \cite{bene2025quantum,wei2025measurement}). Here we study structural properties of SRQCs that go hand-in-hand with the proliferation of MIE and hardness of classical simulation above the critical depth.
	
    Our main focus is on the output distribution $p(x) = |\braket{x|U|0^n}|^2$, where $x \in \{0,1\}^n$, marginals of which we write as $p_C(x_C)$ for any $C \subseteq [n]$, where $x_C \in \{0,1\}^{|C|}$. A succinct way to quantify the non-local dependencies between regions $A, B, C \subseteq [n]$ is via the \textit{conditional mutual information}, or CMI, $I(A;C|B) = \sum_{x_B}p_B(x_B) I(A:C)_{x_B}$, where $I(A;C)_{x_B}$ is the (classical) mutual information between $x_A$ and $x_C$ in the conditional distribution $p_{AC|B}(x_Ax_C|x_B) \coloneqq p_{ABC}(x_Ax_Bx_C)/p_B(x_B)$. The CMI vanishes if $x_A$ and $x_C$ are conditionally independent, $p_{AC|B}(x_Ax_C|x_B) = p_{A|B}(x_A|x_B)p_{C|B}(x_C|x_B)$, and is non-zero otherwise. Informally, long-(short-)ranged CMI is often associated with hardness (easiness) of classical simulation \cite{napp2022efficient, zhang2025classically}.

    As a motivating example, we numerically estimate the distribution of CMI for 2D SRQCs in which all gates are uniformly random two-qubit Clifford gates (Fig.~\ref{fig:distribution_of_r}). For Clifford circuits, $I(A;C|B)$ is quantized to an integer. We consider a $100 \!\! \times \!\! 100$ grid of qubits, with $ABC$ arranged as in Fig.~\ref{fig:distribution_of_r}(a), and simulate $10^5$ random circuit realizations. Our results are tabulated in Fig.~\ref{fig:distribution_of_r}(b). At depth $d = 5$, we find $I(A;C|B) = 0$ in all but one realization, whereas at depth $d = 7$, the CMI has a nonzero mean. This behaviour is consistent with the short- and long-range MIE phases, respectively. However, at $d = 7$ (above the critical depth) we observe an unexpected signature of \textit{global} randomness: The distribution of CMI converges to that of a uniformly random stabilizer state as $n \rightarrow \infty$. (Results for $d = 6$ are less conclusive,  possibly due to finite-size effects---see Tab.\ref{tab:distribution_of_r_full} in Appendix \ref{sec:Z_type_stablizer_numerics}.) 

    This observation suggests that, once qubits are measured, global correlations coexist with the local structure of SRQCs. We now put forward our central conjecture which makes this notion precise. 

	\emph{The Scrooge hypothesis.}---We consider a tripartition $[n]=ABC$ of the qubits and define an ensemble of post-measurement states on region $A$ as follows. Let $x_B$ and $x_C$ be measurement outcomes on $B$ and $C$, let $U_{AB} \sim \mathcal{U}_{AB}$ denote the product of all gates in the backwards light-cone of $AB$, and let $U_C \sim \mathcal{U}_{C}$ denote the product of all remaining gates. We have $U = U_C U_{AB}$ by definition. For each fixed $x_B$ and $U_{AB}$, we define the post-measurement ensemble $\mathcal{E}_A(U_{AB},x_B)$ as
	\begin{align}
		&\mathcal{E}_A(U_{AB},x_B) \nonumber\\
		&= \Bigg\{ \frac{ \bra{x_{BC}} U \dyad{0^n} U^\dagger \ket{x_{BC}}}{\tr_A \! \big( \bra{x_{BC}} U \dyad{0^n} U^\dagger \ket{x_{BC}}\big)} \, \Bigg| \, U_C \sim \mathcal{U}_{C} \Bigg\}.
		\label{eq:ensemble def}
	\end{align}
	This ensemble is independent of $x_C$ owing to the randomization over $U_C$. Its first moment is the density matrix
	\begin{align}
		\rho_A(U_{AB},x_B)& \equiv \E_{\psi_A \sim \mathcal{E}_A} \dyad{\psi_A}\nonumber\\
		& = \frac{ \tr_{C} \! \big( \bra{x_{B}} U \dyad{0^n} U^\dagger \ket{x_{B}} \big)}{\tr_{AC} \! \big( \bra{x_{B}} U \dyad{0^n} U^\dagger \ket{x_{B}}\big)},
	\end{align}
	obtained by measuring $U \dyad{0^n} U^\dagger$ in the state $x_B$ on $B$ and tracing out $C$.
	The measurement outcome $x_B$ is received with probability $p(x_B; U_{AB}) = \tr_{AC} \! \big( \bra{x_{B}} U \dyad{0^n} U^\dagger \ket{x_{B}}\big)$.
	
	Our central hypothesis is that, under suitable conditions on the tripartition $ABC$, the ensemble $\mathcal{E}_A(U_{AB},x_B)$ is well approximated by a certain generalization of the Haar ensemble, called the Scrooge ensemble $\mathcal{S}_\rho$~\cite{jozsa94}. Whereas the average state of the Haar ensemble $\mathbb{E}_{\phi \sim {\rm Haar}} \ket{\phi}\bra{\phi} = I/2^n$ is maximally mixed, the Scrooge ensemble is parametrized by an $n$-qubit density matrix $\rho$, such that its average state is $\mathbb{E}_{\psi \sim {\mathcal{S}_\rho}} \ket{\psi}\bra{\psi} = \rho$. In particular, $\mathcal{S}_\rho$ is defined as the unique, maximally entropic ensemble of pure states with average state $\rho$, where the notion of maximum entropy is discussed in \cite{jozsa94}. More explicitly, 
	to sample a state $|\psi\rangle$ from the Scrooge ensemble $ \mathcal{S}_\rho$, one can first sample $\ket{\phi}$ with probability density $\propto \braket{\phi|\rho|\phi}$, and then let $\ket{\psi} = \sqrt{\rho}\ket{\phi}/\sqrt{\braket{\phi|\rho|\phi}}$.Equivalently, if one takes a purification of $\rho$, i.e.~$\ket{\Phi^\rho_{AA'}}$ on $AA'$ such that $\tr_{A'}[\Phi_{AA'}^\rho] = \rho_A$, then measuring $A'$ with a rank-1 Haar-random POVM generates a Scrooge-random state on $A$ \cite{mcginley2025scrooge}. This characterization in terms of ensembles of post-measurement states will prove important in our understanding of how the Scrooge ensemble emerges in random circuits later on.

    To extend our hypothesis to circuits composed of random Clifford gates, in addition to those of Haar-random gates, we also define a \emph{stabilizer Scrooge ensemble} $\mathcal{D}_{\rho}$ as follows. For any $n$-qubit mixed stabilizer state $\rho$, we sample $\psi \sim \mathcal{D}_{\rho}$ by first sampling a uniformly random $n$-qubit stabilizer state $|\phi\rangle$ such that $\langle \phi|\rho|\phi\rangle\neq 0$, and then letting $|\psi\rangle \propto \rho|\phi\rangle$. The ensemble $\mathcal{D}_\rho$ arises when a POVM composed of uniformly random rank-1 stabilizer projectors is applied to a stabilizer purification of $\rho$.
	
    To be precise, as is standard in studies of pseudorandomness, our hypothesis will pertain to the \textit{$k$-th moments} of the post-measurement ensemble $\mathcal{E}_A(U_{AB}, x_B)$---i.e.,~to quantities of the form
	\begin{align}
		\chi_{\mathcal{E}_A(U_{AB}, x_B)}^{(k)}[O_A] \coloneqq \mathbb{E}_{\psi \sim \mathcal{E}_A(U_{AB}, x_B)} \!\tr\big[O_A \dyad{\psi}^{\otimes k}\!\big]
		\label{eq:k moment def}
	\end{align}
	for some $k$-copy observable $O_A \in \mathcal{B}(\mathcal{H}_A^{\otimes k})$. We posit that, for certain choices of $ABC$ and with high probability over $U_{AB}, x_B$, these moments are close to those of the Scrooge ensemble $\mathcal{S}_\rho$ with background state $\rho_A \equiv \rho_A(U_{AB}, x_B)$,
	\begin{align}
		\big|\chi_{\mathcal{E}_A(U_{AB}, x_B)}^{(k)}[O_A]- \chi^{(k)}_{\mathcal{S}_\rho}[O_A]\big| \leq \varepsilon \cdot \chi^{(k)}_{\mathcal{S}_\rho}[O_A].
		\label{eq:observable_relative_error_def}
	\end{align}
    For random Clifford circuits, we replace $\mathcal{S}_\rho$ with $\mathcal{D}_\rho$.
	Here $\varepsilon$ quantifies the \textit{relative error} of the approximation.

    Our hypothesis is stated precisely as follows:
	\begin{hypothesis*} \hypertarget{hyp:scrooge}{}
		There is a $d^*=O(1)$ such that the following holds. Let $U$ be a depth $d>d^{*}$ 2D brickwork circuit $U$ on a  $\sqrt{n} \times \sqrt{n}$ grid in which each two-qubit gate is Haar random (resp. Clifford random). Partition the qubits as $ABC$ where $C$ contains a $l_C \times l_C$ square of qubits, and let $\rho_A =\rho_A(U_{AB},x_B)$. Fix $k=O(1)$ and a $k$-copy positive semi-definite observable $O_A \geq 0$. Then with probability at least $1-\varepsilon$, where 
		\begin{equation}
			\varepsilon = \textup{poly}(n)e^{-\Omega(\min(l_C, \textup{dist}(A,C)))},
			\label{eq:scroogeeps}
		\end{equation}
        the following holds:
		(i) the $k$-th moment [Eq.~\eqref{eq:k moment def}] of the post-measurement ensemble $\mathcal{E}_A(U_{AB},x_B)$ is
		$\varepsilon$-close in relative error to that of $\mathcal{S}_\rho$ [Eq.~\eqref{eq:observable_relative_error_def}] (resp. $\mathcal{D}_\rho$), and (ii) the background state $\rho_A$ has min-entropy  $S_\infty(\rho_A) = \Omega(|\partial A|)$.
	\end{hypothesis*}
	\noindent Here, $|X|$ denotes the number of qubits in a subregion $X$, $|\partial X|$ denotes the number of qubits on the boundary of a subregion $X$, and $\text{dist}(X,Y)$ denotes the length of the shortest path between two subregions $X$ and $Y$.
	
	\emph{Intuition and analytic evidence.}---We now present an intuitive argument explaining how this hypothesized behaviour arises in SRQCs, and summarize our analytic evidence based on statistical mechanical calculations. 
	
	After applying $U_{AB}$ and measuring $B$ with outcome $x_B$, the remaining unmeasured qubits are in the state $\ket{0_Y}\otimes \ket{\psi_{AX}(U_{AB},{x_B})}$, where $X \subset C$ contains the qubits in $C$ on which $U_{AB}$ acts, and $Y = C \backslash X$. Note that $\ket{\psi_{AX}(U_{AB},{x_B})}$ is a purification of $\rho_A(U_{AB}, x_B)$. Owing to the characterizations of $\mathcal{S}_\rho$ (resp. $\mathcal{D}_\rho$) provided above, our hypothesis amounts to the statement that the low-order moments of $\mathcal{E}_A(U_{AB}, x_B)$ are approximately unchanged if the operations on $C$ are replaced by a Haar-random (resp.~random Clifford) unitary on $X$, followed by an arbitrary (resp.~stabilizer) rank-1 measurement.
	
	To understand how this arises, imagine applying the operations on $C$ sequentially, as in Ref.~\cite{ippoliti2023dynamical}. Start with a qubit $i_1 \in X$, sample the remaining gates in its backwards lightcone and apply them, and then measure $i_1$ with outcome $x_{i_1}$. Write the resulting state as $\ket{0}_{Y_1}\otimes \ket{\psi_{AX_1}(1)}$, with regions $X_1$ and $Y_1$ updated appropriately. Iterating this process over a chosen order of qubits $(i_t)_{t = 1, \ldots, |C|}$  yields a sequence of states $\ket{\psi_{AX_t}(t)}$ each of which are conditioned on the operations in the backwards lightcone of $AB\cup \{i_1, \ldots, i_t\}$, and the last of which evidently has the distribution $\mathcal{E}_A(x_B, U_{AB})$.

    \begin{figure*}[t]
    \subfloat[]{
        \includegraphics[width=0.35\textwidth]{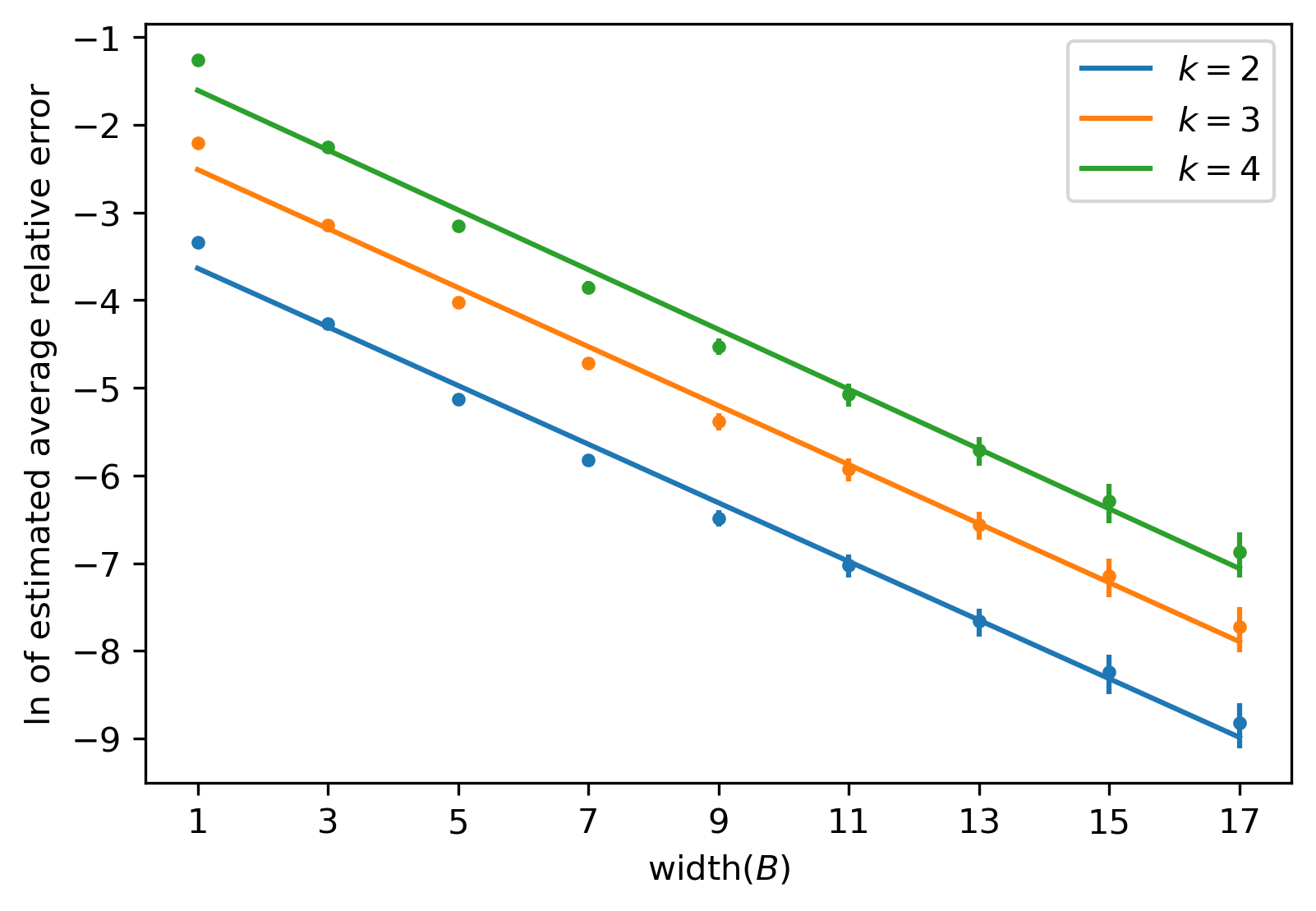}
    }
    \subfloat[]{
        \includegraphics[width=0.35\textwidth]{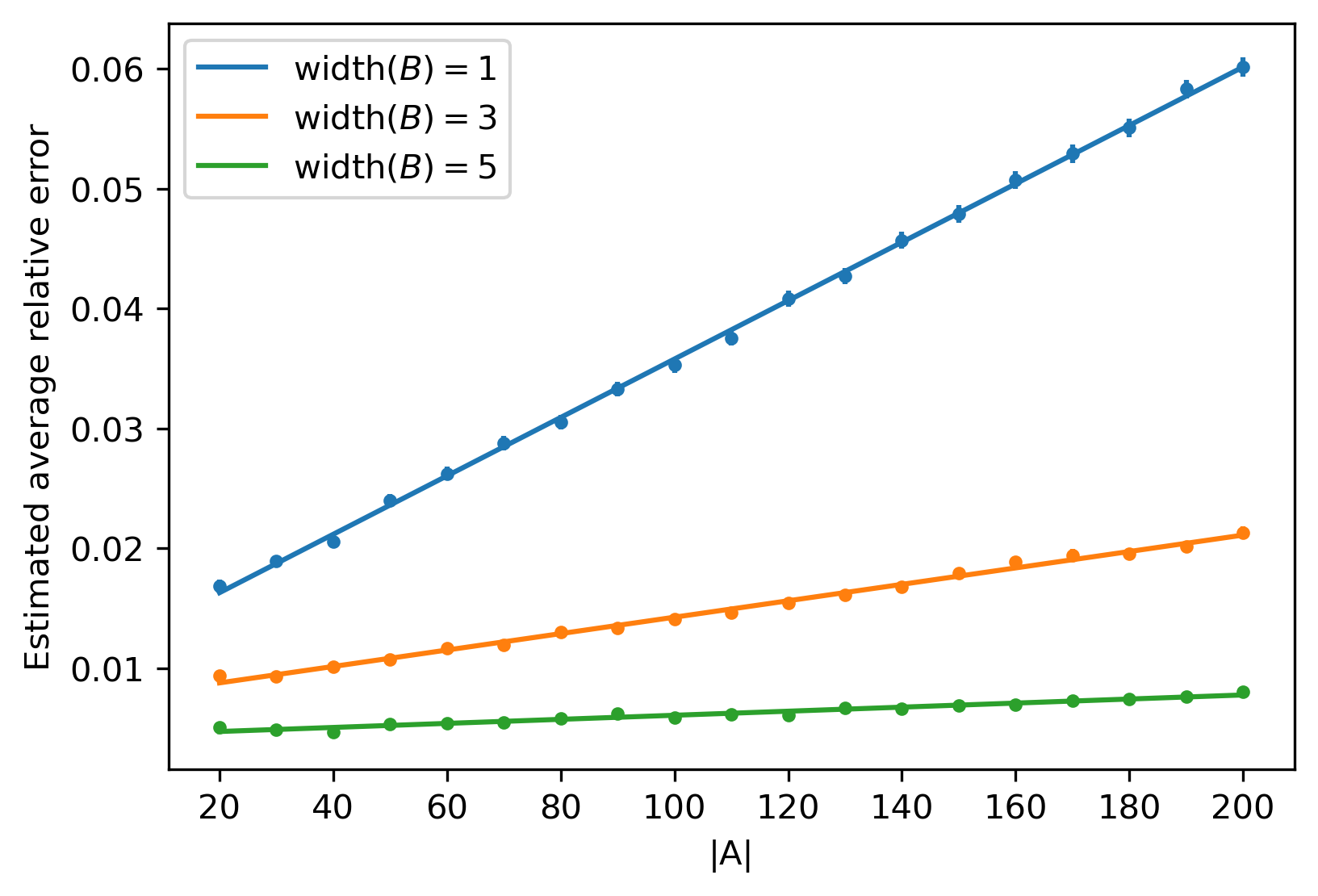}
    }
    \hspace{0cm}
    \subfloat[]{
        \includegraphics[width=0.25\textwidth]{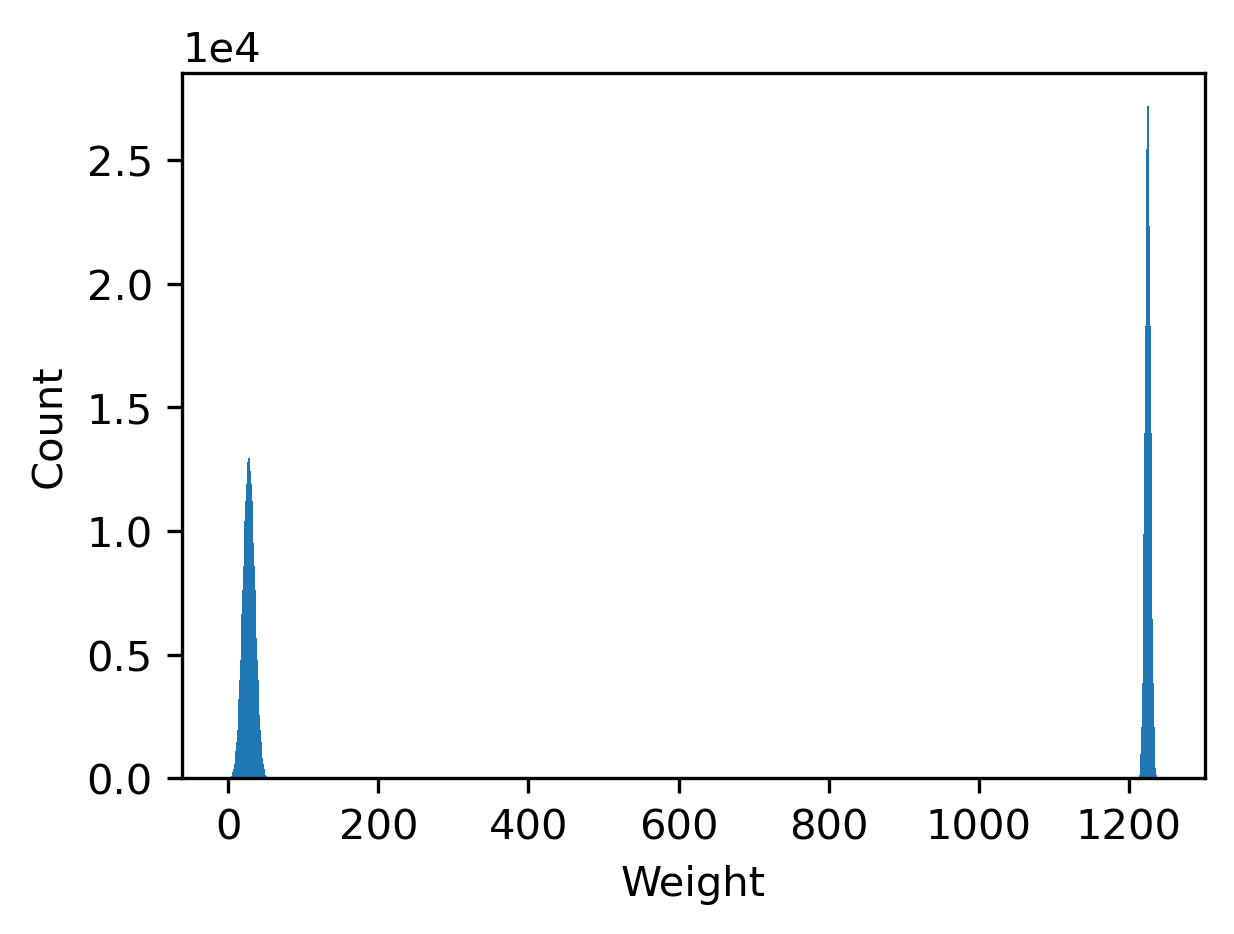}
        \vspace{-14pt}
    }
    \caption{(a,b) Average relative error between the $k$-th diagonal frame potential (DFP) of $\mathcal{E}_A(U_{AB},x_B)$ and $\mathcal{D}_\rho$ as a function of (a) $\text{width}(B)=\text{dist}(A,C)-1$, and (b) $|\partial A|$. Here, $A$ is the top row of qubits,  $C$ is the bottom $6$ rows of qubits, and $B$ is all other qubits; averaging is  over instances of the background state $\rho_A(U_{AB},x_B)$. In (a), we fix $|A|=100$; in (b), we fix $k=2$. Error bars show twice the standard error. (c) Weight distribution of the $Z$-type stabilizers of a depth-$7$ 2D random Clifford circuit acting on a $50\times 50$ grid of qubits. This example shows a bimodal histogram over a total of $2^{19}-1$ non-identity $Z$-type stabilizers, with a low-weight peak of $2^{18}-1$ elements centered at $28$ and a high-weight peak of $2^{18}$ elements centered at $1225$.}
    \label{fig:maintext_plot}
	\end{figure*}
	
	If $\text{dist}(A,C)$ is sufficiently large, we expect that the sequence of marginals $\rho_A(t) \coloneqq \tr_{X_t}[\psi_{AX_t}(t)]$ changes very little in the majority of steps, as quantified by the relative entropy $D(\rho \| \sigma) \coloneqq \tr[\rho(\log \rho - \log \sigma)] \geq 0$. To see why, we observe that
	\begin{align}
		\mathbb{E}[D(\rho_A(t+1)\|\rho_A(t))] = \mathbb{E}[S(\rho_A(t)) - S(\rho_A(t+1))]
		\label{eq:rel ent}
	\end{align}
	where the expectation is over gates and measurement outcomes in step $t+1$. Thus, any change in the marginal states must be accompanied by a proportionate decrease in MIE. Since MIE is long-ranged above the critical depth, and any change in entropy must be non-negative on average, this decrease must be small for all but a negligible fraction of steps. Provided $B$ is large enough such that any transient decay of MIE has subsided, by \eqref{eq:rel ent} we have $\rho_A(t+1)\approx \rho_A(t)$ for the majority of the sequence.
	
	This argument implies that successive states $\ket{\psi_{A X_t}(t)}$ are approximate purifications of the same density matrix $\rho_A(0) \equiv \rho_A(U_{AB}, x_B)$, and therefore related to each other by isometries mapping $X_t$ to $X_{t+1}$. Hence, we can replace (say) the first half of gates and measurements in the sequence by the product of these random isometries without changing the distribution $\mathcal{E}_A(U_{AB}, x_B)$ much. We expect the product of many random isometries to converge to the Haar-measure on $X$, which suffices to generate the Scrooge ensemble on $A$, as explained above.
	
	In the supplementary material, we substantiate this reasoning with an analytical calculation based on mappings between integer moments of the random circuit $U$ and certain classical stat-mech spin models \cite{nahum2017quantum,nahum2018operator}. To deal with the denominator in Eq.~\eqref{eq:ensemble def}, we employ the \textit{replica trick}, where we introduce family of proxy moments  $\chi^{(k,r)}$ for both the post-measurement and Scrooge ensembles, each parametrized by a replica index $r$. These can be mapped to spin models for integer $r \geq 0$, and are constructed such that the correct moments $\chi^{(k)}$ are reproduced in the `replica limit' $r \rightarrow -(k-1)$. This gives us a useful (albeit non-rigorous) technique to probe the behavior of the true moments---one that has proved successful in past studies of post-measurement states \cite{bao2020theory, jian2020measurement, napp2022efficient}.

	We show that the proxy moments for $\mathcal{E}_A(U_{AB}, x_B)$ and $\mathcal{S}_\rho$ each correspond to properties of the spins in $A$ in the same stat-mech model, but with different boundary conditions on $C$. This model features a discrete permutation symmetry, which is respected by both boundary conditions, and becomes spontaneously broken above the phase transition at depth $d^*$. Importantly, there is a finite correlation length in this phase, which controls the extent to which spins in $A$ feel the effect of (symmetric) boundary conditions on $C$. This leads to exponential convergence of the two generalized moments (in relative error), which justifies our hypothesized error Eq.~\eqref{eq:scroogeeps}.
	
	\emph{Numerical evidence for the Scrooge hypothesis in Clifford circuits.}---Let us now return to the mysterious and remarkable agreement between the rightmost two columns in Fig.~\ref{fig:distribution_of_r}(b). We can understand this as a hallmark of the emergence of Scrooge-randomness. Indeed, in Appendix \ref{sec:stabilizer_scrooge_ensemble}, we show that if $A$ is subdivided into regions $A_1A_2$, and $\rho_A = \rho_{A_1} \otimes \rho_{A_2}$ is a sufficiently mixed stabilizer state, then the distribution of the mutual information $I(A_1:A_2)$ in the stabilizer Scrooge ensemble $\mathcal{D}_{\rho_A}$ (corresponding to $I(A;C|B)$ in Fig.~\ref{fig:distribution_of_r}) is well-approximated by the one shown in the rightmost column of Fig.~\ref{fig:distribution_of_r}(b). In this way, the numerically observed CMI distribution is a signature of the Scrooge ensemble.

    We can also utilize Clifford circuits to test the (stabilizer) Scrooge hypothesis more generally, leveraging their efficient numerical simulation.
    As a particular quantitative test,
    we compute the $k$-th diagonal frame potential (DFP) of the post-measurement ensemble $\mathcal{E}_A(U_{AB},x_B)$,
    \begin{equation}
		\text{DFP}^{(k)}(\mathcal{E}) \coloneqq 
		\Exp_{\ket{\psi},\ket{\phi} \sim \mathcal{E}}\bigg[\bigg(\sum_{x\in\{0,1\}^{|A|}}|\langle x|\psi\rangle|^2|\langle x|\phi\rangle|^2\bigg)^k\bigg],
	\end{equation}
    in depth-$7$ 2D brickwork random Clifford circuits. The relative error between the DFP for $\mathcal{E}_A(U_{AB},x_B)$ and for $\mathcal{D}_\rho$ can be used to  bound the typical relative error [Eq.~\eqref{eq:observable_relative_error_def}] for randomly chosen diagonal observables $O$, as described in Appendix \ref{sec:scrooge_numerics}. In Fig. \ref{fig:maintext_plot}, we numerically find that the average relative error in the $k$-th DFP decays exponentially in $\text{width}(B)=\text{dist}(A,C)-1$ (left) and increases mildly in $|\partial A|$ (right), which supports the error scaling predicted in the Scrooge hypothesis [Eq.~\eqref{eq:scroogeeps}]. 

     We can also test more qualitative features of the output distribution $p(x)$. For Clifford $U$, $p(x)$ is determined completely by the $Z$-type stabilizer subgroup of $|\psi\rangle$, i.e., the set $\mathcal{Z}=\{P\in \pm \{I,Z\}^{\otimes n} \, | \, P|\psi\rangle=|\psi\rangle \}$\footnote{In particular, the measured distribution is uniform over an affine space consisting of binary strings $x\in \{0,1\}^n$ such that $P|x\rangle=|x\rangle$ for all $P\in \mathcal{Z}$.}. In Appendix \ref{sec:structure_of_clifford}, we prove that a consequence of the stabilizer Scrooge hypothesis is that $\mathcal{Z}$ admits a generating set containing only two kinds of elements: (i) stabilizers which are each supported on $O(\log(n))$ columns, and (ii) stabilizers which have weight at least $\Omega(n/\log(n))$. We find a similar separation in numerical simulations: the $Z$-type stabilizer subgroup of a typical depth-$7$ 2D  random Clifford circuit admits a generating set consisting of a large number of low-weight, locally-supported generators, and either zero or a small number of high-weight, non-local generators. A representative example is shown in Fig. \ref{fig:maintext_plot}(c), in the case where one high-weight generator occurs. Intuitively, the high-weight $Z$-type stabilizers are responsible for all long-range conditional correlations in the output distribution, and their number is equal to the CMI, $I(A;C|B) = r$, in Fig.~\ref{fig:distribution_of_r}. We also observe their remarkable fragility to noise: the $Z$-type stabilizer subgroup of a reduced state typically admits a generating set comprised solely of low-weight stabilizers after tracing out only a small number of qubits (Appendix \ref{sec:Z_type_stablizer_numerics}).  These results provide further evidence and intuition for the Scrooge hypothesis in the stabilizer setting.
     
	\emph{Implications for noisy SRQC sampling.}---We now show that the Scrooge hypothesis implies an efficient classical simulation of SRQCs in the presence of a vanishing amount of depolarizing noise.
	
	The key property underlying our classical  algorithm is stated in the following lemma. In particular, consider a tripartition $[n]=ABC$ where $B$ shields $A$ from $C$ so that $\mathrm{dist}(A,C)$ is sufficiently large.   Suppose further that we subdivide $A=A_1A_2$ for some sufficiently large region $A_2$ which we trace out. Then, the conditional distribution $p_{A_1|BC}$ is well approximated by $p_{A_1|B}$ as long as $A_1$ is not too big. Roughly speaking, this says that after we marginalize over a sufficiently large region $A_2$, conditional distributions can be computed locally. Intuitively, this follows because the non-local correlations in the Scrooge ensemble are fully global, and hence vanish after tracing out $A_2$. We note that in certain geometries, including the column-by-column one used in our algorithm, the exponential prefactor in $|A_1|$ can be avoided; see the end matter for details.
	\begin{restatable}{lemma}{marginal}
		Suppose the Scrooge hypothesis holds. Consider a depth $d>d^{*}$ 2D brickwork circuit $U$ that acts on a $\sqrt{n}\times \sqrt{n}$ grid of qubits and in which each two-qubit gate is Haar random. Suppose the qubits are partitioned as $[n]=ABC$ where $A=A_1A_2$, and $C$ contains an $l_C\times l_C$ square of qubits. Then
		\begin{align}
			\mathbb{E}_{U}& \mathbb{E}_{ x_Bx_C\sim p_{BC}} \sum_{x_{A_1}}\big|p_{A_1|BC}(x_{A_1}|x_Bx_C)-p_{A_1|B}(x_{A_1}|x_B)\big|\nonumber\\
			& \quad \quad \leq 2^{|A_1|/2}(\mathrm{poly}(n) e^{-\Omega(\min\{l_C,\mathrm{dist}(A,C)\})} \nonumber\\
            &\quad \quad \quad \quad \quad \quad \quad \quad \quad \quad+e^{-\Omega(\min\{|\partial A_2|,|\partial{A}|\})}).
			\label{eq:tracedist}
		\end{align}
		\label{lem:tracecond}
	\end{restatable}
    \vspace{-8mm}
	\noindent The proof of Lemma~\ref{lem:tracecond} is provided in Appendix \ref{app:reducedmoments}.
	
	The lemma suggests that classical algorithms that simulate local dependencies of the output distribution, such as those in Ref.~\cite{napp2022efficient}, will succeed as long as enough qubits are traced out. Indeed, using Lemma~\ref{lem:tracecond}, we prove that a simple ``column-by-column" classical  algorithm succeeds if after each layer of gates every qubit undergoes depolarizing noise with rate $\gamma=\Omega(\log(n)/n))$.
	\begin{restatable}{theorem}{classical} \label{thm:classical-algorithm}
		Assume the Scrooge hypothesis holds.
		For any  $\epsilon=1/\mathrm{poly}(n)$ and depolarizing noise rate $\gamma = \Omega\big( \frac{\log(n/\epsilon)}{n} \big)$, 
		there exists a classical algorithm with runtime $\mathrm{poly}(n)$ which approximately samples from  the output distribution of a  noisy constant-depth two-dimensional random circuit with depth $d>d^{*}$. Here $\epsilon$ is the expected total variation distance between the classically-sampled distribution and true distribution.
	\end{restatable}
    \noindent See the end matter for a detailed description and proof.
	
	Taken together with the arguments from Ref.~\cite{napp2022efficient}, this clarifies our picture of the complexity of 2D SRQCs: above the critical depth $d^{*}$ they are classically hard due to long-range Scrooge-like conditional correlations, but these can be washed out by a vanishing amount of noise, rendering classical simulation easy. In the opposite regime $d\leq d^*$ long-range MIE is believed to be absent and known algorithms such as the one from  Ref.~\cite{napp2022efficient} are thought to provide an efficient classical simulation.
	
	\textit{Acknowledgements.}---M.M.~acknowledges support from Trinity College, Cambridge. T.S. acknowledges support from the Walter Burke Institute for Theoretical Physics at Caltech, and from the U.S. Department of Energy, Office of Science, National Quantum Information Science Research Centers, Quantum Systems Accelerator. The Institute for Quantum Information and Matter is an NSF Physics Frontiers Center. DG acknowledges the support of the Natural Sciences and Engineering Research Council of Canada through grant numbers RGPIN-2019-04198 and RGPIN-2026-04480. We acknowledge the use of generative AI to suggest proof strategies and as an aid in proofreading.

	\let\oldaddcontentsline\addcontentsline
	\renewcommand{\addcontentsline}[3]{}
	\bibliography{references}

@article{arute2019quantum,
  title={Quantum supremacy using a programmable superconducting processor},
  author={Arute, Frank and Arya, Kunal and Babbush, Ryan and Bacon, Dave and Bardin, Joseph C and Barends, Rami and Biswas, Rupak and Boixo, Sergio and Brandao, Fernando GSL and Buell, David A and others},
  journal={Nature},
  volume={574},
  number={7779},
  pages={505--510},
  year={2019},
  publisher={Nature Publishing Group UK London}
}

@article{aliferis2007accuracy,
  title={Accuracy threshold for postselected quantum computation},
  author={Aliferis, Panos and Gottesman, Daniel and Preskill, John},
  journal={arXiv preprint quant-ph/0703264},
  year={2007}
}

@article{bao2024finite,
  title={Finite-time teleportation phase transition in random quantum circuits},
  author={Bao, Yimu and Block, Maxwell and Altman, Ehud},
  journal={Physical Review Letters},
  volume={132},
  number={3},
  pages={030401},
  year={2024},
  publisher={APS}
}

@article{schuster2026fast,
  title={Fast chaos and emergent classicality in charge-conserving random circuits},
  author={Schuster, Thomas and Suzuki, Ryotaro and Mitsuhashi, Yosuke and McGinley, Max and Vardhan, Shreya and Preskill, John},
  journal={Forthcoming},
  year={2026}
}

@article{schuster2025strong,
  title={Strong random unitaries and fast scrambling},
  author={Schuster, Thomas and Ma, Fermi and Lombardi, Alex and Brandao, Fernando and Huang, Hsin-Yuan},
  journal={arXiv preprint arXiv:2509.26310},
  year={2025}
}

@article{ippoliti2023dynamical,
  title = {Dynamical Purification and the Emergence of Quantum State Designs from the Projected Ensemble},
  author = {Ippoliti, Matteo and Ho, Wen Wei},
  journal = {PRX Quantum},
  volume = {4},
  issue = {3},
  pages = {030322},
  numpages = {28},
  year = {2023},
  month = {Aug},
  publisher = {American Physical Society},
  doi = {10.1103/PRXQuantum.4.030322},
  url = {https://link.aps.org/doi/10.1103/PRXQuantum.4.030322}
}

@article{brandao2016local,
  title={Local random quantum circuits are approximate polynomial-designs},
  author={Brandao, Fernando GSL and Harrow, Aram W and Horodecki, Micha{\l}},
  journal={Communications in Mathematical Physics},
  volume={346},
  number={2},
  pages={397--434},
  year={2016},
  publisher={Springer}
}

@article{napp2022efficient,
  title={Efficient classical simulation of random shallow 2D quantum circuits},
  author={Napp, John C and La Placa, Rolando L and Dalzell, Alexander M and Brandao, Fernando GSL and Harrow, Aram W},
  journal={Physical Review X},
  volume={12},
  number={2},
  pages={021021},
  year={2022},
  publisher={APS}
}

@article{jian2020measurement,
  title = {Measurement-induced criticality in random quantum circuits},
  author = {Jian, Chao-Ming and You, Yi-Zhuang and Vasseur, Romain and Ludwig, Andreas W. W.},
  journal = {Phys. Rev. B},
  volume = {101},
  issue = {10},
  pages = {104302},
  numpages = {11},
  year = {2020},
  month = {Mar},
  publisher = {American Physical Society},
  doi = {10.1103/PhysRevB.101.104302},
  url = {https://link.aps.org/doi/10.1103/PhysRevB.101.104302}
}

@article{bao2020theory,
  title = {Theory of the phase transition in random unitary circuits with measurements},
  author = {Bao, Yimu and Choi, Soonwon and Altman, Ehud},
  journal = {Phys. Rev. B},
  volume = {101},
  issue = {10},
  pages = {104301},
  numpages = {26},
  year = {2020},
  month = {Mar},
  publisher = {American Physical Society},
  doi = {10.1103/PhysRevB.101.104301},
  url = {https://link.aps.org/doi/10.1103/PhysRevB.101.104301}
}

@article{nahum2017quantum,
  title = {Quantum Entanglement Growth under Random Unitary Dynamics},
  author = {Nahum, Adam and Ruhman, Jonathan and Vijay, Sagar and Haah, Jeongwan},
  journal = {Phys. Rev. X},
  volume = {7},
  issue = {3},
  pages = {031016},
  numpages = {30},
  year = {2017},
  month = {Jul},
  publisher = {American Physical Society},
  doi = {10.1103/PhysRevX.7.031016},
  url = {https://link.aps.org/doi/10.1103/PhysRevX.7.031016}
}

@article{gullans2020dynamical,
  title = {Dynamical Purification Phase Transition Induced by Quantum Measurements},
  author = {Gullans, Michael J. and Huse, David A.},
  journal = {Phys. Rev. X},
  volume = {10},
  issue = {4},
  pages = {041020},
  numpages = {28},
  year = {2020},
  month = {Oct},
  publisher = {American Physical Society},
  doi = {10.1103/PhysRevX.10.041020},
  url = {https://link.aps.org/doi/10.1103/PhysRevX.10.041020}
}

@misc{zhang2025classically,
      title={Classically Sampling Noisy Quantum Circuits in Quasi-Polynomial Time under Approximate Markovianity}, 
      author={Yifan F. Zhang and Su-un Lee and Liang Jiang and Sarang Gopalakrishnan},
      year={2025},
      eprint={2510.06324},
      archivePrefix={arXiv},
      primaryClass={quant-ph},
      url={https://arxiv.org/abs/2510.06324}, 
}

@misc{kueng2015qubitstabilizerstatescomplex,
      title={Qubit stabilizer states are complex projective 3-designs}, 
      author={Richard Kueng and David Gross},
      year={2015},
      eprint={1510.02767},
      archivePrefix={arXiv},
      primaryClass={quant-ph},
      url={https://arxiv.org/abs/1510.02767}, 
}

@article{marcus1963permanent,
  title={The permanent analogue of the Hadamard determinant theorem},
  author={Marcus, Marvin},
  journal={Bull. Amer. Math. Soc.},
  volume={69},
  number={4},
  pages={494--496},
  year={1963}
}

@article{goldstein2006distribution,
title={On the distribution of the wave function for systems in thermal equilibrium},
author={Goldstein, Sheldon and Lebowitz, Joel L and Tumulka, Roderich and Zanghi, Nino},
journal={Journal of statistical physics},
volume={125},
number={5},
pages={1193--1221},
year={2006},
publisher={Springer},
doi={10.1007/s10955-006-9210-z}
}

@article{chang2025deep,
  title = {Deep Thermalization under Charge-Conserving Quantum Dynamics},
  author = {Chang, Rui-An and Shrotriya, Harshank and Ho, Wen Wei and Ippoliti, Matteo},
  journal = {PRX Quantum},
  volume = {6},
  issue = {2},
  pages = {020343},
  numpages = {36},
  year = {2025},
  month = {Jun},
  publisher = {American Physical Society},
  doi = {10.1103/PRXQuantum.6.020343},
  url = {https://link.aps.org/doi/10.1103/PRXQuantum.6.020343}
}

@book{bollobas2006art,
  title={The art of mathematics: Coffee time in Memphis},
  author={Bollob{\'a}s, B{\'e}la},
  year={2006},
  publisher={Cambridge University Press}
}

@book{friedli2017, place={Cambridge}, title={Statistical Mechanics of Lattice Systems: A Concrete Mathematical Introduction}, publisher={Cambridge University Press}, author={Friedli, Sacha and Velenik, Yvan}, year={2017}}

@article{mark2024maximum,
  title = {Maximum Entropy Principle in Deep Thermalization and in Hilbert-Space Ergodicity},
  author = {Mark, Daniel K. and Surace, Federica and Elben, Andreas and Shaw, Adam L. and Choi, Joonhee and Refael, Gil and Endres, Manuel and Choi, Soonwon},
  journal = {Phys. Rev. X},
  volume = {14},
  issue = {4},
  pages = {041051},
  numpages = {49},
  year = {2024},
  month = {Nov},
  publisher = {American Physical Society},
  doi = {10.1103/PhysRevX.14.041051},
  url = {https://link.aps.org/doi/10.1103/PhysRevX.14.041051}
}

@article{cotler2021emergent,
  title = {Emergent Quantum State Designs from Individual Many-Body Wave Functions},
  author = {Cotler, Jordan S. and Mark, Daniel K. and Huang, Hsin-Yuan and Hern\'andez, Felipe and Choi, Joonhee and Shaw, Adam L. and Endres, Manuel and Choi, Soonwon},
  journal = {PRX Quantum},
  volume = {4},
  issue = {1},
  pages = {010311},
  numpages = {29},
  year = {2023},
  month = {Jan},
  publisher = {American Physical Society},
  doi = {10.1103/PRXQuantum.4.010311},
  url = {https://link.aps.org/doi/10.1103/PRXQuantum.4.010311}
}

@article{collins,
  title={The weingarten calculus},
  author={Collins, Benoit and Matsumoto, Sho and Novak, Jonathan},
  journal={Notices of the American Mathematical Society},
  volume={69},
  number={05},
  pages={1},
  year={2022},
  publisher={American Mathematical Society (AMS)}
}

@article{goldstein2016universal,
title={Universal probability distribution for the wave function of a quantum system entangled with its environment},
author={Goldstein, Sheldon and Lebowitz, Joel L and Mastrodonato, Christian and Tumulka, Roderich and Zangh{\`\i}, Nino},
journal={Communications in Mathematical Physics},
volume={342},
number={3},
pages={965--988},
year={2016},
publisher={Springer},
doi={10.1007/s00220-015-2536-0}
}

@article{heinrich2025anti,
  title={Anti-concentration is (almost) all you need},
  author={Heinrich, Markus and Haferkamp, Jonas and Roth, Ingo and Helsen, Jonas},
  journal={arXiv preprint arXiv:2510.23719},
  year={2025}
}

@article{cui2025unitary,
  title={Unitary designs in nearly optimal depth},
  author={Cui, Laura and Schuster, Thomas and Brandao, Fernando and Huang, Hsin-Yuan},
  journal={arXiv preprint arXiv:2507.06216},
  year={2025}
}

@article{schuster2025random,
  title={Random unitaries in extremely low depth},
  author={Schuster, Thomas and Haferkamp, Jonas and Huang, Hsin-Yuan},
  journal={Science},
  volume={389},
  number={6755},
  pages={92--96},
  year={2025},
  publisher={American Association for the Advancement of Science}
}

@article{dalzell2022random,
  title={Random quantum circuits anticoncentrate in log depth},
  author={Dalzell, Alexander M and Hunter-Jones, Nicholas and Brand{\~a}o, Fernando GSL},
  journal={PRX Quantum},
  volume={3},
  number={1},
  pages={010333},
  year={2022},
  publisher={APS}
}

@article{holevo1973bounds,
  title={Bounds for the quantity of information transmitted by a quantum communication channel},
  author={Holevo, Alexander Semenovich},
  journal={Problemy Peredachi Informatsii},
  volume={9},
  number={3},
  pages={3--11},
  year={1973},
  publisher={Russian Academy of Sciences, Branch of Informatics, Computer Equipment and~…}
}

@article{ippoliti2022solvable,
  doi = {10.22331/q-2022-12-29-886},
  url = {https://doi.org/10.22331/q-2022-12-29-886},
  title = {Solvable model of deep thermalization with distinct design times},
  author = {Ippoliti, Matteo and Ho, Wen Wei},
  journal = {{Quantum}},
  issn = {2521-327X},
  publisher = {{Verein zur F{\"{o}}rderung des Open Access Publizierens in den Quantenwissenschaften}},
  volume = {6},
  pages = {886},
  month = dec,
  year = {2022}
}

@article{jozsa94,
  title={Lower bound for accessible information in quantum mechanics},
  author={Jozsa, Richard and Robb, Daniel and Wootters, William K},
  journal={Physical Review A},
  volume={49},
  number={2},
  pages={668},
  year={1994},
  publisher={APS}
}

@article{harrow2023tdesign,
  title={Approximate unitary t-designs by short random quantum circuits using nearest-neighbor and long-range gates},
  author={Harrow, Aram W and Mehraban, Saeed},
  journal={Communications in Mathematical Physics},
  volume={401},
  number={2},
  pages={1531--1626},
  year={2023},
  publisher={Springer}
}

@article{nahum2018operator,
  title={Operator spreading in random unitary circuits},
  author={Nahum, Adam and Vijay, Sagar and Haah, Jeongwan},
  journal={Physical Review X},
  volume={8},
  number={2},
  pages={021014},
  year={2018},
  publisher={APS}
}

@article{google2025observation,
  title={Observation of constructive interference at the edge of quantum ergodicity},
  journal={Nature},
  volume={646},
  number={8086},
  pages={825--830},
  author={Google Quantum AI and Collaborators},
  year={2025},
  publisher={Nature Publishing Group UK London}
}

@article{harrow2023approximate,
  title={Approximate orthogonality of permutation operators, with application to quantum information},
  author={Harrow, Aram W},
  journal={Letters in Mathematical Physics},
  volume={114},
  number={1},
  pages={1},
  year={2023},
  publisher={Springer}
}

@article{bene2025quantum,
  title={Quantum advantage from measurement-induced entanglement in random shallow circuits},
  author={Bene Watts, Adam and Gosset, David and Liu, Yinchen and Soleimanifar, Mehdi},
  journal={PRX Quantum},
  volume={6},
  number={1},
  pages={010356},
  year={2025},
  publisher={APS}
}

@article{mcginley2025measurement,
  title={Measurement-induced entanglement and complexity in random constant-depth 2D quantum circuits},
  author={McGinley, Max and Ho, Wen Wei and Malz, Daniel},
  journal={Physical Review X},
  volume={15},
  number={2},
  pages={021059},
  year={2025},
  publisher={APS}
}

@article{mcginley2025scrooge,
  title={The Scrooge ensemble in many-body quantum systems},
  author={McGinley, Max and Schuster, Thomas},
  journal={arXiv preprint arXiv:2511.17172},
  year={2025}
}

@article{bravyi2016improved,
  title={Improved classical simulation of quantum circuits dominated by Clifford gates},
  author={Bravyi, Sergey and Gosset, David},
  journal={Physical review letters},
  volume={116},
  number={25},
  pages={250501},
  year={2016},
  publisher={APS}
}

@article{gidney2021stim,
  doi = {10.22331/q-2021-07-06-497},
  url = {https://doi.org/10.22331/q-2021-07-06-497},
  title = {Stim: a fast stabilizer circuit simulator},
  author = {Gidney, Craig},
  journal = {{Quantum}},
  issn = {2521-327X},
  publisher = {{Verein zur F{\"{o}}rderung des Open Access Publizierens
                in den Quantenwissenschaften}},
  volume = 5,
  pages = 497,
  month = jul,
  year = 2021
}

@article{nelson2025limitations,
  title={Limitations of Noisy Geometrically Local Quantum Circuits},
  author={Nelson, Jon and Rajakumar, Joel and Gullans, Michael J},
  journal={arXiv preprint arXiv:2510.06346},
  year={2025}
}

@inproceedings{aharonov2023polynomial,
  title={A polynomial-time classical algorithm for noisy random circuit sampling},
  author={Aharonov, Dorit and Gao, Xun and Landau, Zeph and Liu, Yunchao and Vazirani, Umesh},
  booktitle={Proceedings of the 55th Annual ACM Symposium on Theory of Computing},
  pages={945--957},
  year={2023}
}

@article{lee2025classical,
  title={Classical simulation of noisy random circuits from exponential decay of correlation},
  author={Lee, Su-un and Ghosh, Soumik and Oh, Changhun and Noh, Kyungjoo and Fefferman, Bill and Jiang, Liang},
  journal={arXiv preprint arXiv:2510.06328},
  year={2025}
}

@article{wei2025measurement,
  title={Measurement-induced entanglement in noisy 2D random Clifford circuits},
  author={Wei, Zhi-Yuan and Nelson, Jon and Rajakumar, Joel and Cruz, Esther and Gorshkov, Alexey V and Gullans, Michael J and Malz, Daniel},
  journal={arXiv preprint arXiv:2510.12743},
  year={2025}
}

@article{nelson2024polynomial,
  title={Polynomial-time classical simulation of noisy circuits with naturally fault-tolerant gates},
  author={Nelson, Jon and Rajakumar, Joel and Hangleiter, Dominik and Gullans, Michael J},
  journal={arXiv preprint arXiv:2411.02535},
  year={2024}
}

@article{cheng2023efficient,
  title={Efficient sampling of noisy shallow circuits via monitored unraveling},
  author={Cheng, Zihan and Ippoliti, Matteo},
  journal={PRX Quantum},
  volume={4},
  number={4},
  pages={040326},
  year={2023},
  publisher={APS}
}

@article{mok2026nature,
  title={Nature is stingy: Universality of Scrooge ensembles in quantum many-body systems},
  author={Mok, Wai-Keong and Haug, Tobias and Ho, Wen Wei and Preskill, John},
  journal={arXiv preprint arXiv:2601.00266},
  year={2026}
}
	\let\addcontentsline\oldaddcontentsline
	
	\newpage
	
	\appendix
	\onecolumngrid
	
	\let\oldaddcontentsline\addcontentsline
	\renewcommand{\addcontentsline}[3]{}
	\section*{End Matter}
	\let\addcontentsline\oldaddcontentsline
	
	We now present the proof of Theorem \ref{thm:classical-algorithm}.
	\begin{proof}[Proof of Theorem \ref{thm:classical-algorithm}]
		We shall in fact establish something slightly stronger: an efficient approximate classical simulation even if the depolarizing noise of rate $\gamma$ is \textit{only} applied to all qubits after the last circuit layer (i.e., immediately before measurement). This implies Theorem \ref{thm:classical-algorithm} because we can simulate the depolarizing channels that appear in the middle of the circuit by replacing them with random single-qubit unitaries from the set $\{I,X,Y,Z\}$. For concreteness, we assume the target precision and noise rate satisfy $\epsilon=n^{-\alpha}$ and $\gamma= \beta \log(n/\epsilon)/n$ for absolute constants $\alpha,\beta$.
        
        Our algorithm (Algorithm 1) samples a bitstring $x$ as follows. First, we build a subset $S\subseteq [n]$ by including each qubit $j$ in the subset with probability $\gamma$ independently. Then, we approximately sample a bitstring $x_T\in \{0,1\}^{|T|}$ from the marginal distribution on qubits in $T=[n]\setminus S$,
		\begin{equation}
			p_T(x_T)=\langle x_T|\mathrm{Tr}_{S}\left(U|0^n\rangle \langle 0^n|U^{\dagger}\right)|x_T\rangle.
			\label{eq:pt}
		\end{equation}
        Finally, we append a string of $|S|$ uniformly random bits to obtain the desired sample $x=x_Sx_T$.
		
		The only nontrivial step  above is approximately sampling from the distribution $p_T$. To perform this step, we first argue that $S$ satisfies three elementary conditions with high probability, when $\beta$ is above a constant threshold value.
		
		First, we note $S$ has expected size $\mathbb{E}[|S|]=\gamma n=\beta\log(n/\epsilon)$. Using the (multiplicative) Chernoff bound, we have 
		\begin{equation}
			\mathrm{Pr}\left[||S|-\beta \log(n/\epsilon)|\geq R\beta\log(n/\epsilon)\right]\leq 2e^{-R^2\beta\log(n/\epsilon)/3},
		\end{equation}
		for any $R\in [0,1]$.
		We choose $R=1/2$ and require $\beta$ to be a constant such that the RHS is at most $\epsilon/6$.
        
        Second, let $G'$ be the subgrid consisting of all columns except the rightmost one, let $S'$ be all qubits in $S$ that are in $G'$, and let $\partial S'$ be the boundary of $S'$ (in $G'$). Then by a  similar probabilistic argument to the above, one can choose $\beta$ to be a sufficiently large constant such that $\mathrm{Pr}[|\partial S'|\geq (1/4)\beta \log(n/\epsilon)]\geq 1-\epsilon/6$ \footnote{Let $M$ be any perfect matching of $G'$. For each edge $e\in M$, the probability that $e$ has exactly one endpoint in $S'$ is $2\gamma(1-\gamma)$. Let $F\subseteq M$ be the set of edges in $M$ which have exactly one endpoint in $S'$. Then we have $\mathbb{E}[|\partial S'|]\geq \mathbb{E}[|F|]=2\gamma(1-\gamma)|V|/2$, where $|V|= n-\sqrt{n}$ is the number of vertices in $G'$. For $n$ sufficiently large we have $\gamma\leq 1/2$ and therefore $\mathbb{E}[|\partial S'|]\geq \mathbb{E}[|F|]\geq\gamma (n-\sqrt{n})/2$. Moreover, whether or not each $e\in M$ satisfies $e\in F$ is independent of all others. We can therefore use Chernoff to bound the tail.}. 
        
        Third, for any qubit $i\in [n]$, let $Q_i$ consist of all qubits $j$ such that $\mathrm{dist}(i,j)\leq w^*$.  Let $r=C(w^{*}+d)$ for an absolute constant $C$ chosen such that the lightcones of $Q_i,Q_j$ are disjoint whenever $\mathrm{dist}(i,j)>r$. Let $G_r$ be a graph with vertex set $[n]$ and an edge between two vertices $i,j$ if and only if $\mathrm{dist}(i,j)\leq r$. Let us say that a subset $X\subseteq [n]$ is $r$-connected iff $X$ is connected in $G_r$. Then we may choose $q=O(1)$ (depending only on $\alpha,\beta$) such that
        \[
        \mathrm{Pr}[X\subseteq S \text{ for some $r$-connected set $X$ of size } |X|=q]\leq \epsilon/6.
        \]
        Indeed, the number of $r$-connected sets in $G_r$ of size $q$ is at most
        \[
        n(Dr^2)^{q-1}=n(\mathrm{polylog}(n))^{2(q-1)}
        \]
        for some constant $D$ (see, e.g., Lemma 5 in Ref.~\cite{aliferis2007accuracy}), while the probability that any one of them is contained in $S$ is $(\beta \log(n/\epsilon)/n)^q$.
        By a union bound the probability that $S$ contains any $r$-connected set of size $q$ is then  upper bounded as $O((\mathrm{poly}(\log(n/\epsilon))/n)^{q-1})$ which can be upper bounded by $\epsilon/6$ by taking $q$ to be a sufficiently large constant.
        
        By a union bound, all three of these conditions occur with probability at least $1-\epsilon/2$. It then suffices to show that we can sample from a distribution $\tilde{p}_T$ whose expected total variation distance to Eq.~\eqref{eq:pt} is at most $\epsilon/2$, assuming: (1) $(1/2)\beta\log(n/\epsilon)\leq |S|\leq (3/2)\beta\log(n/\epsilon)$,  (2) $|\partial S'|\geq (1/4)\beta \log(n/\epsilon)$,  (3) $S$ contains no $r$-connected subset of size $q=O(1)$. 
		
		To this end, we use Algorithm 1 [Fig.~\ref{fig:alg}(a)].  The algorithm  approximately samples from the output distribution of a two-dimensional random circuit by proceeding column-by-column from left to right. The first step of the algorithm is to  sample from the marginal distribution of an initial set $I$ consisting of all qubits in $[n]\setminus S$ that lie in the first $L-1=O(\log(n))$ columns. Then, for each subsequent column $j=L,L+1,\ldots, \sqrt{n}$ we sample the qubits in the $j$-th column (that are not in $S$) from a certain conditional distribution. 
		
		To describe the $j$-th step in more detail, it is helpful to define four subsets of qubits, denoted $A_1(j),A_2(j),B(j),C(j)$ (see Fig.~\ref{fig:alg}). 
        The subset $A_1(j)$ consists of all qubits in column $j$ that are not in $S$. $A_2(j)$ consists of all qubits in columns $\geq j+1$ as well as all qubits in $S$. $B(j)$ contains all qubits in the $w^{*}$ columns preceding $j$ that are not in $S$. $B(j)$ also contains all qubits in columns $1,2,\ldots, j-1$ that are within a $w^{*}$ distance of any qubit in $S$. Finally, $C(j)$ contains all qubits in $[n]\setminus S$ that are not in $A_1(j)\cup A_2(j)\cup B(j)$.

        \begin{figure}[t]
		\subfloat[]{
			\begin{minipage}[b]{0.5\textwidth}
				\begin{algorithm}[H]
					\begin{flushleft}
						\caption{Column-by-column sampling}
						\textbf{Input}: Shallow 2D random circuit $U$ on $\sqrt{n}\times \sqrt{n}$ grid of qubits, a set $S\subseteq [n]$ of size $|S|=\Omega(\log(n/\epsilon))$, and two constants $c_1,c_2>0.$\\
						\textbf{Output}:  A binary string $x\in \{0,1\}^{|T|}$, where $T=[n]\setminus S$\\
						\begin{algorithmic}[1]
							\State{$w^* \gets \lceil c_1\log(n/\epsilon)\rceil$}
							\State{$L \gets \lceil c_2\log(n/\epsilon)\rceil $}
							\State{Let $I$ be the set of all qubits in $T$ that lie in the $L-1$ leftmost columns of the grid.}
							\State{Sample $x_I$ from $p_{I}(x_I)$}
							\For{$j$ from $L$ to $\sqrt{n}$}
							\State{$A_1,A_2,B,C \gets A_1(j), A_2(j),B(j),C(j)$ }
							\State{Sample $x_{A_1}$ from $p_{A_1|B}(x_{A_1}|x_B)$}
							\EndFor
							
						\end{algorithmic}
					\end{flushleft}
				\end{algorithm}
			\end{minipage}
		}
		\hspace{0.2cm}
		\subfloat[]{
			\begin{minipage}[b]{0.45\textwidth}
				\begin{tikzpicture}[scale=0.55]
                
					\fill[red, fill opacity=0.7](7.5,0) rectangle (8,10);
					\fill[gray](8,0) rectangle (10,10);
					\fill[yellow] (6,0) rectangle (7.5,10);
					\fill[blue, fill opacity=0.2](-2,0) rectangle (6,10);
					
					\filldraw[yellow] (3,3) circle (25pt);
					\filldraw[gray] (3,3) circle (3pt);

					\filldraw[yellow] (2,8) circle (25pt);
					\filldraw[gray] (2,8) circle (3pt);
					\filldraw[gray] (7.75,1) circle (3pt);

					\filldraw[yellow] (5.5,6) circle (25pt);
					\filldraw[gray] (5.5,6) circle (3pt);

					\filldraw[yellow] (6.5,9) circle (25pt);
					\filldraw[gray] (6.5,9) circle (3pt);
					\draw[<->] (-2,-0.5)--(7.5,-0.5);
					\node at (2.75,-1) {$j-1$};
					
					\draw[->](2,8)--(2.88,8.2);
					\node at (2.4,8.35) {$w^{\star}$};
					\draw[<->](6,10.5)--(7.5,10.5);
					\node at (6.8,11) {$w^{\star}$};
					\draw[<->] (7.5,10.5)--(8,10.5);
					\node at (7.75,11) {$1$};
					
					\begin{scope}[shift={(-12.5,-7.5)}]
						
						\draw[black,fill=white] (10.75,8) rectangle (13,10.75);
						\node at (11.7,10.35){Legend};
						\filldraw[red] (11,9.75) rectangle (12,10);
						\node at (12.5,9.85) {$A_1$};
						\filldraw[gray] (11,9.25) rectangle (12,9.5);
						\node at (12.5,9.35) {$A_2$};
						\filldraw[yellow] (11,8.75) rectangle (12,9);
						\node at (12.5,8.85) {$B$};
						\fill[blue,fill opacity=0.2] (11,8.25) rectangle (12,8.5);
						\node at (12.5,8.35) {$C$};
						;
					\end{scope}
				\end{tikzpicture}
			\end{minipage}
		}
		\caption{The sets $A_1(j),A_2(j),B(j),C(j)$ used in step $j$ of Algorithm 1 (see line 6 of (a)) are depicted in (b) and described in the text. We show that for a given $\epsilon=1/\mathrm{poly}(n)$ it is possible to choose constants $c_1,c_2=O(1)$ such that Algorithm 1 samples from a distribution with expected total variation distance at most $\epsilon/2$ from $p_T$ (cf. Eq.~\eqref{eq:pt}).\label{fig:alg}}
	\end{figure}
		
		In the $j$-th step of the algorithm, we sample all bits in region $A_1(j)$ from the conditional distribution $p_{A_1(j)|B(j)}(x_{A_1(j)}|x_{B(j)})$.
		The overall distribution sampled by Algorithm 1 is therefore,
		\begin{equation}
			\tilde{p}_T(x)=p_{I}(x_{I})\prod_{j=L}^{\sqrt{n}} p_{A_1(j)|B(j)}(x_{A_1(j)}|x_{B(j)}),
		\end{equation}
		while the distribution we are trying to approximately sample can be expressed as,
		\begin{equation}
			p_T(x)=p_{I}(x_{I})\prod_{j=L}^{\sqrt{n}} p_{A_1(j)|B(j)C(j)}(x_{A_1(j)}|x_{B(j)}x_{C(j)}).
		\end{equation}
		The expected total variation distance between the two distributions can be bounded column-by-column via
		\begin{equation}
			\Exp_{U} \big\|p_T-\tilde{p}_T\big\|_1\leq \sum_{j=L}^{\sqrt{n}} \Exp_U \Exp_{x_{B(j)}x_{C(j)}}
            \big\|p_{A_1(j)|B(j)}-p_{A_1(j)|B(j)C(j)}\big\|_1\label{eq:1norm}
		\end{equation}
		where the second expectation is taken over $x_{B(j)}x_{C(j)}\sim p_{B(j)C(j)}$, and
		\begin{equation}
			\big\|p_{A_1(j)|B(j)}-p_{A_1(j)|B(j)C(j)} \big\|_1=\sum_{x_{A_1(j)}} \big|p_{A_1(j)|B(j)}(x_{A_1(j)}|x_{B(j)})-p_{A_1(j)|B(j)C(j)}(x_{A_1(j)}|x_{B(j)}x_{C(j)})\big|.
		\end{equation}
        
        In Claim~\ref{claim: bound} below, we apply Lemma \ref{lem:tracecond} to bound each term corresponding to a given column $j$ by $\epsilon/2\sqrt{n}$. After doing so, the expected total variation distance becomes $\mathbb{E}_{U} \|p_T-\tilde{p}_T\|_1\leq \sum_{j=L}^{\sqrt{n}} \big(\epsilon/2\sqrt{n})\leq \epsilon/2$, as desired.
		
    \begin{claim} \label{claim: bound}
       We can choose $w^*=c_1\log(n/\epsilon)$ and $L=c_2\log(n/\epsilon)$ for suitable constants $c_1,c_2$ and $\beta$ sufficiently large, such that 
       \[
       \Exp_U\Exp_{x_{B(j)}x_{C(j)}\sim p_{B(j)C(j)}} \big\|p_{A_1(j)|B(j)}-p_{A_1(j)|B(j)C(j)}\big\|_1\leq \frac{\epsilon}{2\sqrt{n}} \quad \text{for all } L\leq j\leq \sqrt{n}.
       \]
        \end{claim}
        \begin{proof}
        Fix $j$ and let $A_1=A_1(j), A_2=A_2(j),B=B(j), C=C(j)$. Write the qubits in column $j$ which are not in $S$ in order from top to bottom of the grid as $ A_1=\{Y_1,Y_2,\ldots, Y_m\}$
        and define, for $t\leq m$,
        \[
        Y_{<t}=\{Y_1,Y_2,\ldots, Y_{t-1}\}.
        \]
        By successively conditioning on qubits one by one we have
        \begin{equation}
        \Exp_U\Exp_{x_{B}x_{C}\sim p_{BC}} \|p_{A_1|B}-p_{A_1|BC}\|_1\leq\Exp_U\Exp_{x_{B}x_{C}\sim p_{BC}} \sum_{t=1}^{m} \mathbb{E}_{y\sim p_{Y_{<t}|BC}}\left[\|p_{Y_{t}|BCY_{<t}}-p_{Y_t|BY_{<t}}\|_1\right].
        \label{eq:colsum}
        \end{equation}
        Now to bound the $t$-th term on the RHS we use Lemma \ref{lem:tracecond} with $A_1^{(t)}=Y_t,A_2^{(t)}=A_2\cup \{Y_{t+1},\ldots ,Y_m\},A^{(t)}=A_1^{(t)}\cup A_2^{(t)}, B^{(t)}=B\cup Y_{<t}$ and $ C^{(t)}=C$. To bound the RHS of  Eq.~\eqref{eq:tracedist}, note that by choosing $c_1$ sufficiently large we can ensure that $2^{1/2}\mathrm{poly}(n)e^{-\mathrm{dist}(A^{(t)},C^{(t)})}\leq \epsilon/5m\sqrt{n}$, and by choosing $c_2$ sufficiently large we can ensure $2^{1/2}\mathrm{poly}(n)e^{-l_C}\leq \epsilon/5m\sqrt{n}$.

		In all but the last step ($j=\sqrt{n}$) of the algorithm we have $\min\{|\partial A_2^{(t)}|,|\partial A^{(t)}|\}= \Theta(\sqrt{n})$ and the second term on the RHS of Eq.~\eqref{eq:tracedist} is negligible compared to the first term. For the last step, $j=\sqrt{n}$,  we can use the fact that $\partial S'\subseteq \partial A_2^{(t)}\cap \partial A^{(t)}$ along with the lower bound $|\partial S'|\geq (1/4)\beta \log(n/\epsilon)$ from Condition (2). This gives
		\begin{equation}
			\sqrt{2}e^{-\Omega(\mathrm{min}\{|\partial A_2^{(t)}|,|\partial A^{(t)}|\})}\leq e^{-\Omega(|\partial S'|))}=e^{-\Omega(\beta \log(n/\epsilon))}\leq \epsilon/5m\sqrt{n},    
		\end{equation}
		where we have required $\beta$ to be a sufficiently large constant in the last step. In all cases, the sum in Eq.~\eqref{eq:colsum} is upper bounded as $\sum_{t=1}^{m} (2\epsilon/(5m\sqrt{n})+\exp(-\Omega(\sqrt{n})))\leq \epsilon /2\sqrt{n}$.
        \end{proof}

		Finally, let us discuss the runtime of Algorithm 1. 
        Suppose $O,Q\subseteq [n]$ are any subsets of qubits that are contained in $O(\log(n))$ columns of the 2D grid. We shall use the fact that the marginal distribution $p_O$ and the conditional distribution $p_{O|Q}$ can be sampled with runtime $\mathrm{poly}(n)$ using Matrix Product State techniques.  We will also use the Condition (3) which ensures that the $r$-neighborhoods of qubits in $S$ do not form any connected components of size larger than a constant. This in turn ensures that each lightcone-connected component of the set of qubits $A_1\cup B$ considered in line 7 of the algorithm is contained in $O(\log(n))$ columns of the 2D grid.
        
        We conclude that the runtime of line 4 and line 7 of Algorithm 1--and therefore the overall runtime-- is upper bounded as  $\mathrm{poly}(n)$, since the initial set $I$ is contained in the leftmost $L-1=O(\log(n))$ columns, and each lightcone-connected component of $A_1(j)\cup B(j)$ is also contained in $O(\log(n))$ columns of the grid.
	\end{proof}
	
	\onecolumngrid
	\newpage

	{\centering
		\large\bfseries
		Supplementary Material: Conditional dependence and Scrooge ensembles in shallow random quantum circuits
		\par}

	\tableofcontents
	
	\section{The Scrooge ensemble and shallow 2D Haar-random quantum circuits} \label{sec:scrooge}

    \renewcommand*{\theHequation}{supp.\Alph{section}.\arabic{equation}}
    
	\subsection{The Scrooge ensemble}
	
	In this section, we review the definition of the Scrooge ensemble and the approximate formula for its moments~\cite{mcginley2025scrooge}. The Scrooge ensemble $\mathcal{S}_\rho$ was introduced briefly in the main text. Informally, $\mathcal{S}_\rho$ is the maximally entropic ensemble of pure states over all ensembles whose average state is given by the density matrix $\rho$. 
	
	\begin{definition}[Scrooge ensemble \cite{jozsa94}] \label{def:scrooge}
		Given a density matrix $\rho$ describing the state of $n$ qubits, the corresponding Scrooge ensemble $\mathcal{S}_\rho$ is a continuous distribution of pure quantum states that can be characterized in the following equivalent ways.
		\begin{enumerate}
			\item The ensemble of states obtained by sampling $\ket{\phi}$ from the measure $\mu'(\dif \phi) = 2^n \braket{\phi|\rho|\phi} \mu_{H, 2^n}(\dif \phi)$, where $\mu_{H, d}$ is the Haar measure over normalized states on $\mathbb{C}^{d}$, and returning the `distorted' state \cite{jozsa94}
			\begin{align}
				\ket{\psi_\phi} \coloneqq \frac{\sqrt{\rho}\ket{\phi}}{\|\sqrt{\rho}\ket{\phi}\|}.
			\end{align}
			
			\item The ensemble of normalized wavefunctions $\ket{\psi}$ over $\textup{supp}(\rho) \subseteq \mathbb{C}^{2^n}$ whose measure $\nu_{\mathcal{S}_\rho}$ has probability density  \cite{goldstein2016universal}
			\begin{align}
				\frac{ \dif \nu_{\mathcal{S}_\rho}}{\dif \mu_{H, r}}(\psi) = \frac{r}{\det \rho} \braket{\psi|\rho^{-1}|\psi}^{-r-1} 
			\end{align}
			where $r \coloneqq \textup{rank}(\rho)$.
			
			\item The minimizer of the \emph{accessible information} over all pure state ensembles $\mathcal{E}$ with average state $\mathbb{E}_{\psi \sim \mathcal{E}} \ket{\psi}\bra{\psi} = \rho$ \cite{jozsa94}.
			
			\item The ensemble of post-measurement states on subsystem $A$ obtained when performing a rank-1 Haar-random POVM on subsystem $B$ of any purification $\ket{\Phi_\rho^{AB}}$, i.e.~a bipartite state such that $\tr_B[\Phi_\rho^{AB}] = \rho^A$ \cite{mcginley2025scrooge}. Equivalently, if one applies a Haar-random unitary $U_B$ to $B$, and then performs a projective measurement on $B$ in the computational basis, with outcome $x_B \in \{0,1\}^n$, the post-measurement states on $A$ over the joint randomness of $U_B$ and $x_B$ have the distribution $\mathcal{S}_\rho$.
		\end{enumerate}
		In particular, for the case $\rho = I/2^n$, the Scrooge ensemble is equal to the Haar ensemble, i.e.~the uniform distribution over all normalized states.
	\end{definition}
	In the third characterization of the Scrooge ensemble above, the accessible information $I_{\rm acc}(\mathcal{E})$ for an ensemble of states $\mathcal{E}$ is defined as follows. For any POVM $\mathcal{A} = \{A(x)\}_{x \in X}$ over some outcome space $X$, consider the joint distribution of states and outcomes that arises from sampling $\psi \sim \mathcal{E}$, and then performing the measurement $\mathcal{A}$, whose outcome probability density is given by $p(x|\psi) = \braket{\psi|A(x)|\psi}$. Then, $I_{\rm acc}(\mathcal{E})$ is the supremum of the mutual information between $\phi$ and $x$ over all possible POVMs $\mathcal{A}$ \cite{holevo1973bounds,jozsa94}.
	
	Informally, the fact that the Scrooge ensemble minimizes the accessible information implies that Scrooge-random states are maximally indistinguishable from one another. That is, any party performing measurements on an unknown state $\ket{\psi}$ drawn from the ensemble $\mathcal{S}_\rho$ can only gain a small amount of information about which state they have access to. This minimization of information about the state (or maximization of the entropy of the ensemble) is reminiscent of similar principles that arise in statistical mechanics, and this has led to the suggestion that the Scrooge ensemble ought to emerge in various generic many-body settings---a hypothesis for which a growing body of evidence is emerging \cite{goldstein2006distribution,goldstein2016universal,cotler2021emergent, ippoliti2022solvable,  mark2024maximum, chang2025deep, mcginley2025scrooge, mok2026nature,schuster2026fast}.\\
	
	In order to make predictions about the properties of Scrooge-random states---as is necessary for \cref{lem:tracecond} in the main text---one must be able to evaluate averages of the form $\mathbb{E}_{\psi \sim \mathcal{S}_\rho} f(\psi)$ for appropriate functions $f$. A particularly important class is the integer moments of the ensemble, namely the collection of averages for which $f(\psi)$ is a $k$-th order polynomial of the projector $\dyad{\psi}$ for some $k \in \mathbb{N}$. These can be conveniently organized in a matrix form by defining the $k$-th moment operator
	\begin{align}
		\chi^{(k)}_{\mathcal{S}_\rho} \coloneqq \mathbb{E}_{\psi \sim \mathcal{S}_\rho} \big[ \dyad{\psi}^{\otimes k}\big].
	\end{align}
	Any particular moment $f(\psi)$ can be written in the form 
	$\mathbb{E}_{\psi \sim \mathcal{S}_\rho}[f(\psi)] = \tr[\chi^{(k)}_{\mathcal{S}_\rho}O^{(k)}]$ for some operator $O^{(k)} \in \mathcal{B}(\mathcal{H}^{\otimes k})$ acting on $k$ copies of the Hilbert space $\mathcal{H} \cong \mathbb{C}^{2^n}$.
	
	Unfortunately, exact expressions for such averages are not attainable in general. This contrasts with the Haar ensemble, for which the moments have a well-known and concise expression. Nevertheless, in settings where the background state $\rho$ is highly mixed, well-controlled approximations for integer moments were derived in Ref.~\cite{mcginley2025scrooge}. In particular, we have
	\begin{align}
		(1-\epsilon_{\rho, k}) \chi^{(k)}_{\mathcal{S}_\rho, \text{appr.}} \preceq \chi^{(k)}_{\mathcal{S}_\rho} \preceq (1+\epsilon_{\rho, k}) \chi^{(k)}_{\mathcal{S}_\rho, \text{appr.}} 
		\label{eq:rel error scrooge}
	\end{align}
	where the approximate moments are given by
	\begin{align}
		\chi^{(k)}_{\mathcal{S}_\rho, \text{appr.}} = \rho^{\otimes k}  \sum_{\pi \in S_k} \pi
		\label{eq:scrooge approx moments}
	\end{align}
	and the relative error $\epsilon_{\rho,k}$ is bounded as
	\begin{align}
		\epsilon_{\rho, k} &\leq \mathcal{O}\big(k^2 2^{-S_\infty(\rho)/2}\big) & \text{for any }k \leq 2^{S_\infty(\rho)/3 - 1}.
		\label{eq:rel error scrooge bound}
	\end{align}
	Here, $S_\infty(\rho) = -\log \|\rho\|_\infty$ is the $\alpha \rightarrow \infty$ limit of the $\alpha$-R{\'e}nyi entropies $S_\alpha(\rho) = (1-\alpha)^{-1}\log \tr[\rho^\alpha]$, where $\|\rho\|_\infty$ is the maximum eigenvalue of $\rho$. In Eq.~\eqref{eq:rel error scrooge}, the notation $A \preceq B$ indicates that $(B-A)$ is a positive semi-definite matrix.
	
	\subsection{Reduced moments and conditionals of Scrooge ensembles\label{app:reducedmoments}}
	
	Our proof of Theorem~\ref{thm:classical-algorithm} uses the following theorem on the moments of marginals of output distributions of Scrooge ensembles.
	The theorem was proven in a more general form (at the level of quantum states instead of output distributions) in Ref.~\cite{mcginley2025scrooge}.
	We provide a much simpler proof for the specific case of output distributions below.

	We adopt the following notations in this subsection. For an $n$-qubit state $\rho$, which could be pure, we let $P^\rho(x)=\bra{x}\rho\ket{x}$, $x\in\{0,1\}^n$ denote its output distribution in the computational basis. For a region $A\subseteq[n]$, we use $P^\rho_A(x_A)$, $x_A\in\{0,1\}^{|A|}$ to denote the marginal output distribution of $\rho$ on $A$.

	\begin{theorem} [Marginals of Scrooge output distributions, adapted from Theorem~4 of~\cite{mcginley2025scrooge}] \label{thm: trace out A}
		Let $\psi \sim \mathcal{S}_\rho$ be drawn from the Scrooge ensemble defined by an $n$-qubit background state $\rho$. Let $A\subseteq[n]$ be a subregion and $B=[n]\setminus A$ be its complement. Let $x_B\in\{0,1\}^{|B|}$ such that $P^{\rho}_B(x_B)>0$. Let $\rho_{x_B}=\bra{x_B}\rho\ket{x_B}/P^{\rho}_B(x_B)$ denote the post-measurement state of $\rho$ on $A$ after measuring $x_B$ on $B$. It holds that for every $k\geq 1$, if $k\leq 2^{S_\infty(\rho)/3-1}$ and $k(k-1)\leq 2^{S_k(\rho_{x_B})}$, then the $k$-th moment of $P^\psi_B(x_B)$ is close to the $k$-th moment of $P^\rho_B(x_B)$ up to relative error $\varepsilon_{x_B}$; that is, 
		\begin{equation}
			(1-\varepsilon_{x_B}) P^\rho_B(x_B)^k \leq \E_{\psi \sim \mathcal{S}_\rho} [P^\psi_B(x_B)^k] \leq (1+\varepsilon_{x_B}) P^\rho_B(x_B)^k
		\end{equation}
		where $\varepsilon_{x_B} = k(k-1)/2^{S_k(\rho_{x_B})} + \mathcal{O}(k^{2}/2^{S_\infty(\rho)/2})$, $S_k(\rho_{x_B})$ is $k$-th Renyi entropy of $\rho_{x_B}$, and $S_\infty(\rho)$ is the min-entropy of $\rho$.
	\end{theorem}

    \noindent In practice, when applying the theorem to projected ensembles of random circuits on a subsystem $A$, we will replace $[n] \rightarrow A$, $A \rightarrow A_2$, and $B \rightarrow A_1$ in the above theorem, for the subsystems $A_1, A_2, A$ described in the main text.
	
	\begin{proof}
		Invoking the approximate formula for the Scrooge moments \eqref{eq:scrooge approx moments}, whose accuracy is given by \eqref{eq:rel error scrooge bound}, our aim is to compute
		\begin{equation}
			\E_{\psi \sim \mathcal{S}_\rho} [P^\psi_B(x_B)^k ] \approx \bra{x_B}^{\otimes k} \tr_{A^{\otimes k}}\big( \rho^{\otimes k} \sum_{\pi \in S_k} \pi \big) \ket{x_B}^{\otimes k} \equiv \chi_a,
		\end{equation}
		where the approximation holds up to relative error $\mathcal{O}(k^2/2^{S_\infty(\rho)/2})$ for any $k \leq 2^{S_\infty(\rho)/3-1}$ from Theorem~1 of Ref.~\cite{mcginley2025scrooge}.
		Pulling the projectors onto $x_B$ inside the trace yields,
		\begin{equation} \label{eq: chi noise perm}
			\chi_a
			= \sum_{\pi \in S_k} \prod_{\ell \in \pi} \tr_{A} \big( \langle x_B|\rho |x_B\rangle^{|\ell|} \big),
		\end{equation}
		where $\ell$ indexes the cycles in $\pi$.
		We have $\sum_\ell |\ell| = k$ and $\sum_\ell 1 = k - |\pi|$, where $|\pi|$ is the Cayley distance of the permutation $\pi$ from the identity.
		Note that the projectors $\dyad{x_B}$ act only on $B$; the remaining subregion $A$ is contracted between the adjacent copies of $\rho$.
		
		The identity term in Eq.~(\ref{eq: chi noise perm}) yields our desired quantity, $P^\rho_B(x_B)^k \equiv \chi_\rho$.
		To prove the lemma, we will show that all non-identity terms are small relative to this identity term.
		We proceed via the triangle inequality,
		\begin{equation} \label{eq: diff chi rho}
			\big| \chi_a - \chi_\rho \big| \leq \sum_{\pi \neq \mathbbm{1}} \prod_{\ell \in \pi} \bigg| \tr_A \big( \langle x_B|\rho |x_B\rangle^{|\ell|} \big) \bigg|=\sum_{\pi\neq\mathbbm{1}}\prod_{\ell\in\pi} P^\rho_B(x_B)^{|\ell|} \tr_{A}(\rho_{x_B}^{|\ell|}),
		\end{equation}
		where $\rho_{x_B} \equiv \bra{x_B} \rho \ket{x_B} / P^\rho_B(x_B)$ is the normalized post-measurement state on $A$ after measuring $x_B$ on $B$. Below we upper bound each individual trace in the product, and then use this to upper bound the sum as a whole.

        By the monotonicity of the Renyi entropies of $\rho_{x_B}$, for every $1\leq|\ell|\leq k$, we get
        \begin{equation}
        \tr_A(\rho_{x_B}^{|\ell|})=2^{-(|\ell|-1)S_{|\ell|}(\rho_{x_B})}\leq 2^{-(|\ell|-1)S_{k}(\rho_{x_B})}.
        \end{equation}
		Taking the product over all cycles $\ell \in \pi$ for a given permutation $\pi$, we find
		\begin{equation}
			\prod_{\ell \in \pi} \bigg| \tr \big( \langle x_B|\rho |x_B\rangle^{|\ell|}\big) \bigg|
			\leq 
			\prod_{\ell \in \pi} P^\rho_B(x_B)^{|\ell|}
			2^{-(|\ell|-1) S_k(\rho_{x_B})} = P^\rho_B(x_B)^k 2^{-|\pi| S_k(\rho_{x_B})} = \chi_\rho 2^{-|\pi| S_k(\rho_{x_B})},
		\end{equation}
        where we use $\sum_{\ell \in \pi} (|\ell| - 1) = |\pi|$. This is our final bound on each individual trace.
		
		Our bound on the total summation now follows easily.
		Inserting our upper bounds into Eq.~(\ref{eq: diff chi rho}), we find
		\begin{equation} 
			\big| \chi_a - \chi_\rho \big| \leq \chi_\rho \sum_{\pi \neq \mathbbm{1}} 2^{-|\pi| S_k(\rho_{x_B})} \leq \chi_\rho k(k-1) 2^{-S_k(\rho_{x_B})},
		\end{equation}
		which completes our proof.
		In the final inequality,
		we use the standard geometric series upper bound on the sum over permutations~\cite{harrow2023approximate,schuster2025random}, 
		\begin{equation}
			\sum_{\pi \neq \mathbbm{1}} b^{-|\pi|} \leq \sum_{r=1}^k \left( \frac{k(k-1)}{2b} \right)^r \leq \frac{\frac{k(k-1)}{2b}}{1-\frac{k(k-1)}{2b}} \leq \frac{k(k-1)}{b},
		\end{equation}
		which applies for any $k(k-1)/b \equiv k(k-1)2^{-S_k(\rho_{x_B})} \leq 1$.
	\end{proof}

	We now give the proof of Lemma \ref{lem:tracecond}, restated below.
	\marginal*
	\begin{proof}        
		First we use the Scrooge hypothesis with respect to the tripartition $ABC$ and $k=2$. The Scrooge hypothesis implies that with probability at least $1-\varepsilon$ over $U_{AB}$ and $x_B$, where
        \begin{equation}
            \varepsilon=\mathrm{poly}(n) e^{-\Omega(\min\{l_C,\mathrm{dist}(A,C)\})},
            \label{eq:epsilon}
        \end{equation}
             we get a background state $\rho_A(U_{AB},x_B)$ that satisfies $S_\infty (\rho_A(U_{AB},x_B))\geq \Omega(|\partial A|)$ and  
        \begin{equation}
            \Exp_{\psi\sim \mathcal{E}_A(U_{AB},x_B)}\bigg[\sum_{x_{A_1}} P^{\psi}_{A_1}(x_{A_1})^2\bigg]\leq\Exp_{\psi\sim\mathcal{S}_{\rho_A(U_{AB},x_B)}}\bigg[\sum_{x_{A_1}} P^{\psi}_{A_1}(x_{A_1})^2\bigg](1+\varepsilon),
            \label{eq:scrooge_x_A1}
        \end{equation}
        where the second inequality follows from invoking the Scrooge hypothesis with the observable
		\begin{equation}
        O=\sum_{x_{A_1}}\ket{x_{A_1}}\bra{x_{A_1}}\otimes I_{A_2}\otimes\ket{x_{A_1}}\bra{x_{A_1}}\otimes I_{A_2}.
		\end{equation}
		To ease notation in the following we shall omit dependence on $U_{AB}$, writing $\rho_A(x_B)\equiv \rho_A(U_{AB},x_B)$ and $\mathcal{E}_A(x_B)\equiv \mathcal{E}_A(U_{AB},x_B)$ for example.

		For the moment let us suppose that $U_{AB},x_B$ are such that the above holds. Next we use Theorem \ref{thm: trace out A} to simplify the RHS of \eqref{eq:scrooge_x_A1}, giving
		\begin{equation}
			\Exp_{\psi\sim \mathcal{E}_A(x_B)}\bigg[\sum_{x_{A_1}} P^{\psi}_{A_1}(x_{A_1})^2\bigg]\leq (1+\varepsilon)\sum_{x_{A_1}}(1+\delta(x_{A_1})) P^{\rho_A(x_B)}_{A_1}(x_{A_1})^2,
			\label{eq:deleps}
		\end{equation}
		where
		\begin{equation}
			\delta(x_{A_1})=2^{-\Omega(S_2(\rho_{A_2}(x_{A_1},x_B)))}+2^{-\Omega(|\partial{A}|)},
			\label{eq:dela1}
		\end{equation}
				and $\rho_{A_2}(x_{A_1},x_B)$ is obtained from $\rho_A(x_B)$ by measuring the bits in $A_1$ and obtaining outcome $x_{A_1}$. Note that we adopt \eqref{eq:dela1} as the definition of $\delta(x_{A_1})$ for every $(U_{AB},x_B)$, not just for those corresponding to the event in the Scrooge hypothesis occurring. Then in particular, $\delta(x_{A_1})\leq 1+2^{-\Omega(|\partial{A}|)}\leq 2$ for every $(U_{AB},x_B)$. Now using Cauchy Schwarz gives
\begin{align}
			\Exp_{\psi\sim \mathcal{E}_A(x_B)}\bigg[\sum_{x_{A_1}}\big| P^{\psi}_{A_1}(x_{A_1})-P^{\rho_A(x_B)}_{A_1}(x_{A_1})\big|\bigg]&\leq 2^{|A_1|/2}\Exp_{\psi\sim \mathcal{E}_A(x_B)}\bigg[\bigg(\sum_{x_{A_1}}\big( P^{\psi}_{A_1}(x_{A_1})-P^{\rho_A(x_B)}_{A_1}(x_{A_1})\big)^2\bigg)^{1/2}\bigg]\\
            &\leq 2^{|A_1|/2}\bigg(\Exp_{\psi\sim \mathcal{E}_A(x_B)}\bigg[\sum_{x_{A_1}}\big(P^{\psi}_{A_1}(x_{A_1})-P^{\rho_A(x_B)}_{A_1}(x_{A_1})\big)^2\bigg]\bigg)^{1/2}\\
            &= 2^{|A_1|/2}\bigg(\Exp_{\psi\sim \mathcal{E}_A(x_B)}\bigg[\sum_{x_{A_1}}P^{\psi}_{A_1}(x_{A_1})^2-P^{\rho_A(x_B)}_{A_1}(x_{A_1})^2\bigg]\bigg)^{1/2}
\end{align}
where we used Jensen's inequality in the second line and in the last line we used the fact that $\Exp_{\psi\sim \mathcal{E}_A(x_B)} P^{\psi}(x_{A_1})=P^{\rho_A(x_B)}(x_{A_1})$. Using Eq.~\eqref{eq:deleps}  gives
		\begin{align}
			\Exp_{\psi\sim \mathcal{E}_A(x_B)} \bigg[\sum_{x_{A_1}}\big| P^{\psi}_{A_1}(x_{A_1})-P^{\rho_A(x_B)}_{A_1}(x_{A_1})\big|\bigg]&\leq2^{|A_1|/2}\bigg(\sum_{x_{A_1}}(\varepsilon+(1+\varepsilon)\delta(x_{A_1}))P^{\rho_A(x_B)}_{A_1}(x_{A_1})^2\bigg)^{1/2}\\
            &\leq 2^{|A_1|/2}\bigg(\varepsilon+(1+\varepsilon)\sum_{x_{A_1}}\delta(x_{A_1})P^{\rho_A(x_B)}_{A_1}(x_{A_1})\bigg)^{1/2}.
			\label{eq:expect}
		\end{align}
		
		Note that (for a fixed $x_B$), drawing $\psi\sim \mathcal{E}_A(x_B)$ and then drawing $x_{A_1}\sim P^{\psi}_{A_1}(x_{A_1})$ we obtain the same distribution over $x_{A_1}$ as drawing $x_C\sim p_{C|B}(\cdot|x_B)$ and then drawing $x_{A_1}\sim p_{A_1|BC}(\cdot|x_Bx_C)$. Moreover, note that 
		\[
		P^{\rho_A(x_B)}_{A_1}(x_{A_1})=p_{A_1|B}(x_{A_1}|x_B).
		\]
		Using these facts we can write the LHS of Eq.~\eqref{eq:expect} as
		\begin{equation}
			\Exp_{\psi\sim \mathcal{E}_A(x_B)}\bigg[\sum_{x_{A_1}}\big| P^{\psi}_{A_1}(x_{A_1})-P^{\rho_A(x_B)}_{A_1}(x_{A_1})\big|\bigg]=\Exp_{x_C\sim p_{C|B}(\cdot|x_B)}\bigg[\sum_{x_{A_1}}\big| p_{A_1|BC}(x_{A_1}|x_Bx_C)-p_{A_1|B}(x_{A_1}|x_B)\big|\bigg].
		\end{equation}
		Therefore (for any fixed $x_B,U_{AB}$ satisfying the conditions described earlier) we have 
		\begin{equation}
			\Exp_{x_C\sim p_{C|B}(\cdot|x_B)}\bigg[\sum_{x_{A_1}}\big| p_{A_1|BC}(x_{A_1}|x_Bx_C)-p_{A_1|B}(x_{A_1}|x_B)\big|\bigg]\leq 2^{|A_1|/2}\bigg(\varepsilon+(1+\varepsilon)\sum_{x_{A_1}}\delta(x_{A_1})p_{A_1|B}(x_{A_1}|x_B)\bigg)^{1/2}.
		\end{equation}
		Now, we use the law of total expectation to take the expected value with respect to $U$ and $x_B\sim p_B$. Also using the fact that $\Exp[x^{1/2}]\leq \Exp[x]^{1/2}$ for any nonnegative random variable $x$, we get
		\begin{align}
			&\Exp_{U}\Exp_{x_Bx_C\sim p_{BC}}\bigg[\sum_{x_{A_1}}\big| p_{A_1|BC}(x_{A_1}|x_Bx_C)-p_{A_1|B}(x_{A_1}|x_B)\big|\bigg]\\
			\leq& 2^{|A_1|/2}\bigg(\varepsilon+(1+\varepsilon)\Exp_{U}\Exp_{x_B\sim p_B}\bigg[\sum_{x_{A_1}}\delta(x_{A_1})p_{A_1|B}(x_{A_1}|x_B)\bigg]\bigg)^{1/2}+2\varepsilon\\
			=&2^{|A_1|/2}\bigg(\varepsilon+(1+\varepsilon)\Exp_{U}\Exp_{x_{A_1}x_B\sim p_{A_1B}}[\delta(x_{A_1})]\bigg)^{1/2}+2\varepsilon.
			\label{eq:beforescrooge}
		\end{align}

		Now we can use the Scrooge hypothesis again, this time with tripartition $A_2B'C$ where $B'=A_1\cup B$. It says in particular that with probability at least 
		$1-\varepsilon'$ over $U$ and measurement outcomes $x_{A_1},x_{B}$ we have
		\begin{equation}
			S_{\infty}(\rho_{A_2}(x_{A_1},x_{B}))=\Omega(|\partial A_2|),
			\label{eq:sinf}
		\end{equation}
		where 
		\begin{equation}
			\varepsilon'=\mathrm{poly}(n) e^{-\Omega(\min\{l_C,\mathrm{dist}(A_2,C)\})}.
			\label{eq:epsprime}
		\end{equation}
		Therefore 
		\begin{equation}
			\Exp_{U}\Exp_{x_{A_1}x_B\sim p_{A_1B}}[\delta(x_{A_1})]\leq 2\varepsilon'+(1-\varepsilon')2^{-\Omega(\mathrm{min}\{|\partial A_2|,|\partial A|\})},
			\label{eq:doubleE}
		\end{equation}
		where we used Eqs.~(\ref{eq:dela1},\ref{eq:sinf}) and the fact that $S_2(\sigma)\geq S_{\infty}(\sigma)$ for any quantum state $\sigma$. Plugging Eq.~\eqref{eq:doubleE} into Eq.~\eqref{eq:beforescrooge} gives
		\begin{align}
			&\Exp_{U}\Exp_{x_Bx_C\sim p_{BC}}\bigg[\sum_{x_{A_1}}\big| p_{A_1|BC}(x_{A_1}|x_Bx_C)-p_{A_1|B}(x_{A_1}|x_B)\big|\bigg]\\
			\leq& 2^{|A_1|/2}\bigg(O(\varepsilon+\varepsilon'+2^{-\Omega(\mathrm{min}\{|\partial A_2|,|\partial A|\})})\bigg)^{1/2}+2\varepsilon\\
			\leq& 2^{|A_1|/2}\mathrm{poly}(n)\left( e^{-\Omega(\min\{l_C,\mathrm{dist}(A,C)\})}+ e^{-\Omega(\min\{l_C,\mathrm{dist}(A_2,C)\})}\right)+2^{|A_1|/2}e^{-\Omega(\mathrm{min}\{|\partial A_2|,|\partial A|\})},
		\end{align}
		where we used Eqs. (\ref{eq:epsprime},\ref{eq:epsilon}) and the fact that $\sqrt{a_1+a_2+\cdots+a_k}\leq \sqrt{a_1}+\cdots+ \sqrt{a_k}$ for any nonnegative numbers $a_1,\ldots, a_k$. Finally, note that since $A_2\subseteq A$ we have $\mathrm{dist}(A,C)\leq \mathrm{dist}(A_2,C)$ and therefore
		\[
		\Exp_{U}\Exp_{x_Bx_C\sim p_{BC}} \sum_{x_{A_1}}\big| p_{A_1|BC}(x_{A_1}|x_Bx_C)-p_{A_1|B}(x_{A_1}|x_B)\big|\leq  2^{|A_1|/2}(\mathrm{poly}(n) e^{-\Omega(\min\{l_C,\mathrm{dist}(A,C)\})}+e^{-\Omega(\min\{|\partial A_2|,|\partial{A}|\})}).
		\]
	\end{proof}
	\section{Analytical evidence for the Scrooge hypothesis in shallow 2D Haar-random quantum circuits}
	
	In this section, we provide analytical arguments that support our hypothesis regarding the emergence of the Scrooge ensemble in SRQCs. Our approach will be based on the method of replicas, which has been used to analyze the dynamics of random quantum circuits with measurements in several settings \cite{bao2020theory, jian2020measurement, napp2022efficient}. Owing to the need to take a certain `replica limit' (familiar from the study of disordered systems in statistical physics), these arguments fall short of a rigorous mathematical proof, but nevertheless provide strong evidence for our conjecture.
	
	\subsection{Setup}
	
	\subsubsection*{Circuit architecture}
	
	Our hypothesis is that the Scrooge hypothesis will hold for any natural choice of circuit architecture. However, in order to simplify our calculation later on, we will work with a particular coarse-grained architecture, in a similar spirit to previous works \cite{bene2025quantum, mcginley2025measurement}. This architecture has the particular advantage that the lightcone distance is exactly one (in units of the coarse-grained lattice spacing), which will avoid some tedious complications later on.
	
	The coarse-grained architectures we consider are parametrized by an undirected graph $G = (V, E)$ of bounded degree, and a positive integer $\xi$, which serves as an effective circuit depth. Each vertex $v \in V$ hosts $\xi\times \text{deg}(v)$ qubits, where $\text{deg}(v) = |E(v)|$, where  $E(v) \coloneqq \{e \in E : v \in e \}$ is the set of edges adjacent to $v$. The collection of qubits at a vertex $v$, which we denote $Q(v)$, are partitioned into non-overlapping subsets $Q(v,e) \subset Q(v)$ for each edge $e \in E(v)$, where each $Q(v, e)$ contains $\xi$ qubits. We write $n = \xi\sum_{v \in V}  \text{deg}(v)$ for the total number of qubits, which should be distinguished from the total number of vertices $|V|$.
	
	The circuits in the ensemble are composed of two layers. In the first layer, a Haar-random $2\xi$-qubit unitary is independently sampled for each edge $e=\{v_1, v_2\} \in E$, and applied to the qubits in $Q(v_1, e) \cup Q(v_2, e)$. In the second layer, a $\xi\times \text{deg}(v)$-qubit Haar-random unitary is independently sampled for each $v \in V$, and applied to the qubits in $Q(v)$. In principle, all of these Haar-random gates can be replaced by gates sampled from a $k$-design for a suitably large (but system size-independent) integer $k$; however we will assume fully Haar-random gates for simplicity.
	
	\subsubsection*{Defining the post-measurement ensembles}
	
	Let $ABC$ be a tripartition of the graph $(V, E)$ for which $\text{dist}(A, C) > 1$, so that $A$ and $C$ are lightcone-separated. Our hypothesis pertains to the ensemble of post-measurement states on $A$ after the circuit $U$ is applied, and regions $B$ and $C$ are measured with outcomes $x_B$ and $x_C$, respectively. We can break up the circuit as $U = U_C U_{AB}$, where $U_{AB}$ is the unitary containing all the gates in the backwards light cone of $AB$. We define $\partial B \subset C$ to be the set of vertices in the support of $U_{AB}$ that are not contained in $AB$.  The state after applying $U_{AB}$ and measuring on $B$ only is
	\begin{align}
		\psi_{A \partial B}(U_{AB}, x_B) &= \frac{\bra{x_B}U_{AB} \dyad{0^{A B  \partial B}}U_{AB}^\dagger \ket{x_B}}{p(x_B|U_{AB})} \\
        & \text{where }p(x_B|U_{AB}) = \tr[\bra{x_B}U_{AB} \dyad{0^{A B  \partial B}}U_{AB}^\dagger\ket{x_B}],
		\label{eq:inter def}
	\end{align}
	where $p(x_B|U_{AB})$ is the measurement output probability conditioned on $U_{AB}$ having been chosen. The subsequent application of $U_C$ and the measurements on $C$ can be described using POVM operators supported on $\partial B$,
	\begin{align}
		F_{\partial B}(U_C, x_C) \coloneqq \bra{0_{C\backslash \partial B}}U_C^\dagger \dyad{x_C} U_C \ket{0_{C\backslash \partial B}}.
		\label{eq:POVM def}
	\end{align}
	which are positive semi-definite and satisfy $\sum_{x_C}F_{\partial B}(U_C, x_C) = I_{\partial B}$. The final post-measurement state can then be written
	\begin{align}
		\psi_A(U, x_{BC}) &= \frac{\tr_C[F_{\partial B}(U_C, x_C) \psi_{A\partial B}(U_{AB}, x_B)]}{p(x_C|U,x_B)} & \text{with }p(x_C|U,x_B) &= \tr[\big(I_A\otimes F_{\partial B}(U_C, x_C)\big) \psi_{A\partial B}(U_{AB}, x_B)].
		\label{eq:post meas state POVM}
	\end{align}
	Here, $p(x_C|U,x_B)$ is the probability of measuring $x_C$ on $C$ conditioned on having measured $x_B$ on $B$. As explained in the main text, our hypothesis is that, with high probability over $(U_{AB}, x_B)$, the ensemble of post-measurement states over the randomness induced by $U_{C}, x_C$ approximates the Scrooge ensemble with respect to the background state $\rho_A(U_{AB}, x_B)$.

	\subsubsection*{Geometry of the tripartition $ABC$}
	
	The accuracy to which the post-measurement ensemble approximates the Scrooge ensemble naturally depends on the geometry of the tripartition $ABC$, and we provided an informal description of the form of this dependence in the main text, Eq.~\eqref{eq:scroogeeps}. Here we will make this hypothesis more concrete, which necessitates some formal assumptions on the geometry. These are by no means the most permissive conditions that could be dealt with, but will help us analyse the problem more formally. In particular, for concreteness we will let $G$ be a subgraph of the square lattice, whose vertices $V$ embedded in the plane as $V = \mathbb{Z}^2 \subset \mathbb{R}^2$ in the canonical way. For this purpose, a subset of vertices $A \subset V$ will be said to be simply connected if the set $X_A \coloneqq \bigcup_{v \in A} X_v$ is a simply connected region of $\mathbb{R}^2$, where $X_v = [i_v-1/2,i_v+1/2]\times [j_v-1/2,j_v+1/2]$ for the vertex $v$ embedded as $(i_v, j_v)$.
	
	Our conditions on the geometry of the partition depend on two tunable length parameters, $r_{AC}$, and $l_C$, which will appear in our expression for the relative error. 
	\begin{enumerate}
		\item Qubits are arranged on a graph $G = (V, E)$ which forms a connected subgraph of the square lattice $\mathbb{Z}^2$, with an edge between every pair of vertices a unit distance apart.
		\item $C$ is a connected subgraph of $G$, and each connected component of $A$, denoted $A_i$ for $i = 1, \ldots, n_A$ is simply connected.
		\item The graph distance between $A$ and $C$ is lower bounded by $\text{dist}(A, C) \geq r_{AC}$.
		\item A reference vertex $v^\star \in C$ and a simply connected region $R_C \subset C$ that contains $v^\star$ have been chosen.
		\item (Minimum cut.) If $E_\gamma \subset E$ is a connected set of edges that intersects the boundary of $R_C$ or the boundary of $A\cup B$ such that, when $E_\gamma$ is removed from $G$, one of the components $A_i$ becomes disconnected from  $v^\star$ or one of the other components $A_j, j \neq i$, then $|E_\gamma| \geq l_C$.
	\end{enumerate}
	While it is possible that some of these conditions could be relaxed, these will be sufficient to justify the following hypothesis (albeit non-rigorously): that the relative error between the moments of the post-measurement ensemble $\mathcal{E}(U_{AB}, x_B)$ and the Scrooge ensemble with respect to the background state $\rho_A(U_{AB}, x_B)$ is at most
	\begin{align}
		\epsilon \leq \text{poly}(n)e^{-\Omega(l_C)} + \text{poly}(|\partial A|)e^{-\Omega(r_{AC})}
		\label{eq:scroogeeps full}
	\end{align}
	with probability $\geq 1 - \mathcal{O}(\epsilon)$. 
	
	The above is a more general form of the hypothesis stated in the main text, Eq.~\eqref{eq:scroogeeps}, where we assumed $G$ to be a $\sqrt{n} \times \sqrt{n}$ square grid, and that $C$ contains a $l_C \times l_C$ square region. In that case, $R_C$ would be the $l_C \times l_C$ square region, and $v^*$ a point in the center of $R_C$. In this case, any cut $E_\gamma$ that intersects the boundary of $R_C$ and encircles $v^*$ (thereby separating it from $A$) must have length at least $l_C$. Moreover, since $G$ is a square grid and $\text{dist}(A, C) \geq r_{AC}$, we immediately have that any cut separating one component $A_i$ from another $A_j$ must either be contained entirely in $A \cup B$ (in which case it does not intersect the boundary of $R_C$ or the boundary of $A \cup B$), or must have length at least $r_{AC}$. This explains the form of Eq.~\eqref{eq:scroogeeps}.
    
    Intuitively, the parameter $l_C$ characterizes the `thickness' of $C$. For example, in the row-by-row geometry shown in Fig.~\ref{fig:distribution_of_r}(a), $l_C$ can be chosen to be the number of rows in $C$. If the full graph $(V, E)$ contains bottlenecks between different components $A_i$, $A_j$ (i.e.~a small number of edges through which any path connecting $A_i$ to $A_j$ must pass), then $l_C$ will be small, and so the error becomes large. As an extreme example, in a quasi-one-dimensional system of dimension $r \times L$, with $r = o(\log L)$, the error will not be small since $l_C \leq r$.
	
	\subsection{Moments of constant-depth random circuits  \label{subsec:output moments}}
	
	We now introduce some preliminary technical tools that will be needed to analyse the statistical properties of these states, in particular regarding the computation of moments of random quantum circuits. We will work mainly with the coarse-grained architecture, though similar considerations apply more generally.
	
	Given an ensemble of random unitaries $U \sim \mathcal{U}$, the $k$th moments of the output state $\ket{\psi} = U \ket{0^n}$ are given by
	\begin{align}
		\chi_{\mathcal{U}}^{(k)} \coloneqq \mathbb{E}_{\psi} [\dyad{\psi}^{\otimes k}] \equiv \mathbb{E}_{U \sim \mathcal{U}}[(U\dyad{0^n}U^\dagger)^{\otimes k}].
	\end{align}
	For the coarse-grained random circuits introduced above, the final layer involves applying a Haar-random gate to the qubits $Q(v)$ for each vertex $v$. By Schur-Weyl duality, the moments can therefore be written in the form
	\begin{align}
		\chi_{\mathcal{U}}^{(k)} = \sum_{\pi_1, \ldots , \pi_{|V|} \in S_k} b[\pi_1, \ldots, \pi_{|\mathcal{G}|}] \bigotimes_{g \in \mathcal{G}} V_{\mathcal{H}_g}(\pi_g).
	\end{align}
	Here, $\pi_g \in S_k$ are permutations of $k$ elements, which are represented by linear operators $V_{\mathcal{H}_g}(\pi_g) \in \mathcal{B}(\mathcal{H}_g^{\otimes k})$ that exchange the $k$ copies of the Hilbert space for all qubits $i \in g$, and $b[\pi_1, \ldots, \pi_n]$ is a collection of $(k!)^{|\mathcal{G}|}$ coefficients. For simplicity of discussion, we will assume that every $g \in \mathcal{G}$ has the same size, and we will define $q= 2^{|g|}$ as the local Hilbert space dimension for each gate. We will then use $V_{q}(\pi_g)$ as a shorthand to denote permutation operators on $(\mathbb{C}^q)^{\otimes k}$.
	
	For $k \leq q$, the operators $V_q(\pi_g)$ are linearly independent, which means the coefficients $b[\pi_1, \ldots, \pi_{|\mathcal{G}|}]$ are uniquely specified. We can then construct a dual set of operators $\text{Wg}_q(\pi_g)$, which are linear combinations of permutation operators with the property that $\tr[\text{Wg}_q(\pi) V_q(\pi')] = \delta_{\pi, \pi'}$. The moments can be rewritten in terms of these dual operators as
	\begin{align}
		\chi_{\mathcal{U}}^{(k)} = \alpha_{\mathcal{U}}^{(k)} \sum_{\vec{\pi}} p[\vec{\pi}] \bigotimes_{g \in \mathcal{G}} \text{Wg}_q[\pi_g],
	\end{align}
	where we use the shorthand $\vec{\pi} = (\pi_1, \ldots, \pi_{|\mathcal{G}|})$, and $\alpha_{\mathcal{U}}^{(k)}$ is a constant chosen such that the coefficients $p[\vec{\pi}]$ obey the normalization condition $\sum_{\vec{\pi}} p[\vec{\pi}] =1$. The coefficients can now be directly related to specific averages over $\mathcal{U}$ as
	\begin{align}
		\alpha_{\mathcal{U}}^{(k)}p[\vec{\pi}] = \mathbb{E}_{\psi}\left(\tr[\dyad{\psi}^{\otimes k}\cdot \bigotimes_{g \in \mathcal{G}}V_q(\pi_g)]\right).
		\label{eq:spin weights general}
	\end{align}
	Given a particular circuit architecture, the right hand side of the above expression can in principle be computed using Weingarten calculus \cite{collins}. While closed-form expressions cannot be obtained in full generality, for the coarse-grained architecture on the graph $G = (V, E)$ defined above these these coefficients can be exactly computed. Since the final layer involves a random unitary being applied to every qubit within the same vertex $v \in V$, the only configurations that contribute to the moments are those for which $\vec{\pi}$ is uniform within each vertex. Thus, we can label configurations by $\{\pi_v\}_v \in S_k^V$. The corresponding weights are given by
	\begin{align}
		p[\vec{\pi}] \propto \prod_{\{v_1, v_2\}\in E}\left( \sum_{\tau \in S_k} (2^{-\xi})^{d(\tau, \pi_{v_1}) + d(\tau, \pi_{v_2})} \right) \geq 0
	\end{align}
	where $d(\pi,\sigma) = k - \#\text{cycles}(\pi \sigma^{-1})$ is the Cayley distance between two permutations $\pi, \sigma \in S_k$. The non-negativity of these weights allows us to interpret $p[\vec{\pi}]$ as the Boltzmann weight $e^{-\mathcal{H}[\vec{\pi}]}$ of a classical statistical mechanics model, with configuration space $S_k^{\times V} \ni \vec{\pi}$, and the free energy $\mathcal{H}[\vec{\pi}]$ taking the form
	\begin{align}
		\mathcal{H}[\vec{\pi}] &= \sum_{\{v_1, v_2\}\in E} h(\pi_{v_1}^{-1}\pi_{v_2}), & \text{where }h(\sigma) = -\log\left(\sum_{\tau \in S_k}(2^{-\xi})^{d(\tau, \sigma) + d(\tau, I)} \right) \; \forall \sigma \in S_k.
		\label{eq:moment energy}
	\end{align}
	The function $h(\sigma)$ can be shown to depend only on the conjugacy class of $\sigma$ in $S_k$. From this, we can infer a $S_k \times S_k$ symmetry of this model, whose action is given by left and right group multiplication by uniform permutations,
	\begin{align}
		\pi_v &\mapsto \tau \pi_v \tau' & \text{ for } (\tau, \tau') \in S_k \times S_k.
	\end{align}
	In the special case $k = 2$, the permutations can only take on 2 different values, $\pi_v \in \{I, S\}$, where $I$ is the identity and $S$ swaps the two elements. The energy function can then be written as
	\begin{align}
		\mathcal{H}_{k=2}[\vec{\pi}] &= \beta_\xi\sum_{\{v_1, v_2\}\in E} \delta(\pi_{v_1} =\pi_{v_2}) & \text{with }J = \log\left(\frac{2^{2\xi}+1}{2^{\xi+1}}\right), \label{eq:Ising energy}
	\end{align}
	which is the free energy of a nearest-neighbor ferromagnetic Ising model at inverse temperature $\beta_\xi$. The Ising model undergoes a transition from a disordered to a long-range ordered phase as $\xi$ is increased past some $\mathcal{O}(1)$ critical value. This transition, where the local variables $\pi_v$ become positively correlated with one another at long distances, is associated with the long-range MIE transition in the underlying random circuit \cite{bao2024finite}. For higher $k$, and for other circuit architectures beyond the coarse-grained model, we similarly expect the spin model with weights \eqref{eq:spin weights general} to be in long-range ordered phase once the depth exceeds some constant critical value. As we shall see, the emergence of Scrooge-randomness can be connected to the universal statistical properties of this long-range ordered phase. Thus, while we will use the  coarse-grained architecture as a concrete example where explicit calculations can be made, we expect our conclusions to hold for any circuit architectures that features a long-range MIE phase, e.g.~2D brickwork circuits.
	
	In the following, we will also need to evaluate averages over subsets of gates. The same principles can be applied to these objects. For example, the moments of the state obtained after applying $U_{AB}$ only are
	\begin{align}
		\chi_{\mathcal{U}_{AB}}^{(k)} = \mathbb{E}_{U_{AB} \sim \mathcal{U}_{AB}}[(U_{AB}\dyad{0^{A B  \partial B}}U_{AB}^\dagger)^{\otimes k}] \propto \sum_{\vec{\pi}_{AB\partial B}}e^{-\mathcal{H}[\vec{\pi}_{AB\partial B}]}\bigotimes_{\substack{ v \in A\cup B\cup \partial B} }\text{Wg}_v[\pi_v]
		\label{eq:moment AB}
	\end{align}
	where $\partial B$ is the set of vertices in $C$ that share an edge with a vertex in $B$, and we now only sum over permutations for each vertex in $A \cup B \cup \partial B$, which is the support of $U_{AB}$. Similarly, we can define integer moments of the POVM operators \eqref{eq:POVM def}, which can be computed using the same techniques as the previous subsection.
	\begin{align}
		\mathbb{E}_{U_C} \sum_{x_C} F_{\partial B}(U_C, x_C)^{\otimes k}  
		\propto \sum_{\vec{\pi}_C} e^{-\mathcal{H}[\vec{\pi}_C]} V_{\partial B}(\vec{\pi}_{\partial B})
		\label{eq:POVM moment}
	\end{align}
	where $\mathcal{H}[\vec{\pi}_C]$ is energy function \eqref{eq:moment energy} that appeared in the evaluation of the moments of the output state. The sum is over all choices of permutations on each vertex $v \in C$, but the operator content of the right hand side only depends on the restriction to $\partial B \subset C$.  In the last equality, we have used the fact that $\braket{x_v^{\otimes k}|\text{Wg}_v(\pi_v)|x_v^{\otimes k}}$ is a positive constant independent of $\pi_v$.
	
	\subsection{Moments of post-measurement states and the replica trick}
	
	Let $O^A \in \mathcal{B}(\mathcal{H}_A^{\otimes k})$ be a positive semi-definite operator acting on $k$ copies of the Hilbert space of $A$. For a given $U_{AB}, x_B$, the corresponding moment in the post-measurement ensemble $\mathcal{E}_A(U_{AB}, x_B)$ was defined in the main text as \eqref{eq:k moment def}. This can be written as
	\begin{align}
		\chi^{(k)}_\mathcal{E}[U_{AB}, x_B, O^A] &\coloneqq \mathbb{E}_{U_C}\sum_{x_C} p(x_C|U,x_B) \tr[O_A \big(\psi_{A}(x_{BC}, U)\big)^{\otimes k}]. \nonumber\\
		&= \mathbb{E}_{U_C}\sum_{x_C} p(x_C|U,x_B)^{1-k} \tr[\big(O_A \otimes F_{\partial B}(U_C, x_C)^{\otimes k}\big) \psi_{A\partial B}(x_{B}, U_{AB})^{\otimes k}],
	\end{align}
	Due to the fourth characterization of the Scrooge ensemble in \cref{def:scrooge}, the moments of the Scrooge ensemble for the background state $\rho^A(U_{AB}, x_B)$ can be obtained by replacing the POVM defined by $F_{\partial B}(U_C, x_C)$ with a rank-1 Haar-random POVM over $\partial B$, giving
	\begin{align}
		\chi^{(k)}_{\text{Scr}}[U_{AB}, x_B, O^A] 
		&= d_{\partial B}\int \dif \mu_H(\phi_{\partial B}) p(\phi_{\partial B}|U_{AB},x_B)^{1-k} \tr[(O_A\otimes \phi_{\partial B}^{\otimes k}) \psi_A(U_{AB}, x_B, \phi_{\partial B})^{\otimes k}]
	\end{align}
	where $d_{\partial B}$ is the Hilbert space dimension on $\partial B$, $\dif \mu_H(\phi_{\partial B})$ is the Haar measure over $\mathbb{C}^{2^{|\partial B|}}$, and we define $p(\phi_{\partial B}|U_{AB}, x_B) \coloneqq \tr[\bra{\phi_{\partial B}} \psi_{A\partial B}(U_{AB}, x_B) \ket{\phi_{\partial B}}]$ by analogy to Eq.~\eqref{eq:post meas state POVM}. 
	Evidently, the dependence of these moments on $U_{AB}, x_B$ is fully determined by the partially conditioned state $\psi_{A \partial B}(U_{AB}, x_B)$ [Eq.~\eqref{eq:inter def}]. Thus, we can suppress the labels $U_{AB}, x_B$, and write $\chi^{(k)}_\mathcal{E}[\psi_{A \partial B}, O^A]$ for an arbitrary state $\psi_{A \partial B}$, and similar for $\chi^{(k)}_{\text{Scr}}[\psi_{A \partial B}, O^A]$.
    
	Unfortunately, the averages over $U_C$ and $\phi_{\partial B}$ in the above cannot be evaluated analytically, since neither integrand can be written as a polynomial function of $U_C$ and $\phi_{\partial B}$. Instead, we introduce a family of related quantities, labelled by non-negative integers $r$
	\begin{subequations}
		\begin{align}
			\chi_\mathcal{E}^{(k,r)}[\psi_{A \partial B}, O^A] &\coloneqq \frac{\mathbb{E}_{U_C}\sum_{x_C} \tr[\big(O_A \otimes F_{\partial B}(U_C, x_C)^{\otimes k}\big) \psi_{A \partial B}^{\otimes k}]\tr[\big(I_A \otimes F_{\partial B}(U_C, x_C)\big) \psi_{A \partial B}]^{r}}{\mathbb{E}_{U_C}\sum_{x_C} \tr[\big(I_A \otimes F_{\partial B}(U_C, x_C)\big) \psi_{A \partial B}]^{k+r}} \label{eq:post moment kr} \\
			\chi_{\text{Scr}}^{(k,r)}[\psi_{A \partial B}, O^A] &\coloneqq \frac{\int \dif \mu_H(\phi_{\partial B}) \tr[\big(O_A \otimes \phi_{\partial B}^{\otimes k}\big) \psi_{A \partial B}^{\otimes k}]\tr[\big(I_A \otimes \phi_{\partial B}\big) \psi_{A \partial B}]^{r} }{\int \dif \mu_H(\phi_{\partial B})\tr[(I_A \otimes \phi_{\partial B}) \psi_{A \partial B}]^{k+r}} \label{eq:scr moment kr}
		\end{align}
		\label{eq:gen moments kr}
	\end{subequations}
	The numerator and denominator in both of these expressions are each averages of polynomial functions of $U_C$ and $\phi_{\partial B}$, and hence can be related to integer moments of random unitaries which have explicit expressions (see the previous subsection). After evaluating these quantities for non-negative integer values of $r$, one could in principle analytically continue the resulting expressions to negative values of $r$, which would allow the desired quantities to be reproduced via
	\begin{subequations}
		\begin{align}
			\chi^{(k)}_\mathcal{E}[\psi_{A \partial B}, O^A] &= \lim_{r\rightarrow -k+1}  \chi_\mathcal{E}^{(k,r)}[U_{AB}, x_B, O^A], \\
			\chi^{(k)}_{\text{Scr}}[\psi_{A \partial B}, O^A] &= \lim_{r\rightarrow -k+1}  \chi_{\text{Scr}}^{(k,r)}[U_{AB}, x_B, O^A].
		\end{align}
		\label{eq:analytic continuation}
	\end{subequations}
	Performing this final analytic continuation step, known as the `replica trick' would require us to have a closed-form expression for these generalized moments. In practice, these are not obtainable, except for the simplest of models. Thus, we will only be able to characterize these generalized moments for non-negative $r = 0, 1, \ldots$, and not the required limit. Nevertheless, away from any phase transitions we will find that the ratio of the generalized moments  $\chi_\mathcal{E}^{(k,r)} / \chi_{\text{Scr}}^{(k,r)}$ is close to 1 for all non-negative $r$, suggesting that the dependence on $r$ is weak overall. Thus, we expect the ratio to remain close to 1 in the relevant limit $r \rightarrow 1-k$. This is a non-rigorous part of our argument, but one that mirrors successful approaches in previous works \cite{bao2020theory, jian2020measurement, napp2022efficient}.

	\subsection{Qualitative argument for convergence to the Scrooge ensemble}
	
	Using the framework described above, we now give an overview of our argument for convergence of the moments of the post-measurement ensemble towards those of the Scrooge ensemble, deferring a quantitative  discussion to Section \ref{sec:error quant}. We address the case $r = 0$, $k = 2$ first as it is the simplest nontrivial example. Inserting the expression \eqref{eq:POVM moment} for the moments of the POVM operators into Eq.~\eqref{eq:post moment kr}, we can write the generalized moment in question in the following form
	\begin{align}
		\chi^{(2,0)}_\mathcal{E}[\psi_{A \partial B}, O^A] &= \frac{\sum_{\vec{\pi}_C} e^{-\mathcal{H}[\vec{\pi}_C]} \tr[(O_A\otimes V_{\partial B}(\vec{\pi}_{\partial B})) \psi_{A \partial B}^{\otimes 2}]}{\sum_{\vec{\pi}_C} e^{-\mathcal{H}[\vec{\pi}_C]} \tr[(I_A^{\otimes 2}\otimes V_{\partial B}(\vec{\pi}_{\partial B})) \psi_{A \partial B}^{\otimes 2}]}
		= \frac{\sum_{\vec{\pi}_{\partial B}}p[\vec{\pi}_{\partial B}] \tr[(O_A\otimes V_{\partial B}(\vec{\pi}_{\partial B})) \psi_{A \partial B}^{\otimes 2}] }{\sum_{\vec{\pi}_{\partial B}}p[\vec{\pi}_{\partial B}] \tr[(I_A^{\otimes 2}\otimes V_{\partial B}(\vec{\pi}_{\partial B})) \psi_{A \partial B}^{\otimes 2}]} \label{eq:post moment 20}
	\end{align}
	where, following Eq.~\eqref{eq:Ising energy}, $\mathcal{H}[\vec{\pi}_C]$ is the free energy of the Ising model (restricted to vertices $v \in C$), with the two possible spin states corresponding to the identity and swap permutations. In the final equality in Eq.~\eqref{eq:post moment 20}, we have defined the probabilities
	\begin{align}
		p_{\partial B}[\vec{\pi}_{\partial B}] = \frac{\sum_{\vec{\pi}_C \sim \vec{\pi}_{\partial B}}e^{-\mathcal{H}[\vec{\pi}_C]}}{\sum_{\vec{\pi}_C'} e^{-\mathcal{H}[\vec{\pi}_C']}},
		\label{eq:boundary prob}
	\end{align}
	where the sum in the numerator is restricted to configurations $\vec{\pi}_C$ that coincide with $\vec{\pi}_{\partial B}$ on the boundary; this can be viewed as the distribution of boundary spins $v \in \partial B$ of the Ising model on the region $C$. A similar expression for the generalized moments of the Scrooge ensemble can also be obtained
	\begin{align}
		\chi^{(2,0)}_{\text{Scr}}[\psi_{A \partial B}, O^A] &= \frac{\sum_{\tau \in S_2} \tr[\big(O_A \otimes V_{\partial B}(\tau)\big) \psi_{A \partial B}^{\otimes 2}] }{\sum_{\tau \in S_2} \tr[\big(I_A^{\otimes 2} \otimes V_{\partial B}(\tau)\big) \psi_{A \partial B}^{\otimes 2}]} \label{eq:scr moment 20} \\ &
		= \frac{\sum_{\tau \in S_2} \tr[O_A V_{A}(\tau) \rho_{A}^{\otimes 2}] }{\sum_{\tau \in S_2} \tr[V_{A}(\tau) \rho_{A}^{\otimes 2}]} \label{eq:scr moment 20 A}
	\end{align}
	where the last equality follows from the fact that $\ket{\psi^{\otimes 2}_{A \partial B}}$ is invariant under the action of $V_A(\tau) \otimes V_{\partial B}(\tau)$. The expression \eqref{eq:scr moment 20} mirrors the right hand side of \eqref{eq:post moment 20}, with the difference that the average over configurations $\vec{\pi}_{\partial B}$ is restricted to those that are perfectly uniform, $\pi_{\partial B} = \tau^{\times \partial B}$, rather than being averaged over the boundary distribution \eqref{eq:boundary prob}.
	
	To show that the generalized moments (\ref{eq:post moment 20}, \ref{eq:scr moment 20}) converge towards one another, we will establish two key properties. First, in the ferromagnetic phase of the Ising model, which is associated with the long-ranged MIE phase, we expect that the distribution \eqref{eq:boundary prob} can be written as an equal-weight mixture of two different short-ranged correlated distributions $p[\vec{\pi}_{\partial B}] = \frac{1}{2}\sum_{\tau} p[\vec{\pi}_{\partial B}|\tau]$, labelled by $\tau \in \{I, S\}$. Typical configurations of $p[\vec{\pi}_{\partial B}|\tau]$ have a majority of spins in the state $\tau$, and the subset $X = \{v \in \partial B : \sigma_v \neq \tau\}$ will be made up of small `minority islands' (this will be made more explicit later on). We can then write $V_{\partial B}(\vec{\pi}_{\partial B}) = V_{X}(S) V_{\partial B}(\tau)$, and using the fact that $\ket{\psi_{A \partial B}^{\otimes 2}}$ is invariant under the action of uniform permutations $V_{A}(\tau) \otimes V_{\partial B}(\tau)\ket{\psi_{A \partial B}^{\otimes 2}} = \ket{\psi_{A \partial B}^{\otimes 2}}$, we have
	\begin{align}
		\tr[\big(O_A \otimes V_{\partial B}(\vec{\pi}_{\partial B})\big)\psi^{\otimes 2}_{A \partial B}] &= \tr[\big(O_A V_A(\tau) \otimes V_{X}(S)\big)\rho_{A X}^{\otimes 2}] 
		\label{eq:perm shift A}
	\end{align}
	where $\rho_{A X} = \tr_{\partial B \setminus X}[\psi_{A \partial B}]$. Subsequently, we will argue that for typical $\vec{\pi}_{\partial B}$ (where the minority region $X$ is made up of a dilute collection of small patches), the expectation value on the right hand side of \eqref{eq:perm shift A} approximately factorizes between $A$ and $X$,
	\begin{align}
		\tr[\big(O_A V_A(\tau) \otimes V_{X}(S)\big)\rho_{A X}^{\otimes 2}]  &\approx \tr[O_A V_A(\tau) \rho_{A}^{\otimes 2}] \tr[ V_{\bar{X}}(S)\rho_{ \bar{X}}^{\otimes 2}],
		\label{eq:corr factorize}
	\end{align}
	with high probability over $U_{AB}$, $x_B$, provided $\text{dist}(A,C)$ is large enough. This property reflects the fact that in the partially conditioned state $\psi_{A \partial B}(U_{AB}, x_B)$, information about the state of $A$ is encoded non-locally in $\partial B$, meaning that $A$ exhibits weak correlations with small subsystems $X \subset \partial B$. This same phenomenon can be related back to our statement in the main text that each successive measurement on $C$, which acts on a small region of $\partial B$, does not disturb the state on $A$. We note in passing that this non-local encoding of information has been argued to underpin the volume law phase in monitored quantum circuits \cite{gullans2020dynamical}, which is closely related to the long-ranged MIE phase in our current setting \cite{mcginley2025measurement}.
	
	The above statements will be made more quantitative later, but at this point we can appreciate at a high level how the two properties introduced here imply convergence of the moments $\chi_{\mathcal{E}}^{(2,0)}$ and $\chi_{\text{Scr}}^{(2,0)}$. Since a configuration $\vec{\pi}_{\partial B}$ is fully specified by a choice of majority spin $\tau$ and a minority region $X \subseteq \partial B$, we can re-express the sums in the numerator and denominator of Eq.~\eqref{eq:post moment 20} in terms of averages over $\tau$ (which is uniformly random over $\{I, S\}$), and $X \subseteq \partial B$, whose distribution $p[X]$ is independent of $\tau$ due to the global Ising symmetry $I \leftrightarrow S$. Then,
	\begin{align}
		\chi^{(2,0)}_\mathcal{E}[\psi_{A \partial B}, O^A] &= \frac{\sum_{\tau \in \{I, S\}} \sum_{X \subset \partial B}p[X] \tr[(O_A V_A(\tau) \otimes V_{X}(S)) \psi_{A \partial B}^{\otimes 2}] }{\sum_{\tau \in \{I, S\}} \sum_{X\subset \partial B}p[X] \tr[(V_A(\tau) \otimes V_{X}(S)) \psi_{A \partial B}^{\otimes 2}] } & \text{by Eq.~\eqref{eq:perm shift A}} \nonumber\\
		&\approx  \frac{\sum_{\tau \in \{I, S\}} \tr[O_A V_A(\tau)  \rho_A^{\otimes 2}] \sum_{X \subset \partial B}p[X] \tr[V_{\bar{X}}(S) \rho_{\bar{X}}^{\otimes 2}] }{\sum_{\tau \in \{I, S\}} \tr[V_A(\tau)  \rho_A^{\otimes 2}] \sum_{X \subset \partial B}p[X]  \tr[V_{\bar{X}}(S) \rho_{X}^{\otimes 2}] } & \text{by Eq.~\eqref{eq:corr factorize}} \nonumber\\
		&= \chi^{(2,0)}_{\text{Scr}}[\psi_{A \partial B}, O^A].
	\end{align}
	In going to the second line, we invoke the approximation \eqref{eq:corr factorize} with $O_A$ in the numerator, and with $O_A$ replaced by $I_A^{\otimes k}$ in the denominator.
	
	We can also develop a similar line of reasoning for larger values of $k$, which allows us to make claims regarding higher moments of the post-measurement ensemble, and also larger values of $r$, which adds credibility to our claim that the correct analytically continued moments \eqref{eq:analytic continuation} should also converge towards one another. For a general $k$ and $r$, note that
	\begin{align}
		\chi_{\mathcal{E}}^{(k,r)}[U_{AB}, x_B, O_A] = \chi_{\mathcal{E}}^{(k+r,0)}[U_{AB}, x_B, O_A\otimes I_A^{\otimes r}],
        \label{eq:replica shift}
	\end{align}
	and similar for the generalized Scrooge moments. Thus, by showing convergence of $\chi_{\mathcal{E}}^{(k,0)}$ for arbitrary positive semi-definite observables with a higher number of copies $k$, we immediately obtain the same for higher $r$. We therefore work with $r = 0$ without loss of generality. These moments can be written in a form analogous to Eqs.~(\ref{eq:post moment 20}, \ref{eq:scr moment 20}),
	\begin{align}
		\chi^{(k,0)}_\mathcal{E}[\psi_{A \partial B}, O^A] &= \frac{\sum_{\vec{\pi}_{\partial B} \in S_k^{\partial B}}p^{(k)}[\vec{\pi}_{\partial B}] \tr[(O_A\otimes V_{\partial B}(\vec{\pi}_{\partial B})) \psi_{A \partial B}^{\otimes k}] }{\sum_{\vec{\pi}_{\partial B} \in S_k^{\partial B} }p^{(k)}[\vec{\pi}_{\partial B}] \tr[(I_A^{\otimes k}\otimes V_{\partial B}(\vec{\pi}_{\partial B})) \psi_{A \partial B}^{\otimes k}] }
		\label{eq:post moment k0} \\
		\chi^{(k,0)}_{\text{Scr}}[\psi_{A \partial B}, O^A] &= \frac{\sum_{\tau \in S_k} \tr[O_A V_{A}(\tau) \rho_{A}^{\otimes k}] }{\sum_{\tau \in S_k} \tr[V_{A}(\tau) \rho_{A}^{\otimes k}]}
		\label{eq:scr moment k0}
	\end{align}
	where now the variables $\pi_v$ take values in $S_k$, and thus can be interpreted as $(k!)$-state spins, and $p^{(k)}[\vec{\pi}_{\partial B}]$ is the distribution of spin configurations on the boundary $\partial B \subset C$ of a classical statistical mechanics model on $C$ whose free energy is given by Eq.~\eqref{eq:moment energy}. Deep enough in the long-ranged MIE phase, this model will exhibit long-ranged ferromagnetic order, just as we had for the $k = 2$ Ising case. This arises through spontaneous breaking of the $S_k \times S_k$ symmetry exhibited by the free energy \eqref{eq:moment energy} (generated by global transformations $\pi_v \mapsto \sigma \pi_v \sigma'$, for $\sigma, \sigma' \in S_k$) down to a diagonal $S_k$ subgroup. Analogously to the $k=2$ case, in this phase we expect that $p^{(k)}[\vec{\pi}_{\partial B}]$ can be written as an equal-weight mixture of $k!$ different distributions $p[\vec{\pi}_{\partial B}|\tau]$, typical configurations of which are dominated by a particular majority permutation $\tau \in S_k$. Letting $X$ be the set of sites on which $\vec{\pi}_v \neq \tau$, we can again write $V_{\partial B}(\vec{\pi}_{\partial B}) = V_{\partial B}(\tau)V_X(\vec{\pi}'_X)$, where $\vec{\pi}'_X$ is a list of permutations over $X$ only. Then, by the same arguments, convergence of the higher moments amounts to showing that for small enough regions $X$, we have
	\begin{align}
		\tr[\big(O_A \otimes V_{\partial B}(\vec{\pi}_{\partial B})\big)\rho_{A \partial B}^{\otimes k}]  = \tr[\big(O_A V_A(\tau) \otimes V_{X}(\vec{\pi}_{X})\big)\rho_{A X}^{\otimes k}]  &\approx \tr[O_A V_A(\tau) \rho_{A}^{\otimes k}] \tr[ V_{X}(\vec{\pi}_{X})\rho_{X}^{\otimes k}].
		\label{eq:corr factorize higher}
	\end{align}
	In the following section, we will present a more detailed and quantitative version of this argument, being careful to characterise the \textit{relative} error between the moments as accurately as possible.
	
	\subsection{Quantifying the error \label{sec:error quant}}
	
	\textit{Proxy expression for relative error.---}While the necessary use of the replica trick means that we will not be able to rigorously verify \eqref{eq:scroogeeps full}, our ambition is to demonstrate that this is a plausible upper bound on the \textit{relative} error between the moments, on the basis that the ratio of the proxy quantities \eqref{eq:gen moments kr} likely behaves similarly to that of the true moments in the appropriate limit \eqref{eq:analytic continuation}.  Following the overall logic outlined in the previous section, we will identify a set of quantities that serves as a plausible proxy for the relative error, which we can directly analyse in terms of the statistical mechanics of the classical spin models introduced in Section \ref{subsec:output moments}.
	
	Starting from the expression for the generalized moments of the post-measurement ensemble \eqref{eq:post moment k0}, we recall that the distribution $p^{(k)}[\vec{\pi}_{\partial B}]$ is a marginal of the distribution of spins $p^{(k)}[\vec{\pi}_{C}]$ on $C \supset \partial B$ [cf.~Eq.~\eqref{eq:boundary prob}]. As previously explained, we expect that this distribution can be decomposed as an equal-weight mixture of $k!$ sectors, corresponding to the different symmetry-broken free energy minima, each of which is distinguished by its majority spin configuration $\tau \in S_k$. We will make this decomposition explicit by partitioning the spin configurations on $C$ into $k!$ mutually exclusive subsets $\mathcal{G}_\tau \subset S_k^C$, labelled by $\tau \in S_k$, such that $S_k^C = \bigcup_{\tau \in S_k} \mathcal{G}_\tau$. These subsets will be chosen to satisfy
	\begin{align}
		\vec{\pi}_C \in \mathcal{G}_\tau &\Longleftrightarrow \tau' \cdot \vec{\pi}_C \in \mathcal{G}_{\tau' \tau} & \forall \tau, \tau' \in S_k
		\label{eq:gtau}
	\end{align}
	where $\tau' \cdot \vec{\pi}_C$ denotes the configuration in which the spin at vertex $v \in C$ is in the state $\tau'\pi_v$. Then, we can write the distribution of boundary spin configurations as a mixture of $k!$ conditional distributions
	\begin{align}
		p^{(k)}[\vec{\pi}_{\partial B}] &= \frac{1}{k!}\sum_{\tau \in S_k} p^{(k)}[\vec{\pi}_{\partial B}|\pi_C \in \mathcal{G}_\tau] & \text{where } p^{(k)}[\vec{\pi}_{\partial B}|\pi_C \in \mathcal{G}_\tau] &\coloneqq k!\sum_{\vec{\pi}_C \sim \vec{\pi}_{\partial B}} p^{(k)}[\vec{\pi}_{C}] \mathbbm{1}(\vec{\pi}_{C} \in \mathcal{G}_\tau).
		\label{eq:g prob}
	\end{align}
	By virtue of the property \eqref{eq:gtau}, $p^{(k)}[\vec{\pi}_{\partial B}|\mathcal{G}_\tau]$ defines a valid and properly normalized probability distribution for each $\tau \in S_k$, as desired. We defer an explicit choice of these subsets until later, but for now suppose that  $\mathcal{G}_\tau$ have been specified, and have the effect of `picking out' configurations $\vec{\pi}_C$ that have a majority state $\tau$. 
	
	The subsets $\mathcal{G}_\tau$ can be used to define a corresponding set of operators on $\partial B$ via
	\begin{align}
		W_{\partial B}(\tau) &\coloneqq \sum_{\vec{\pi}_C} p^{(k)}[\vec{\pi}_C] \mathbbm{1}(\vec{\pi}_{C} \in \mathcal{G}_\tau) V_{\partial B}(\vec{\pi}_{\partial B}),
		\label{eq:WB def}
	\end{align}
	which by virtue of Eq.~\eqref{eq:gtau} satisfy $V_{\partial B}(\tau') W_{\partial B}(\tau) = W_{\partial B}(\tau' \tau)$. Note that, by permutation invariance, the generalized moments (\ref{eq:post moment k0},\ref{eq:scr moment k0}) only depend on the part of $O_A$ supported on the symmetric subspace, namely $\Pi^{\rm sym}_{A}O_A \Pi^{\rm sym}_{A}$, where $\Pi^{\rm sym}_A = \frac{1}{k!}\sum_{\tau S_k}V_A(\tau)$. Thus, without loss of generality we can assume $\Pi^{\rm sym}_A O_A  = O_A \Pi^{\rm sym}_A = O_A$, and in turn $V_A(\tau)O_A = O_A$ for all $\tau \in S_k$. Then, by the arguments of the previous section, in particular Eq.~\eqref{eq:corr factorize higher}, our expectation is that the ratio
	\begin{align}
		\Upsilon_{k}(\psi_{A \partial B}, O_A) \coloneqq  \frac{\tr[\big(O_A\otimes W_{\partial B}(I)\big)\psi^{\otimes k}_{A \partial B}]}{\tr[O_A  \rho_A^{\otimes A}] \tr[W_{\partial B}(I) \rho_{\partial B}^{\otimes k}]}
		\label{eq:ups def}
	\end{align}
	should be close to unity. Demonstrating this would allow us to upper bound the relative error between the $(k,0)$th generalized moments as
	\begin{align}
		\epsilon^{(k)} &= \left| \frac{\chi_{\mathcal{E}}^{(k,0)}[\psi_{A\partial B}, O_A] }{\chi_{\rm Scr}^{(k,0)}[\psi_{A\partial B}, O_A]} - 1 \right| = \left| \frac{ \Upsilon_{k}(\psi_{A \partial B}, O_A) }{ \Upsilon_{k}(\psi_{A \partial B}, \Pi_A^{\rm sym}) } - 1 \right| \leq \frac{1+\epsilon_O}{1-\epsilon_\Pi} - 1 
        \label{eq:error upsilon} \\
		\text{where }\epsilon_O &\coloneqq \left| \Upsilon_{k}(\psi_{A \partial B}, O_A) - 1\right| \nonumber\\ \nonumber
		\epsilon_\Pi &\coloneqq \left| \Upsilon_{k}(\psi_{A \partial B}, \Pi^{\rm sym}_A) - 1\right|.
	\end{align}
	In the above, we have exploited the permutation invariance of $O_A$ and $\Pi_A^{\rm sym}$.

	Recall that the state $\psi_{A \partial B}$, and its marginals which appear in \eqref{eq:ups def}, is conditioned on the random variables $U_{AB}$, $x_B$. Thus, the quantities $\epsilon_{O, \Pi}$ are random variables, and our aim now is to show that these are small with high probability over $U_{AB}, x_B$. A natural way to show this is by characterising the moments of $\Upsilon_k(\psi_{A \partial B}, O_A)$ over $U_{AB}, x_B$. However, since the states appearing in the definition of $\Upsilon_k$ are (marginals of) post-measurement states, we cannot directly compute  these moments, and we must instead resort to a replica trick for the second time. (This is to be expected since we are averaging over $(U_C, x_C)$ and $(U_{AB}, x_B)$ in separate steps, hence the two separate uses of the replica trick.) Let us define the unnormalized state
	\begin{align}
		\dyad{\tilde{\psi}_{A \partial B}(x_B)} \coloneqq \bra{x_B}U_{AB}\dyad{0}U_{AB}^\dagger\ket{x_B},
		\label{eq:psi unnormalized def}
	\end{align}
	which differs from its normalized counterparts (without tildes) only by a scalar multiple of $p_{x_B} = \tr[\tilde{\psi}_{A \partial B}(x_B)]$. Using the shorthand $\braket{M_A \otimes M_{\partial B}} = \tr[(M_A \otimes M_B)\tilde{\psi}_{A \partial B}(x_B)^{\otimes k}]$ and $\braket{M_A} \equiv \braket{M_A \otimes I_{\partial B}^{\otimes k}}$, we observe that the family of quantities
	\begin{align}
		\Xi^{(t)}_{k,s}[O_A] = \frac{\big(\mathbb{E}_{U_{AB}} \sum_{x_B} p_{x_B}^{t(ks-k-1)}\braket{O_A}^{s-1} \braket{W_{\partial B}(I)}^{s-1} \braket{O_A\otimes W_{\partial B}(I)}^{ts+1}  \big)\big(\mathbb{E}_{U_{AB}} \sum_{x_B} p_{x_B}^{(t+1)(2ks-k-1)+1}\big)}{\big(\mathbb{E}_{U_{AB}} \sum_{x_B} p_{x_B}^{(t+1)(ks-k-1)+1} \braket{O_A}^{s(t+1)} \big)\big(\mathbb{E}_{U_{AB}} \sum_{x_B} p_{x_B}^{(t+1)(ks-k-1)+1} \braket{W_{\partial B}(I)}^{s(t+1)} \big)},
		\label{eq: xi def}
	\end{align}
	has the property that
	\begin{align}
		\mathbb{E}_{U_{AB}, x_B}\big[\Upsilon_{k}[\psi_{A \partial B}(U_{AB}, x_B), O_A]^{1-s} \big] = \lim_{t \rightarrow -1} \Xi^{(t)}_{k,s}[O_A].
		\label{eq: moments ups}
	\end{align}
	As before, we have constructed $\Xi^{(t)}_{k,s}[O_A]$ such that it can be related to positive integer moments of $U_{AB}$ when $s, t$ are non-negative integers satisfying $s \geq t + 1$. We will argue that $\Xi^{(t)}_{k,s}$ is close to unity for non-negative values of $t$, irrespective of $s$. This provides evidence that the true moments \eqref{eq: moments ups} are near unity as well. In turn, this bound will be used to justify a probabilistic statement regarding the relative deviation of the moments of the post-measurement and Scrooge ensembles, of the kind stated in the \hyperlink{hyp:scrooge}{Scrooge Hypothesis} in the main text. This is because, for any $\epsilon < 1/2$, we have
	\begin{align}
		\text{Pr}\Big(|\Upsilon_k - 1| > \varepsilon\Big) &\leq \text{Pr}\Big(|\Upsilon^{-1}_k - 1| > \varepsilon/2\Big) \nonumber\\ & \leq 4\varepsilon^{-2} \mathbb{E}\big[(\Upsilon^{-1}_k - 1)^2\big] \nonumber\\ &\leq 4\varepsilon^{-2}\left(\left| \mathbb{E}\big[\Upsilon^{-2}_k\big] - 1\right| +  2\left| \mathbb{E}\big[\Upsilon^{-1}_k\big] - 1\right| \right).
        \label{eq:upsilon conc ineq}
	\end{align}
	Thus, if we accept $\Xi^{(t)}_{k,s}$ as a replica proxy for the $(1-s)$th moments of $\Upsilon_k$ [in light of Eq.~\eqref{eq: moments ups}], then if we can bound its deviation away from unity as
    \begin{align}
		\big|\Xi^{(t)}_{k,s}[O_A] - 1 \big| \leq \delta
        \label{eq:xi deviation}
	\end{align}
    for some $\delta> 0$, we can reasonably infer a concentration inequality on $\Upsilon_k$ in the form $\text{Pr}(|\Upsilon_k-1|<\varepsilon)\leq 12 \delta/\varepsilon^2$. In turn, by Eq.~\eqref{eq:error upsilon}, we would have $\text{Pr}(|\epsilon^{(k)}-1|<\varepsilon)\leq 48 \delta/\varepsilon^2$, where $\epsilon^{(k)}$ is the relative error between the moments $\chi_{\mathcal{E}}^{(k,0)}$ and $\chi_{\text{Scr}}^{(k,0)}$. In particular, for
    \begin{align}
        \varepsilon_{\rm scr} = (48\delta)^{1/3} 
		\label{eq:proxy error}
    \end{align}
    we obtain
    \begin{align}
        \text{Pr}(|\epsilon^{(k)}-1|<\varepsilon_{\rm scr})\leq \varepsilon_{\rm scr}.
        \label{eq:epsilon conc ineq}
    \end{align}
    Given that $\chi_{\mathcal{E}}^{(k,0)}$ and $\chi_{\text{Scr}}^{(k,0)}$ are related to the desired moments $\chi_{\mathcal{E}}^{(k)}$ and $\chi_{\text{Scr}}^{(k)}$ via a second replica trick [Eqs.~(\ref{eq:gen moments kr}, \ref{eq:replica shift})], we will use \eqref{eq:epsilon conc ineq} to justify the statement made in the Scrooge hypothesis, with the proxy error given by Eqs.~(\ref{eq:xi deviation}, \ref{eq:proxy error}).
    
	\textit{Bounding the proxy error.---}Note that each of the averages in Eq.~\eqref{eq: xi def} is in the form of a moment over $U_{AB} \sim \mathcal{U}_{AB}$, with $m = (t+1)(2ks-k-1) + 1$ copies. Using the formulae for moments of random circuits derived in Section \ref{subsec:output moments}, we have 
	\begin{align}
		\sum_{x_B} \mathbb{E}_{U_{AB}}[\tilde{\psi}_{A \partial B}(x_B)^{\otimes m}] &= \sum_{x_B}\tr[(I_{A \partial B} \otimes \dyad{x_B})^{\otimes m} \chi_{\mathcal{U}_{AB}}^{(m)}] \nonumber\\ &\propto \sum_{\vec{\pi}_{AB \partial B}} e^{-\mathcal{H}[\vec{\pi}_{AB \partial B}]} \text{Wg}_A(\pi_A) \otimes \text{Wg}_{\partial B}(\pi_{\partial B})
		\label{eq:unnormalized average}
	\end{align}
	which in principle can be used to evaluate all the necessary averages in Eq.~\eqref{eq: xi def}. Since the general expression \eqref{eq: xi def} is cumbersome, it is instructive to first consider the simplest non-trivial case of $t = 0, s = 1, k = 2$, for which $m = 2$, and thus the average over spins $\pi_{AB \partial B}$ is with respect to the Ising model, Eq.~\eqref{eq:Ising energy}.
	\begin{align}
		\Xi_{2,1}^{(0)}[O_A] &= \frac{\big(\mathbb{E}_{U_{AB}} \sum_{x_B} \tr[(O_A\otimes W_{\partial B}(I))\tilde{\rho}^{\otimes 2}_{A\bar{X}}] \big)\big(\mathbb{E}_{U_{AB}} \sum_{x_B} p_{x_B}^{2}\big)}{\big(\mathbb{E}_{U_{AB}} \sum_{x_B}  \tr[O_A\tilde{\rho}_{A}^{\otimes 2}] \big)\big(\mathbb{E}_{U_{AB}} \sum_{x_B} \tr[ W_{\partial B}(I)\tilde{\rho}_{\bar{X}}^{\otimes 2}] \big)} \nonumber\\ &= \frac{\sum_{\vec{\pi}_{AB\partial B}} e^{-\mathcal{H}[\vec{\pi}_{AB\partial B}]}  f_O(\pi_{A}) p[\pi_{\partial B}|\mathcal{G}_I]}{\sum_{\vec{\pi}_{AB\partial B}} e^{-\mathcal{H}[\vec{\pi}_{AB\partial B}]}  f_O(\pi_{A})\mathbbm{1}(\vec{\pi}_{\partial B} = I ) } \frac{\sum_{\vec{\pi}_{AB\partial B}} e^{-\mathcal{H}[\vec{\pi}_{AB\partial B}]} \mathbbm{1}(\vec{\pi}_{A} = I )\mathbbm{1}(\vec{\pi}_{\partial B} = I ) }{\sum_{\vec{\pi}_{AB\partial B}} e^{-\mathcal{H}[\vec{\pi}_{AB\partial B}]} \mathbbm{1}(\vec{\pi}_{A} = I )p[\pi_{\partial B}|\mathcal{G}_I] } \nonumber\\
		&= \frac{\mathbb{E}_{\pi_{ABC}}[f_O(\pi_{A})|\vec{\pi}_{C} \in \mathcal{G}_I]}{\mathbb{E}_{\pi_{ABC}}[f_O(\pi_{A}) | \pi_C = I]} \frac{\mathbb{P}_{\pi_{ABC}}[\vec{\pi}_{A} = I |\vec{\pi}_{C} = I ]}{\mathbb{P}_{\pi_{ABC}}[\vec{\pi}_{A} = I |\vec{\pi}_{C} \in \mathcal{G}_I]}
		\label{eq:xi ising}
	\end{align}
	Here, $f_O(\pi_A) \coloneqq \tr[\text{Wg}_A(\vec{\pi}_A) O_A]$, and $\mathbb{E}_{\pi_{ABC}}[\cdot | \cdot]$ denotes a conditional expectation value with respect to the Ising model on $ABC$. In the above we have used $\tr[\text{Wg}(\vec{\pi}_{\partial B})W_I] = p[\vec{\pi}_{\partial B}|\tau] = \sum_{\pi_C \sim \pi_{\partial B}} p[\pi_C]g_\tau(\pi_C)$. We see that the quantity in question maps to a ratio of conditional probabilities and expectations in the Ising model. In particular, these are functions of the distribution of spins on $A$ conditioned on two different events in $C$, namely $\pi_C = I$ and $\pi_C \in \mathcal{G}_I$. Since the set of configurations $\mathcal{G}_I \subset S_2^C$ is to be chosen to contain configurations that are mostly in the state $I$, and the regions $A$ and $C$ are separated by $\text{dist}(A,C)$, we expect that the two conditional distributions on $A$ will be approximately the same, which would imply that $\Xi_{2,1}^{(0)}$ is correspondingly close to unity. Indeed, using various rigorous methods from classical statistical mechanics, we can prove the following bound:
	\begin{proposition}
		\label{prop:replica bound}
		For the coarse-grained model with depth parameter $\xi > \xi^*$, where $\xi^*$ is some threshold independent of system size, and with a partition $ABC$ satisfying the conditions 1--5, the proxy error $\Xi_{2,1}^{(0)}$ [Eq.~\eqref{eq: xi def}] is bounded as
		\begin{align}
			\left|\Xi_{2,1}^{(0)}[O] - 1\right| \leq |C| e^{-\Omega(l_C)} + \textup{min}(|\partial A|, |\partial C|) e^{-\Omega(\textup{dist}(A, C))}
			\label{eq:error ising}
		\end{align}
		for any positive semi-definite observable $O$ that can be written in the form $\sum_x c_x \dyad{x}^{\otimes 2}$ for computational basis states $\ket{x}$ and coefficients $c_x \geq 0$.
	\end{proposition}
	This rigorous result is limited to the case $k = 2, s = 1, t = 0$, since our proof relies on certain correlation inequalities in the classical spin models described in Section \ref{subsec:output moments}, which are only known to hold for the Ising case ($m = 2$ replicas in total). Nevertheless, our overall argument can be formulated for higher replica indices, and we will argue that the same bound should apply to $\Xi^{(t)}_{k,s}$ for all $k,s,t = O(1)$.
	
	Our proof of Eq.~\eqref{eq:error ising} proceeds as follows. First, for diagonal $O$, we can set $f_O(\pi_A) = 1$, so we need only consider the second ratio in Eq.~\eqref{eq:xi ising}. Since the spins interact only via nearest-neighbour interactions, applying a `hard' conditioning $\pi_C = I$ reduces the distribution of the remaining spins $\pi_{AB}$ to that of a collection of Ising models---one for each connected component of $AB$, which we denote $\Lambda_i$ for $i = 1, \ldots, r_B$, each having fixed boundary conditions $\pi_{\partial \Lambda_i} = I$. Conditioning on the event $\pi_C \in \mathcal{G}_I$ instead corresponds to a `softer' boundary condition where we only assert that the spins on $C$ are `mostly' in the state $I$. (In particular, as will be made precise when the subsets $\mathcal{G}_\tau$ are specified later on, the configurations $\pi_C$ selected by this soft condition are those which, after a certain `cleaning up' procedure is applied to the region $R_C \subseteq C$, leaves the spin at $v^\star$ in the state $I$.) We want to show that the probability of the event $\pi_A = I$ is approximately the same for the hard and soft conditions.
	
\begin{figure}
		\centering
		\includegraphics[width=\textwidth]{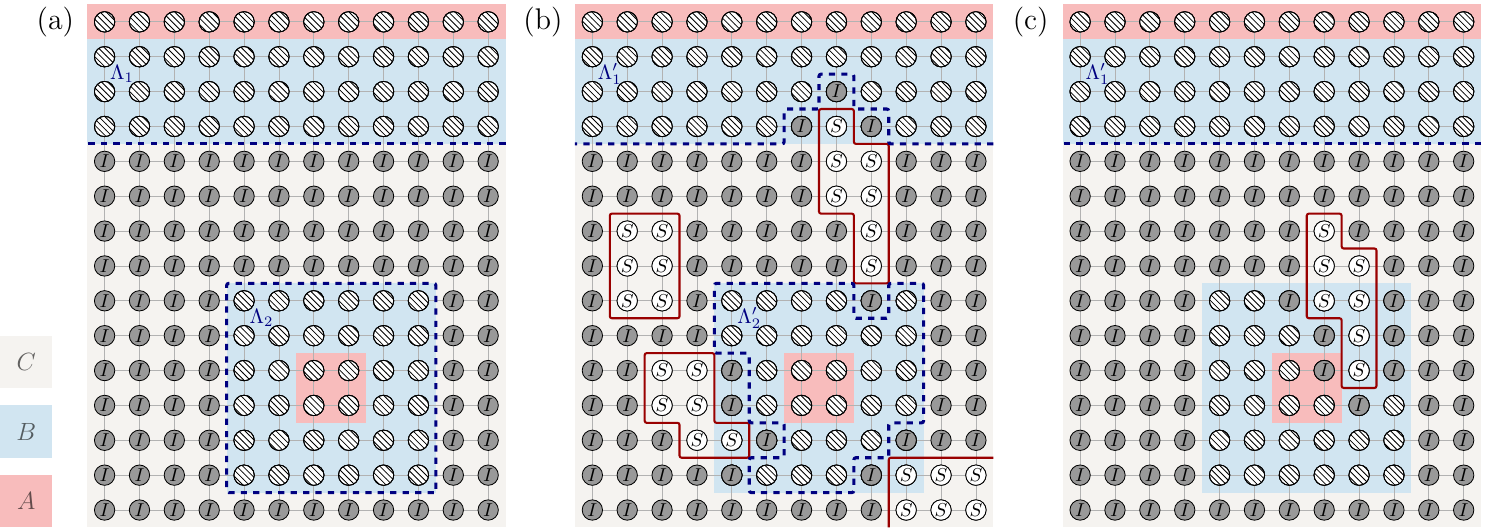}
		\caption{Spin configurations contributing to the conditional probabilities and expectations arising in the expression \eqref{eq:xi ising}, for a particular choice of tripartition $A$ (red), $B$ (blue), and $C$ (light grey). (a): Conditioning on $\pi_C = I$ means all spins in $C$ are frozen to the state $I$ (grey circles). These frozen spins define effective boundary conditions for Ising models on each connected component $\Lambda_i \subset AB$ (boundaries of which are shown with blue dashed lines). Within each component, spins are free to fluctuate (black and white striped circles). (b): After conditioning on the event $\vec{\pi}_C \in \mathcal{G}_I$, a typical configuration has most spins in the state $I$, but with some small minority islands in the state $S$ (white circles), separated from the $I$ spins by domain walls (dark red lines). As long as these islands are sufficiently small and dilute, it is possible to find subregions $\Lambda_i' \subseteq \Lambda_i$ whose boundary $\partial \Lambda_i'$ (dark blue dashed line) is surrounded by spins in the state $I$. (c) A bad configuration involves a domain wall that intersects both  $C$ and the boundary of $A$. 
        In this case, the domain wall fixes one of the spins that neighbours $A$ to the state $S$, and so we cannot find a $\Lambda_i'$ with the desired properties.}
		\label{fig:inner circuit}
	\end{figure}

	After conditioning on $\mathcal{G}_I$, we show that it is highly likely that there are a collection of regions $\Lambda_i'$, such that 1) $A_i \subseteq \Lambda_i' \subseteq \Lambda_i$, where $A_i = A \cap \Lambda_i$, and 2) all the spins on the boundary $\partial \Lambda_i'$ are in the state $I$, see Fig.~\ref{fig:inner circuit}(b). Conditioning further on this event, the resulting distribution of spins on $A$ can then be described as a product of Ising models on these subregions $\Lambda_i'$, still having fixed ($I$) boundary conditions. 
	
	The two terms in our expression \eqref{eq:error ising} can now be readily interpreted. The second term corresponds to the probability that no $\Lambda_i'$ with the above properties can be found for some component $i \in \{1, \ldots, r_B\}$. This only occurs if there is a `domain wall' (i.e.~a boundary between two regions on which the spins are in different states) that touches both $C$ and the boundary of $A$, see Fig.~\ref{fig:inner circuit}(c). In the ferromagnetic phase associated with long-range MIE, domain walls become exponentially rare in their length, and since every such domain wall must pass through a point on $\partial A$, this accounts for the resulting form.
	
	The first term corresponds to the event that the spin configuration on $C$ does not favour $I$ globally, which renders its categorization into the set $\mathcal{G}_I$ physically unmeaningful. Since the categorization of configurations $\pi_C$ into subsets $\mathcal{G}_\tau$ will be based on the restriction of the configuration to $R_C$, this issue can only arise if there is a domain wall that surrounds a significant portion of $R_C$ (in particular, including the vertex $v^\star$), such that $R_C$ is mostly in one configuration, while $A$ is mostly in a different configuration. The minimum length of such a domain wall will be bounded by the parameter $l_C$, hence the form \eqref{eq:error ising}.
    
	\subsection{Generalizing to higher replica indices and contour analysis}
	
	Having outlined the structure of our argument within the context of the simple case $k = 2, s = 1, t = 0$, we now seek to generalize and substantiate this argument, and so return to the more general expression \eqref{eq: xi def}. Observe that each of the four averages over $U_{AB}$ in \eqref{eq: xi def} can be related to the $m$-th moments of the state obtained after applying $U_{AB}$ only, $\chi^{(m)}_{\mathcal{U}_{AB}} = \mathbb{E}_{U_{AB} \sim \mathcal{U}_{AB}}[(U_{AB}\dyad{0^{A B  \partial B}}U_{AB}^\dagger)^{\otimes m}]$. Here, $m = (t+1)(2ks-k-1)+1$ is the total replica number, which tends to 1 in the appropriate limit \eqref{eq: moments ups}. Thanks to Eq.~\eqref{eq:unnormalized average}, these moments can be mapped to averages over configurations of a statistical mechanics model, with $(m!)$-state spins on each vertex in $AB$, whose free energy is given by  \eqref{eq:moment energy}. For example, since $\tr[\text{Wg}_{A\partial B}(\pi_{A \partial B})] = \mathbbm{1}(\pi_{\partial B} = I)$, the second average in the numerator of \eqref{eq: xi def} can be written as
	\begin{align}
		\mathbb{E}_{U_{AB}} \sum_{x_B} p_{x_B}^{m} \propto \sum_{\vec{\pi}_{AB\partial B} }  e^{-\mathcal{H}[\vec{\pi}_{AB\partial B}]} \mathbbm{1}(\vec{\pi}_A = I) \mathbbm{1}(\vec{\pi}_{\partial B} = I)
		\label{eq:collision average AB}
	\end{align}
	which we observe to be proportional to the partition function with spins on $A$ and $\partial B$ pinned to the identity state.
	
\begin{figure}
		\includegraphics[width=\textwidth]{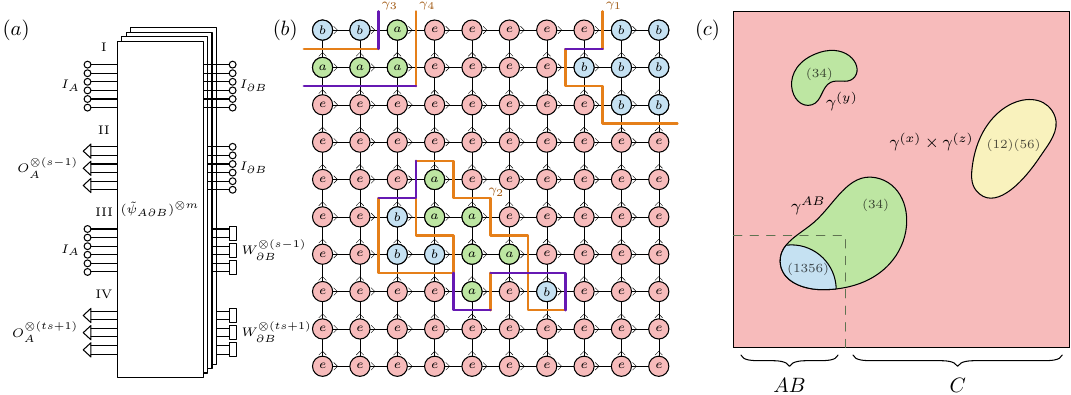}
		\caption{(a) Structure of operators $L_A^{(m)}$ and $L^{(m)}_{\partial B}$ [Eqs.~(\ref{eq:LA def}, \ref{eq:LB def})], which are used along with Eq.~\eqref{eq:unnormalized average} to obtain the expression \eqref{eq:xi LAB}. Circles represent $I_A$, triangles represent the $k$-copy observable $O_A$, and rectangles the $k$-copy operator $W_{\partial B}(I)$. (b) Example configuration $\pi$ for $S_3$-valued spins, along with the corresponding contours $\Gamma[\pi]$. The orientation of each edge is indicated with an arrow. For simplicity we choose a particular configuration that only features spins in the states $e = (1)(2)(3)$, $a = (123)$, and $b = (132)$, and so the relative permutations of the contours $\sigma_\gamma$ also take the values $(123)$ (orange lines) or $(132)$ (purple lines). Note that contours $\gamma_3$ and $\gamma_4$ are separate, even though they are nested inside one another. (b) Coarse-grained illustration of a spin configuration of the kind appearing in \eqref{eq:cond exp G}, along with its contours, which we now classify into $AB$-type and $c$-type. Here, we choose $k = 2$, $s = 2$, $t = 0$, so that the total number of copies is $m = 6$, with copies $x = \{1,2\}$ forming the sole subcomponent of II, and $y = \{3,4\}$ and $z = \{5,6\}$ forming subcomponents of IV. The subgroup $T_C$ of permutations allowed in $C$ is $S_2^{x} \times S_2^{y} \times S_2^{z}$. The contour that intersects both $C$ and $AB$ is a $AB$-contour, the boundary of the $(34)$-region constitutes a $y$-contour, while the boundary of the $(12)(56)$ region should be thought of as two separate contours: one $x$-contour and one $z$-contour.}
		\label{fig:contours}
	\end{figure}

	The other three averages can similarly be related to partition functions, but now with more complicated boundary conditions on $A$ and $\partial B$ which can be specified in terms of the following operators.
	\begin{alignat}{5}
		L_A^{(m)} =& I_A^{\otimes t(ks-k-1)} & &\otimes  O_A^{\otimes (s-1)}& \otimes & \; I_A^{\otimes k(s-1)}  &\otimes & \; O_A^{\otimes (ts+1)} \label{eq:LA def} \\
		L_{\partial B}^{(m)} =&\underbrace{I _A^{\otimes t(ks-k-1)}}_{\text{I}} && \otimes  \underbrace{I_A^{\otimes k(s-1)}}_{\text{II}}& \otimes & \underbrace{W_{\partial B}(I)^{\otimes (s-1)}}_{\text{III}}  &\otimes & \underbrace{W_{\partial B}(I)^{\otimes (ts+1)}}_{\text{IV}} \label{eq:LB def}
	\end{alignat}
	Here, we are being careful about the consistency of  ordering of each factor in the tensor products between these two expressions, by subdividing the $m$ copies into four components I, II, III, IV, consisting of $t(ks-k-1)$, $k(s-1)$, $k(s-1)$, and $k(ts+1)$ copies, respectively. This is illustrated in Figure \ref{fig:contours}(a). The latter three components themselves subdivide into groups of $k$ copies that are acted on by the same factor of $O_A$ or $W_{\partial B}(I)$; these `subcomponents' we denote $\text{II}_{c = 1, \ldots, (s-1)}$, $\text{III}_{c = 1, \ldots, (s-1)}$, and $\text{IV}_{c = 1, \ldots, ts+1}$. Writing $f_O(\vec{\pi}_A) \coloneqq \tr[L_A^{(m)}\text{Wg}(\vec{\pi}_A)]$ and $g^{(m)}_{\partial B}(\vec{\pi}_{\partial B}) \coloneqq \tr[L_{\partial B}^{(m)}\text{Wg}(\vec{\pi}_{\partial B})]$, by analogy to Eq.~\eqref{eq:collision average AB}, we have
	\begin{align}
		\Xi^{(t)}_{k,s}[O_A] = \frac{\mathbb{E}_{\vec{\pi}_{AB\partial B}}[f_O(\vec{\pi}_A)g^{(m)}_{\partial B}(\vec{\pi}_{\partial B})] \mathbb{E}_{\vec{\pi}_{AB\partial B}}[\mathbbm{1}(\vec{\pi}_A = I) \mathbbm{1}(\vec{\pi}_{\partial B} = I)] }{\mathbb{E}_{\vec{\pi}_{AB\partial B}}[f_O(\vec{\pi}_A)\mathbbm{1}(\vec{\pi}_{\partial B} = I)]\mathbb{E}_{\vec{\pi}_{AB\partial B}}[\mathbbm{1}(\vec{\pi}_A = I)g^{(m)}_{\partial B}(\vec{\pi}_{\partial B})]}
		\label{eq:xi LAB}
	\end{align}
	where $\mathbb{E}_{\vec{\pi}_{AB\partial B}}$ denotes an average over configurations of the $m!$-state spin model with free energy \eqref{eq:moment energy}. 
	
	Note that, by definition [Eq.~\eqref{eq:WB def}], $W_{\partial B}(I)$ is a linear combination of permutations $\pi \in S_k$, and thus the operator $L_{\partial B}^{(m)}$ will likewise be a linear combination of permutations that belong to a particular subgroup of $S_m$---namely those that  act trivially on copies in I and II, and only permute copies within the same subcomponents of  II, and IV. We denote this subgroup $T_C \subset S_m$, which is isomorphic to $S_k^{\times s(t+1)}$. 
	Explicitly,
	\begin{align}
		\pi \in T_C \;\Rightarrow\; \pi &= \prod_{c=1}^{s(t+1)}\omega^{(c)}_\pi & \text{with }&\omega^{(c)}_\pi \in S_k|_{c} \label{eq:perm sub C}
	\end{align}
	where $S_k|_c \subset S_m$ is the group of permutations that act only on a particular subcomponent $c \in \text{III} \cup \text{IV}$. By virtue of the orthogonality of permutation operators and their duals $\tr[V_A(\vec{\pi}_A) \text{Wg}_A(\vec{\pi}'_A)] = \delta_{\vec{\pi}_A = \vec{\pi}'_S}$, and Eqs.~(\ref{eq:g prob},\ref{eq:WB def}), we then have
	\begin{align}
		g^{(m)}_{\partial B}(\vec{\pi}_{\partial B}) &= \mathbbm{1}(\vec{\pi}_{\partial B} \in T_{C}) \prod_{c=1}^{s(t+1)} p_{\pi_{BC}}[\vec{\omega}_{\partial B}^{(c)}| \omega_C^{(c)} \in \mathcal{G}_I].
	\end{align}
	The expectation values in \eqref{eq:xi LAB} can then be written in terms of averages over spin configurations on the whole system $\vec{\pi}_{ABC} \in S_m^{ABC}$, with various restrictions on which permutations are allowed in the region $C$.
	\begin{align}
		\Xi^{(t)}_{k,s}[O_A] &= \frac{\mathbb{E}_{\vec{\pi}_{ABC}}[f_O(\vec{\pi}_A) | \mathcal{G}^{(C)}] }{\mathbb{E}_{\vec{\pi}_{ABC}}[f_O(\vec{\pi}_A) | \mathcal{I}^{(C)}]} \frac{\mathbb{P}_{\vec{\pi}_{ABC}}[\vec{\pi}_A = I | \mathcal{I}^{(C)}]}{\mathbb{P}_{\vec{\pi}_{ABC}}[\vec{\pi}_A = I  | \mathcal{G}^{(C)}] }
		\label{eq:xi cond exp}
	\end{align}
	Here, the expectation values and probabilities are with respect to $(m!)$-state spin models with free energy $\mathcal{H}[\vec{\pi}_{ABC}]$ given by Eq.~\eqref{eq:moment energy}, and different conditions imposed on $C$. Namely, for any function on the configuration space $f(\vec{\pi})$, we have
	\begin{align}
		\mathbb{E}[f(\vec{\pi})|\mathcal{I}^{(C)}] &= \frac{\sum_{\vec{\pi}_{ABC}} e^{-\mathcal{H}[\vec{\pi}_{ABC}] }f(\vec{\pi}_{ABC}) \mathbbm{1}(\vec{\pi}_C = I)  }{\sum_{\vec{\pi}_{ABC}} e^{-\mathcal{H}[\vec{\pi}_{ABC}] } \mathbbm{1}(\vec{\pi}_C = I)} \label{eq:cond exp I} \\
		\mathbb{E}[f(\vec{\pi})|\mathcal{G}^{(C)}] &= \frac{\sum_{\vec{\pi}_{ABC}} e^{-\mathcal{H}[\vec{\pi}_{ABC}] }f(\vec{\pi}_{ABC}) \mathbbm{1}(\vec{\pi}_C \in T_C)\prod_c \mathbbm{1}(\omega_{\pi_C}^{(c)} \in \mathcal{G}_I)  }{\sum_{\vec{\pi}_{ABC}} e^{-\mathcal{H}[\vec{\pi}_{ABC}] } \mathbbm{1}(\vec{\pi}_C \in T_C)\prod_c \mathbbm{1}(\omega_{\pi_C}^{(c)} \in \mathcal{G}_I) } \label{eq:cond exp G}
	\end{align}
	We can think of $\mathcal{I}_{C}$ as a `hard' condition where all spins on $C$ are fixed to $I$, and $\mathcal{G}_C$ as a `soft' boundary condition where spins in $C$ are have the product structure defined by $T_C$ [Eq.~\eqref{eq:perm sub C}], and each sub-configuration $\omega_{\pi_C}^{(c)}$ is in the subset $\mathcal{G}_I$. Provided that the functions $g_\tau$ are chosen appropriately, these weights favour configurations in which the majority of spins are in the state $I$. Thus, both boundary conditions will have the same overall effect of pinning the spin configurations to be mostly $I$, but they do so in quantitatively different ways. In the following, we will show that the marginal distributions on $A$ for these two boundary conditions are close, allowing us to bound $\Xi^{(t)}_{k,s}[O_A]$ close to unity. \\
	
	\textit{Analysis via contours.---}Having mapped the generalized moments $\Xi^{(t)}_{k, s}[O_A]$ to expectation values of a classical statistical mechanics model, we now describe a combinatorial way of analysing this model through the method of contours---this will also provide a way to make an explicit choice for the subsets $\mathcal{G}_\tau$ which are so far unspecified.
	
	In the case $m = 2$, we have the Ising model, for which the well-known method of Peierls' contours can be used---see e.g.~Ref.~\cite{friedli2017} for an introduction to this method. In brief, one observes that Ising spin configurations can be fully characterised (up to a global spin flip) by the subset of `open' edges, namely those $(v, v') \in E$ for which $\pi_v \neq \pi_{v'}$. This subset can be decomposed into connected components, which are called \textit{contours} and denoted $\gamma$. Since the Boltzmann factor is unchanged by this spin flip, the partition function can be written as a sum over collections of contours $\Gamma = \{\gamma_1, \ldots, \gamma_{n_\Gamma}\}$, with certain compatibility constraints between the contours. This is a special instance of a more general class of classical statistical mechanics models known as \textit{polymer models}, which describe distributions over collections of geometric objects with independent weights and a pairwise exclusion rule.
	
	We can similarly map the $(m!)$-state spin model with free energy \eqref{eq:moment energy} to a polymer model for  $m > 2$, thanks to the global $S_m \times S_m$ permutation symmetry. Since the state space for each site is larger, we will need to keep track of more information than just the locations of bonds where spins are not aligned. With this in mind, we assign an arbitrary directionality to every edge $e \in E$, so that $e = (v_1, v_2)$ is an ordered pair. For a given $\pi \in \Omega_V = S_m^{\times |V|}$, we write $E_\pi = \{(v_1, v_2) \in E : \pi_{v_1} \neq \pi_{v_2}\}$ for the set of open edges, and $E_\pi^* = \{e^* : e \in E_\pi\}$ the corresponding set of dual edges. We say that  $\gamma = (E_\gamma, \sigma_\gamma)$ is a \textit{contour} of $\pi$ if $E_\gamma^*$ is a maximally connected component of $E_\pi^*$, and $\sigma_\gamma : E_\gamma \rightarrow S_k \setminus \{I\}$ is the function defined by $\sigma_\gamma(e) = \pi_{v_1}^{-1}\pi_{v_2}$ for each $e = (v_1, v_2) \in E_\gamma$. The values of $\sigma_\gamma(e)$ will be referred to as relative permutations. A complete set of contours for $\pi$ can be constructed by taking all maximally connected components of $E_\pi^*$ (i.e.~all components with respect to face-connectivity) and constructing the corresponding contours; we write the resulting collection as $\Gamma = \{\gamma_1, \ldots, \gamma_{n_\Gamma}\}$. Since this process of constructing contours from spin configurations is uniquely specified, and the product $\pi_v^{-1} \pi_{v'}$ can be inferred from $\Gamma$ for any pair of vertices $v, v' \in V$, we have a $(m!)$-to-one mapping $\mathcal{C} : \pi \mapsto \Gamma$, whose image we denote $\Delta_V \coloneqq \{\Gamma : \exists \pi \in \Omega_V, \mathcal{C}[\pi] = \Gamma\}$.
	
	Let $\mathcal{K}_V = \{\gamma : \{\gamma\} \in \Delta_V\}$ be the set of contours that can appear as singletons in the image of $\mathcal{C}$. These contours have to satisfy the condition that, for every closed path of edges on the primal lattice $p = \big((v_0, v_1), (v_1, v_2), \ldots, (v_{l-1},v_0)\big)$, the
	product of relative permutations around this path must give identity, that is $\prod_{e \in p \cap E_\gamma }^{\rightarrow} \sigma^{\omega_e}_\gamma(e) = I$, where the product is in the order specified by $p$, and $\omega_e = \pm 1$ depending on whether the edge $(v,v') \in p$ is oriented in the same direction as the path $p$. For each valid contour $\gamma \in \mathcal{K}_V$, the graph $(V, E \setminus E_\gamma)$ has $r_\gamma \geq 2$ connected components which we denote $\{V^\gamma_i\}_{i=1}^{r_\gamma}$. If $\gamma = (E_\gamma, \sigma_\gamma)$ satisfies these criteria, then we can find a representative spin configuration $\pi_v$ satisfying $\mathcal{C}[\pi_\gamma] = \{\gamma\}$, which is constant on each component: $v, v' \in V^\gamma_i \Rightarrow [\pi_\gamma]_v = [\pi_\gamma]_{v'}$. There are $m!$ such configurations, all related by $\pi_v \mapsto \tau \pi_v$ for some $\tau \in S_m$.
	
	We will say that two contours $\gamma, \gamma' \in \mathcal{K}_V$ are compatible if no dual edge in $E_{\gamma'}^*$ is connected to a dual edge in $E_\gamma^*$, in which case we write $\gamma \sim \gamma'$. Since contours are constructed out of maximally connected components of the set of open dual edges, we immediately have that if $\{\gamma, \gamma'\} \subseteq \Gamma$ for some $\Gamma \in \Delta_V$, then $\gamma \sim \gamma'$. Additionally, if $\Gamma = \{\gamma_1, \ldots, \gamma_{n_\Gamma}\} = \mathcal{C}[\pi]$ is the set of contours for some configuration, and $\gamma_\star \in \mathcal{K}_V$ is a valid contour that is pairwise compatible with the other contours $\gamma_\star \sim \gamma_i\, \forall i \in [n_\Gamma]$, then we can construct a configuration $\pi'$ with contours $\Gamma \cup \{\gamma_\star\}$. To do this, take a representative state $\pi_\star : \mathcal{C}[\pi_\star] = \{\gamma_\star\}$, and apply the vertex-wise group product $[\pi']_v = [\pi_{\star}]_v [\pi_{\gamma'}]_v$.  The inverse of this procedure removes a contour $\gamma_i \in \Gamma = \mathcal{C}[\pi]$ by left multiplying $\pi$ by the inverse of a representative configuration for $\{\gamma_i\}$. These operations allow us to interrelate different valid contour configurations,
	\begin{align}
		\mathcal{C}[\pi] = \Gamma = \{\gamma_i\}_{i=1}^{n_\Gamma}\; \wedge \; \pi_i : \mathcal{C}[\pi_i] = \{\gamma_i\}\; \Longleftrightarrow \; \mathcal{C}[\pi'] = \Gamma \setminus \{\gamma_i\} \quad \text{where }[\pi']_v = [\pi_i]_{v}^{-1} [\pi]_v
		\label{eq:contour removal}
	\end{align}
	Applying these contour addition and removal processes recursively, one can infer that $\Gamma = \{\gamma_i\}_{i=1}^{n_\Gamma}$ is a valid set of contours if and only if $\gamma_i \in \mathcal{K}_V\,\forall i \in [n_\Gamma]$ and $\gamma_i \sim \gamma_j$ for all $1 \leq i < j \leq n_\Gamma$.
	
	Due to the invariance of the free energy \eqref{eq:moment energy} for the $m!$-state spin model under left multiplication by a uniform permutation, the Boltzmann weight for a given spin configuration depends only on its contours,
	\begin{align}
		e^{-\mathcal{H}[\pi]} &= w(\mathcal{C}[\pi]) \coloneqq \prod_{\gamma \in \mathcal{C}[\pi]} w(\gamma_i) & \text{where }w(\gamma_i) = \prod_{e \in E_\gamma} \exp\left(-h\big(\sigma_{\gamma_i}(e)\big) \right).
		\label{eq:contour weights}
	\end{align}
	Here, $h(\sigma)$ is the bond energy function given in Eq.~\eqref{eq:moment energy}. Then, if $g(\pi) = g(\mathcal{C}[\pi])$ is a function that only depends on the contours of spin configurations $\pi \in \Omega_V$, averages with respect to the spin model can be expressed in terms of a polymer model, configurations of which are specified by sets of compatible contours, with weights given above. Specifically,
	\begin{align}
		\mathbb{E}_{\pi}[g(\pi)] = \frac{\sum_{\pi \in \Omega_V} e^{-\mathcal{H}[\pi]} g(\pi)}{\sum_{\pi \in \Omega_V} e^{-\mathcal{H}[\pi]}} &= \frac{\sum_{\Gamma \in \Delta_V } w(\Gamma) g(\Gamma)}{\sum_{\Gamma \in \Delta_V } w(\Gamma)} \nonumber\\ & = \frac{\sum_{n =0}^\infty \sum_{\{\gamma_1, \ldots, \gamma_n\} \subset \mathcal{K}_V} \left(\prod_i w(\gamma_i) \prod_{j > i} \mathbbm{1}(\gamma_i \sim \gamma_j) \right) g(\{\gamma_1, \ldots, \gamma_n\})}{\sum_{n =0}^\infty \sum_{\{\gamma_1, \ldots, \gamma_n\} \subset \mathcal{K}_V} \left(\prod_i w(\gamma_i) \prod_{j > i} \mathbbm{1}(\gamma_i \sim \gamma_j) \right)}
		\label{eq:avg polymer}
	\end{align}
	
	To evaluate the conditional expectation values of the kind \eqref{eq:cond exp G}, some modifications to the above need to be made on account of the fact that the spin configurations in $C \subset V$ are constrained to the subset $T_C$, and each sub-component $\omega^{(c)}_C$ must be in $\mathcal{G}_I$. We write $\Omega_V[\mathcal{G}]$ for the set of valid spin configurations in this case. In essence, the difference will be that contours entirely contained in $C$ will be split up into $c$\textit{-contours} $\gamma^{(c)}$, which are specific to a particular subcomponent $c = 1, \ldots, s(t+1)$.
	
	Our modified procedure for generating contours from a given configuration $\pi \in \Omega_V[\mathcal{G}]$ proceeds as follows. For each $c = 1, \ldots, s(t+1)$, we declare an edge $(v_1, v_2) \in E_C$ to be $c$-open if $\omega_{v_1}^{(c)} \neq \omega_{v_2}^{(c)}$, and $E_\pi^{(c)} \subseteq E_C$ will denote the set of all $c$-open edges. The pair $\gamma^{(c)} = (E_{\gamma^{(c)}}, \sigma_{\gamma^{(c)}})$ is said to be a $c$-contour of $\pi$ if $E_{\gamma^{(c)}}$ is a maximally connected component of $E_\pi^{(c)}$ that is not connected to $\partial C$, and the function $\sigma_{\gamma^{(c)}} : E_{\gamma^{(c)}} \rightarrow S_k \setminus \{I\}$ is now given by  $\sigma_{\gamma^{(c)}}(e) = [\omega_{v_1}^{(c)}]^{-1}\omega_{v_2}^{(c)}$ for $e = (v_1, v_2) \in E_{\gamma^{(c)}}$. Writing $\Gamma^{(c)} = \{\gamma^{(c)}_i\}_{i}$ for the set of all $c$-contours, we repeat this procedure for $c = 1, \ldots, s(t+1)$, and define $\Gamma^C \coloneqq \bigcup_{c=1}^{s(t+1)} \Gamma^{(c)}$. At this stage we remark that $\Gamma^{(c)}$ can be inferred solely from the variables $[\omega^{(c)}_v]_{v \in C}$, namely the part of the permutations in $C$ that act on subcomponent $c$ [Eq.~\eqref{eq:perm sub C}].
	
	We now construct the remaining contours. First, for each $\gamma^C \in \Gamma^{C}$, we apply a contour removal operation to $\pi$ via the operation $[\pi]_v \mapsto [\pi_{\gamma^{C}}]_v^{-1}\cdot[\pi]_v$, where $\pi_{\gamma^{C}}$ is a representative spin configuration having $\gamma^{C}$ as its only contour. We write $\pi'$ for the resulting configuration. Then, we apply the original contour map $\mathcal{C}$ to $\pi'$. This generates a set of $AB$\textit{-contours}, $\Gamma^{AB} = \{\gamma_1^{AB}, \ldots, \gamma_{n_{\Gamma^{AB}}}^{AB} \}$. None of the contours in $\Gamma^{AB}$ can be subsets of $E_C$, since otherwise we could decompose them as $c$-contours, which by construction have been removed. We finally write $\hat{\Gamma} = \Gamma^{AB} \cup \Gamma^C$ for the full set of contours of $\pi$.

	This construction defines a map $\mathcal{C}_{\mathcal{G}_C} : \pi \mapsto \hat{\Gamma}$ which has domain $\Omega_V[\mathcal{G}_C]$ and whose image we denote $\Delta_V[\mathcal{G}_C]$. The map is $\big((k!)^{s(t+1)}\big)$-to-one, with configurations $\pi$ and $\tau \cdot \pi$ having the same image for any uniform permutation $\tau \in S_k^{\times s(t+1)}$. One can verify that the Boltzmann weight of a spin configuration $\pi$ is given by the same function of its contours as before, Eq.~\eqref{eq:contour weights}, which is a product of weights for each individual $\gamma \in \hat{\Gamma}$. Writing $\mathcal{K}_V[\mathcal{G}_C]$ for the set of all possible $\gamma$ that appear in some $\hat{\Gamma} \in \Delta_V[\mathcal{G}_C]$, we can also show that $\Delta_V[\mathcal{G}_C]$ consists of all subsets of $\mathcal{K}_V[\mathcal{G}_C]$ that are pairwise compatible with the following rules. If $\gamma_1$ and $\gamma_2$ are $c$- and $c'$-contours, respectively, with $c \neq c'$, then $\gamma_1 \sim \gamma_2$. If  $\gamma_1$ and $\gamma_2$ are both $AB$-type, then compatibility is the same as before: $\gamma_1 \sim \gamma_2$ iff $E_{\gamma_1}^*$ is not connected to any dual edge in $E_{\gamma_2}^*$. If $\gamma_1$ is $c$-type and $\gamma_2$ is $AB$ type, then we define $E_{\gamma_2}^{(c)}$ to be the set of edges $e \in E_{\gamma_2}$ for which $\sigma_{\gamma_2}(e) = I^{(c)}\times \tau$, where $I^{(c)}$ is the identity permutation on subcomponent $c$, and $\tau$ is some permutation acting on the other copies. Then, $\gamma_1 \sim \gamma_2$ if for every $e \in E_{\gamma_2}^{(c)}$, the corresponding dual edge $e^*$ is not adjacent to $E_{\gamma_1}^*$.
	
	As a result of the above construction, if $g(\pi)$ is a function that is invariant under transformations $\pi_v \mapsto \tau \pi_v$ for all $\tau \in T_C$, then it can be written as a function of the contours $\hat{\Gamma} = \mathcal{C}_{\mathcal{G}_C}[\pi]$, and in turn, we can evaluate the expectation value conditioned on $\pi_C\in T_C$,
	\begin{align}
		\mathbb{E}[g(\pi) | \pi_C \in T_C] = \frac{\sum_{\vec{\pi}_{ABC}} e^{-\mathcal{H}[\vec{\pi}_{ABC}]}  \mathbbm{1}(\pi_C \in T_C)g(\pi)}{\sum_{\vec{\pi}_{ABC}} e^{-\mathcal{H}[\vec{\pi}_{ABC}]}  \mathbbm{1}(\pi_C \in T_C)}  =  \frac{\sum_{\hat{\Gamma} \in \Delta_V[\mathcal{G}_C]} w(\hat{\Gamma})g(\hat{\Gamma})}{\sum_{\hat{\Gamma} \in \Delta_V[\mathcal{G}_C]} w(\hat{\Gamma})},
		\label{eq:avg GC contour}
	\end{align}
	where $w(\hat{\Gamma})$ is defined in the same way as before, Eq.~\eqref{eq:contour weights}. This will be useful for evaluating the conditional expectation \eqref{eq:cond exp G} later on.
	
	\subsection{Typical geometry of contours}
	
	Following the ideas of Peierls' famous proof for the Ising model, we will use the contour removal operation \eqref{eq:contour removal} to bound the distribution of sizes of contours in the polymer model for $m > 2$. This will then give us the ability to characterize the structure of typical configurations that appear in the averages (\ref{eq:cond exp I}, \ref{eq:cond exp G}). Here we will treat only the constrained model [Eq.~\eqref{eq:avg GC contour}], but the same arguments apply, \textit{mutatis mutandis}, to the original polymer model \eqref{eq:avg polymer}.\\
	
	\textit{Contour length distribution.---}For a given set of edges $E_0$ whose dual forms a connected set, let $\mathcal{K}[E_0] \subset \mathcal{K}[\mathcal{G}_C]$ be the set of possible contours of the form $\gamma = (E_0, \sigma_\gamma)$.  and the constrained model We can upper bound the total weight of all contours in $\mathcal{K}[E_0]$ by ignoring the constraints on $\sigma_\gamma$ (other than the fact that $\sigma_\gamma(e) \neq I$ for $e \in E_0$), giving
	\begin{align}
		\sum_{\gamma \in \mathcal{K}[E_0]} w(\gamma) &\leq \left(\sum_{\sigma_e \in S_m \setminus \{I\}} e^{-h(\sigma_e)} \right)^{|E_0|} \leq e^{-\beta |E_0|} & \text{where }\beta \geq \xi \ln 2 - \ln(2m^2)
		\label{eq:contour weight bound}
	\end{align}
	where we recall $\xi$ as the number of qubits within each block of the coarse grained circuit (with $\text{deg}(v)\times \xi = 4\xi$ qubits per vertex). Here we use $e^{-h(\sigma)} \leq 2^{1-\xi(m-|\sigma|)}$, where $|\sigma|$ is the number of cycles of $\sigma$, and by Eq.~(13) of Ref.~\cite{harrow2023approximate}, we have $\sum_{\sigma \in S_m \setminus \{I\}} b^{m-|\sigma|} \leq e^{m^2b/2} -1$, which is at most $m^2b$ for  for $m^2b \leq 1$.
	
	Simultaneously, since $E_0^*$ has to be a connected set and each dual edge has degree at most 6 (i.e.~is face-connected to at most 6 other dual edges), we can bound the size of $\mathcal{K}[l, e_\star] \coloneqq \big\{ E_0 : |E_0| = l; e^\star \in E_0; E_0^* \text{ is connected} \big\}$, the set of connected subsets of dual edges $E_0^*$ of cardinality $l$ and including a particular edge $e_\star$. Indeed, the number of such subsets for a bounded-degree graph is known to scale at most exponentially with $l$. For instance, from Ref.~\cite{bollobas2006art}, pp.~130, we have
	\begin{align}
		|\mathcal{K}[l, e_\star]| &\leq \exp[l(1 + \ln 5)] \label{eq:contour number bound}
	\end{align}
	Now, we can bound the probability that $ \mathcal{C}[\pi]$ contains a contour $\gamma$ of length at least $l_0$, and which includes a particular edge $e_\star$. By a union bound, we have 
	\begin{align}
		\mathbb{P}_\pi\left[\bigvee_{l \geq l_0}\big(\mathcal{C}[\pi] \cap \mathcal{K}[l,e_\star] \neq \emptyset\big) \right] &\leq \sum_{l=l_0}^{|E|} \sum_{\gamma \in \mathcal{K}[l,e_\star]} \frac{\sum_{\hat{\Gamma} \in \Delta_V[\mathcal{G}_C]} \mathbbm{1}(\gamma \in \hat{\Gamma}) w(\hat{\Gamma}) }{\sum_{\hat{\Gamma} \in \Delta_V[\mathcal{G}_C]} w(\hat{\Gamma})} \nonumber\\ &= \sum_{l=l_0}^{|E|} \sum_{\gamma \in \mathcal{K}[l,e_\star]} \frac{\sum_{\hat{\Gamma}' \in \Delta_V[\mathcal{G}_C]} \prod_{\gamma' \in \hat{\Gamma}'}\mathbbm{1}(\gamma' \sim \gamma)  w(\hat{\Gamma}')w(\gamma) }{\sum_{\hat{\Gamma} \in \Delta_V[\mathcal{G}_C]} w(\hat{\Gamma})} \nonumber\\ &\leq \sum_{l=l_0}^{|E|} \sum_{\gamma \in \mathcal{K}[l,e_\star]} w(\gamma) \leq 2e^{-\alpha l_0} & \text{for }l_0 \geq \alpha^{-1}
		\label{eq:short contours}
	\end{align}
	where $\alpha \geq \xi \ln 2 - \ln(10m^2)-1$, which is positive assuming $\xi$ exceeds some $\mathcal{O}(1)$ threshold. In going to the second line, we have used the fact that any configuration $\Gamma$ that contains $\gamma$ can be written as the union of the singleton $\{\gamma\}$ with a configuration $\Gamma' \in \Delta_V$, whose contours $\gamma' \in \Gamma'$ are all compatible with $\gamma$. We can upper bound this sum by removing the compatibility restriction, hence the inequality in the third line. The above will allow us to rule out the existence of long contours with high probability.\\
	
	\textit{Identifying good configurations.---}We start by specifying a `goodness' condition on the space of global spin configurations $\pi \in \Omega_V[\mathcal{G}_C]$, which can inferred from the collection of contours $\hat{\Gamma} = \mathcal{C}_{\mathcal{G}_C}[\pi]$. We will then use our assumptions regarding the geometry of the system and the partition $ABC$ to lower bound the probability that $\pi$ is good. Our definition is as follows.
	
	\begin{flushright}
		\begin{minipage}{0.97\textwidth}
			\emph{A contour $\gamma \in \mathcal{K}_V[\mathcal{G}_C]$ is \textup{bad} if its edges $E_\gamma$ intersect $C \setminus R_C$, and either 1) $E_\gamma$ intersects $\partial A$, or 2) the subgraph $(V, E \backslash E_\gamma)$ does not have a connected component that contains all of $\{v_\star\} \cup A$.}
			\hfill\refstepcounter{equation}\label{eq:bad def}(\theequation)
		\end{minipage}
	\end{flushright}
	A configuration $\pi \in \Omega_V[\mathcal{G}_C]$ is said to be bad if $\hat{\Gamma} = \mathcal{C}_{\mathcal{G}_C}[\pi]$ contains a bad contour; otherwise, we say that $\pi$ (equivalently $\Gamma$) is \textit{good}.
	
	We can distinguish two types of bad contours. The first are those that connect $A$ and $C$, as shown in Fig.~\ref{fig:inner circuit}(c). By assumption 3, these must have length at least $r_{AC}$, and since they contain at least one edge of $|\partial A|$ and one edge of $|\partial C|$, their weight is upper bounded by $2\text{min}(|\partial A|, |\partial C|)e^{-\alpha r_{AC}}$. The remaining bad contours are those that do not intersect $A$, but separate a component $A_i$ from either another component $A_j$, $j \neq i$, or the reference vertex $v^\star$. By assumption 5, these have length at least $l_C$. Invoking Eq.~\eqref{eq:short contours} again, we find
	\begin{align}
		\mathbb{P}[\pi \text{ is bad}] \leq |G|e^{-\Omega(l_C)} + \text{min}(|\partial A|, |\partial C|)e^{-\Omega(r_{AC})}
		\label{eq:prob bad}
	\end{align}
	provided $\xi$ exceeds a $\mathcal{O}(1)$ threshold value.
	
	\textit{Specifying the subsets $\mathcal{G}_\tau$.---}The above goodness condition will be used to motivate a particular choice for the subsets $\mathcal{G}_\tau \subset S_k^C$, which are yet to be specified. Recall that each sub-configuration $\omega^{(c)} \in S_k^{C}$ must be categorized into different subsets $\mathcal{G}_\tau$, which determines whether $\omega^{(c)}$ contributes to the conditional expectation \eqref{eq:cond exp G}. These subsets must be chosen in a way such that conditions \eqref{eq:gtau} are satisfied. Our aim is to choose $\mathcal{G}_\tau$ to `pick' configurations where the spin majority is $\tau$.
	
	Our representation of spin configurations $\pi$ in terms of $AB$- and $c$-contours allows us to formalise this approach very straightforwardly. Recall that $\Gamma^{(c)}$ is a function of $[\omega^{(c)}_v]_{v \in C}$ only, and thus we are free to categorize $\omega^{(c)}$ using data in $\Gamma^{(c)}$. Therefore, we can identify $\Gamma^{(c)}[R_C] \coloneqq \{\gamma \in \Gamma^{(c)} : E_\gamma \subseteq E_{R_C} \}$, namely the set of all $c$-contours that are entirely contained in $R_C$. We then apply contour removal operations \eqref{eq:contour removal} to generate a new configuration $\pi_C'$, in which all contours in $\Gamma^{(c)}[R_C]$ are removed. We then set
	\begin{align}
		\mathcal{G}_\tau = \big\{ \pi : \pi'_{v^*} = \tau \big\}
		\label{eq:gtau choice}
	\end{align}
	This evidently satisfies the criteria \eqref{eq:gtau}: there are $(k!)^{s(t+1)}$ spin configurations $\pi$ with a given set of contours $\hat{\Gamma}[\pi] = \hat{\Gamma}'$, and for each $\tau$ there is exactly one such configuration that satisfies the condition $\bigwedge_{c} \omega^{(c)}_\pi \in \mathcal{G}_\tau$.  
	
	Since each of the $(k!)^{s(t+1)}$ spin configurations associated with a given $\hat{\Gamma}'$ are equally likely, we can also infer the following useful fact. Let $f(\pi) = f(\mathcal{C}_{\mathcal{G}_C}[\pi] )$ be a function that only depends on the contours of $\pi \in \Omega_V[\mathcal{G}_C]$. Then,
	\begin{align}
		\mathbb{E}_\pi\left[f(\pi)\Bigg|\pi_C \in T_C, \bigwedge_c \omega^{(c)}_\pi \in \mathcal{G}_I \right] = \frac{1}{(k!)^{s(t+1)}} \mathbb{E}_\pi\left[f(\pi)\big|\pi_C \in T_C\right]
		\label{eq:cont factorize}
	\end{align}
	In particular, recalling the definition of the conditional expectation \eqref{eq:cond exp G}, and using the fact that the goodness of a configuration only depends on its contours, we have
	\begin{align}
		\mathbb{P}_\pi[\pi \text{ is bad}|\mathcal{G}^{(C)}] &=\mathbb{P}[\pi \text{ is bad}] \leq  \delta_{\rm bad}
		\label{eq:bad prob GC}
	\end{align}
	where $\delta_{\rm bad}$ is the expression in Eq.~\eqref{eq:prob bad}.
	
	\subsection{Proof of Proposition \ref{prop:replica bound}}
	
	For a given function $g : \Omega_A \rightarrow \mathbb{R}$, we define the quantity
	\begin{align}
		r[g] \coloneqq \frac{\mathbb{E}[g(\pi_A)|\pi_C \in \mathcal{G}_I]}{\mathbb{E}[g(\pi_A)|\pi_C = I]}
		\label{eq:fO def}
	\end{align}
	Our aim is to derive two-sided bounds on $r[g]$ for the cases $g = f_O$, where $f(\pi_A) = \tr[\text{Wg}(\pi_A) O]$, and also $g = I_A$, where $I_A(\pi_A) = \mathbbm{1}(\pi_A = I)$. By Eq.~\eqref{eq:xi cond exp}, this can be used to bound the deviation of the proxy error \eqref{eq:proxy error}, since 
	\begin{align}
		\varepsilon_{\rm proxy} = |\Xi^{(t)}_{k,s}[O] - 1| = \left|\frac{r[f_O]}{r[I_A]} - 1\right|.
		\label{eq:proxy error fO}
	\end{align}
	Using Eq.~\eqref{eq:bad prob GC}, we have
	\begin{align}
		r[g] &= (1 -\mathbb{P}[\pi \text{ is bad}]) r_{\text{good}}[g] + \mathbb{P}[\pi \text{ is bad}]\frac{\mathbb{E}[g(\pi_A)|\pi_C \in \mathcal{G}_I \wedge \pi \text{ is bad}]}{\mathbb{E}[g(\pi_A)|\pi_C = I]} \\
		\text{where } r_{\rm good}[g] &\coloneqq \frac{\mathbb{E}[g(\pi_A)|\pi_C \in \mathcal{G}_I \wedge \pi \text{ is good}]}{\mathbb{E}[g(\pi_A)|\pi_C = I]}
		\label{eq:fO bad good}
	\end{align}
	In light of Eq.~\eqref{eq:prob bad}, we expect the first term to dominate, and accordingly we focus on computing $r_{\text{good}}[g]$ to start with.  
	
	First, for each configuration $\pi$, we follow the procedure from our definition of $\mathcal{G}_\tau$ to remove all contours entirely contained in $C \setminus R_C$; we write $\pi'$ for the resulting configuration, and $\hat{\Gamma}$ for its contours. Evidently, $\pi$ and $\pi'$ agree on $A$. For each contour $\gamma \in \hat{\Gamma}$, we identify the unique maximally connected component of $(V, E \backslash E_\gamma)$ that contains $v^\star$, and declare this to be the exterior of $\gamma$, denoted $\text{Ext}(\gamma)$, with the interior as the complement $\text{Int}(\gamma) = V \setminus \text{Ext}(\gamma)$.  We define the exterior and interior of a set of contours $\Gamma'$ be $\text{Ext}(\Gamma') \coloneqq \bigcap_{\gamma \in \Gamma'} \text{Ext}(\gamma)$ and $\text{Int}(\Gamma') \coloneqq \bigcup_{\gamma \in \Gamma'} \text{Int}(\gamma)$, respectively. In particular, the exterior of the full configuration $\text{Ext}(\hat{\Gamma})$ is a region on which the spin configuration $\pi'$ is uniform. Due to our choice of $\mathcal{G}_\tau$ [Eq.~\eqref{eq:gtau}], if we condition on $\pi_C \in \mathcal{G}_I$, then we must have $\pi_v' = I$ for all $v \in \text{Ext}(\hat{\Gamma})$.
	
	We now partition the contours of $\pi'$ as $\hat{\Gamma} = \hat{\Gamma}^{\rm in} \cup \hat{\Gamma}^{\rm out}$, where $\hat{\Gamma}^{\rm out}$ are the contours that intersect $C \setminus R_C$, and $\hat{\Gamma}^{\rm in}$ are those entirely contained in $AB$. Due to the definition \eqref{eq:bad def}, $\pi$ is good if and only if $\hat{\Gamma}^{\rm out}$ is good. This in turn is equivalent to $\text{Ext}(\hat{\Gamma}^{\rm out}) \supseteq A$, since every $\gamma \in \hat{\Gamma}^{\rm out}$ intersects $C \setminus R_C$, and $\{v^\star\}\cup A$ must be contained in a single connected component of  $(V, E \setminus E_\gamma)$.  Moreover, since $\pi_{v^\star}$ is fixed to identity, the spin configuration on $A$ is a unique function of $\hat{\Gamma}$, which we write as $\mathcal{C}^{-1}_A[\hat{\Gamma}]$.  Its distribution can then be written
	\begin{align}
		\mathbb{P}[\pi_A |\pi_C \in \mathcal{G}_I \wedge \pi \text{ is good}] = \frac{\sum_{\hat{\Gamma}^{\rm out} \in \text{good}} w(\hat{\Gamma}^{\rm out})
			\sum_{\hat{\Gamma}^{\rm in} \sim \hat{\Gamma}^{\rm out}} w(\hat{\Gamma}^{\rm in}) \mathbbm{1}\left( \mathcal{C}^{-1}_A(\hat{\Gamma}^{\rm in}\cup \hat{\Gamma}^{\rm out}) = \pi_A\right) }{\sum_{\hat{\Gamma}^{\rm out} \in \text{good}} w(\hat{\Gamma}^{\rm out})
			\sum_{\hat{\Gamma}^{\rm in} \sim \hat{\Gamma}^{\rm out}} w(\hat{\Gamma}^{\rm in}) }
		\label{eq:prob piA good}
	\end{align}
	The sums over $\hat{\Gamma}^{\rm out}$ are over good contour configurations only, while the sums over $\hat{\Gamma}^{\rm in}$ are over contours that are compatible with$\hat{\Gamma}^{\rm out}$, but with no extra goodness constraint.
	
	For a given $\hat{\Gamma}^{\text{out}} \in \text{good}$, we now define
	\begin{align}
		D \coloneqq \Big\{v \in \text{Ext}(\hat{\Gamma}^{\rm out}) : \big(v \in C\big) \vee \big(v \sim^* \text{Int}(\hat{\Gamma}^{\rm out}) \big) \Big\}
	\end{align}
	where $\sim^*$ denotes face-connectivity of vertices. In words, $D$ is the set of vertices in $\text{Ext}(\hat{\Gamma}^{\rm out})$ that are either in $C$ and adjacent to $B$, or are face-connected to $\text{Int}(\hat{\Gamma}^{\rm out})$. In Fig.~\ref{fig:inner circuit}(b), we illustrate an example good configuration, and the region $D$ are those spins shaded in dark gray. We now argue that for any choice of $\hat{\Gamma}^{\rm in}$ compatible with $\hat{\Gamma}^{\rm out}$, we have $\pi_v = I$ for all $v \in D$.
	
	By construction, the exterior of any contour in $\hat{\Gamma}^{\rm in}$ must contain $C$, and so we have $C \cap \text{Ext}(\hat{\Gamma}) = C \cap \text{Ext}(\hat{\Gamma}^{\rm out})$. Because the spin configuration on $\text{Ext}(\hat{\Gamma})$ is uniform and equal to $I$, this proves our claim for the case $v \in C$. For the second case $v \sim^* \text{Int}(\hat{\Gamma}^{\rm out})$, we decompose $\text{Int}(\hat{\Gamma}^{\rm out})$ into maximally face-connected components $\{J_i\}_{i=1}^{r}$---these are the regions surrounded by red lines in Fig.~\ref{fig:inner circuit}(b). Let $D_i$ be the subset of vertices in $D$ that are face-connected to $J_i$. Then, one can verify that $D_i$ is path-connected on the primal lattice---that is, any two vertices $v, v' \in D_i$ can be connected via a sequence of edges $p = \big\{(v, v_{1}), (v_1, v_2), \ldots, (v_{|p|}, v')\big\} \subset E$, with $v_i \in D_j \forall i = 1, \ldots, |p|$. Each edge $(v_t, v_{t+1})$ neighbours exactly two faces, one of which (denoted $f_t$) must contain a vertex $v'' \in J_i$ by virtue of the fact that $v_t, v_{t+1} \in D_i$. Without loss of generality, we can assume that $v_t$ is edge-connected to $v''$. Then, there must be a contour $\gamma \in \hat{\Gamma}^{\rm out}$ that contains the edge $(v_t, v'')$. Since compatible contours cannot be face-connected, this means that $(v_t, v_{t+1})$ cannot appear in any contour $\gamma \in \hat{\Gamma}^{\rm in}$. Since $v_t, v_{t+1} \in \text{Ext}(\hat{\Gamma}^{\text{out}})$, we also know that $(v_t, v_{t+1})$ cannot appear in any contour $\gamma \in \hat{\Gamma}^{\rm out}$, and so $\pi_{v_{t}} = \pi_{v_{t+1}}$. Recursing, we have $\pi_v = \pi_{v'}$ for all $v, v' \in D_j$. Since $\hat{\Gamma}^{\rm out}$ consists of contours that intersect $C$, at least one vertex in $D_j$ must be contained in $C$, and the corresponding spin must be in the state $I$ by the previous argument. This proves our claim.
	
	Now, for any vertex $a \in A$, and any $v \in C$, we argue that any path $p$ that connects $a$ to $v$ must intersect $D$. Since this is obvious for $v \in D$, we suppose $v \notin D$. Then, $v$ must be contained in some component $J_i \subset \text{Int}(\hat{\Gamma}^{\rm out})$. We now use the goodness condition $A \subseteq \text{Ext}(\hat{\Gamma}^{\rm out})$ derived above, meaning that $a \in \text{Ext}(\hat{\Gamma}^{\rm out})$. Thus, $p$ must cross the boundary between $J_i$ and $\text{Ext}(\hat{\Gamma}^{\rm out})$, which necessitates an intersection with some $v \in D_i \subseteq D$.
	
	Now define $\text{Int}(D)$ as the set of vertices that are not in $D$, and are connected to $A$ via a path that does not intersect $D$. Since $D \cap A = \emptyset$, we have $\text{Int}(D) \supseteq A$, and by the above argument, $\text{Int}(D) \subseteq A \cup B$. Then, we write $\{\Lambda'_i\}_i$ for the maximally connected components of $\text{Int}(D)$. These are the regions surrounded by dashed lines in Fig.~\ref{fig:inner circuit}. 
	
	Letting $A_j$ be the connected components of $A$, each $A_j$ must be contained in exactly one $\Lambda'_i$. Since all spins in $D$ are fixed to $I$, the spin configuration on $A_j$ is uniquely determined by the contours $\gamma \in \hat{\Gamma}^{\rm in}$ that are supported on the corresponding component $\Lambda'_i$. The distribution of these contours exactly the same as that of the spin model \eqref{eq:moment energy} on the region $\Lambda'_i$, with fixed boundary conditions. Hence, after conditioning on any particular set of good outer contours $\hat{\Gamma}^{\text{out}}$, the marginal distribution of spins on $A$ is the same as a collection of independent spin models on components $\Lambda'_i$, surrounded by spins pinned to $I$. Explicitly, this means the distribution \eqref{eq:prob piA good} simplifies to
	\begin{align}
		\mathbb{P}[\pi_A| \pi_C \in \mathcal{G}_I \wedge \pi \text{ is good}] = \sum_{\hat{\Gamma}^{\rm out} \in \text{good}} \mathbb{P}[\hat{\Gamma}^{\rm out}| \pi \text{ is good}]  \mathbb{P}[\pi_{A}| \pi_{V \setminus \text{Int}(D)} = I]. 
	\end{align}
	Thus, we have
	\begin{align}
		r_{\text{good}}[g] = \sum_{\hat{\Gamma}^{\rm out} \in \text{good}} \mathbb{P}[\hat{\Gamma}^{\rm out}| \pi \text{ is good}] \frac{\mathbb{E}[g(\pi_A)| \pi_{V \setminus \text{Int}(D)} = I]}{\mathbb{E}[g(\pi_A)| \pi_{C} = I]}
		\label{eq:fO good D}
	\end{align}
	While the argument so far has been general, we now turn to the specific case covered in Proposition \ref{prop:replica bound}, namely $k = 2, s = 1, t = 0$, corresponding to the Ising model with $m = 2$ total replicas, and $O = \sum_{x} c_x\dyad{x}^{\otimes 2}$, with $c_x \geq 0$. Then, in the case where $g = f_O$, we have $f_O(\pi_A)$ equal to a constant independent of $\pi_A$; thus $r[f_O] = 1$ trivially. In the other case, $g = I_A$, we can employ the FKG inequality for the ferromagnetic Ising model to bound $r[I_A]$. For the Ising model on a graph $G = (V, E)$, with free energy $\mathcal{H}_{J,h}[\pi] = \sum_{\{v_1, v_2\} \in E}J_{v_1,v_2} \pi_{v_1} \pi_{v_2}  + \sum_{v \in V} h_v \pi_v$,  where we associate $I$ with the value $+1$ and $S$ with the value $-1$, suppose that $J_{v_1, v_2} \geq 0$ for all $\{v_1, v_2\} \in E$. The FKG inequality states that
	\begin{align}
		\mathbb{E}_{\pi}[f(\pi) g(\pi)] \geq \mathbb{E}[f(\pi)]\mathbb{E}[g(\pi)],
		\label{eq:FKG}
	\end{align}
	for any increasing functions $f, g : \Omega \rightarrow \mathbb{R}$. Here, $f$ being increasing means that for any $\pi, \pi' \in \Omega$ such that $\pi_v \geq \pi_{v'}\, \forall v$, we have $f(\pi) \geq f(\pi')$. Two important consequences of this inequality are 
	\begin{align}
		\mathbb{P}[\pi_A = I | \pi_X = I] &\geq \mathbb{P}[\pi_A = I | \pi_X = \eta_X] & \forall \eta_X \in \{\pm 1\}^X \label{eq:FKG1} \\
		\mathbb{P}[\pi_A = I | \pi_X = I \wedge \pi_Y = I] &\geq \mathbb{P}[\pi_A = I | \pi_X = I]  \label{eq:FKG2}
	\end{align}
	Recalling that $I_A(\pi_A) = \mathbbm{1}(\pi_A = I)$, we can combine Eq.~\eqref{eq:FKG1} with Eq.~\eqref{eq:fO def} to infer $r[I_A] \leq 1$, since $\mathbb{E}[\pi_A = I|\pi_C \in \mathcal{G}_I] = \sum_{\eta_C \in \{\pm 1\}^C} \mathbb{P}[\pi_C = \eta_C] \mathbb{P}[\pi_A|\pi_C = \eta_C] \leq \mathbb{P}[\pi_A|\pi_C = I]$. Combining  Eq.~\eqref{eq:FKG2} with Eq.~\eqref{eq:fO good D}, we get $r_{\rm good}[I_A] \geq 1$, since, by virtue of the fact that $V \setminus \text{Int}(D) \supseteq C$, we have $\mathbb{P}[\pi_A = I|\pi_{V \setminus \text{Int}(D)} = I] = \mathbb{P}[\pi_A = I|\pi_{C} = I \wedge \pi_{AB \setminus \text{Int}(D)} = I] \geq  \mathbb{P}[\pi_A = I|\pi_{C} = I]$. Then, since the second term in \eqref{eq:fO bad good} is non-negative for the case $g = I_A$, we have
	\begin{align}
		(1 - \mathbb{P}[\pi \text{ is bad}]) \leq r[I_A] \leq 1
	\end{align}
	which, together with Eqs.~(\ref{eq:prob bad}, \ref{eq:proxy error fO}), proves Proposition \ref{prop:replica bound}. \hfill $\square$
	
	\subsection{Heuristic argument for higher replica indices and non-diagonal observables}
	
	While the rigorous bound we have obtained (Proposition \ref{prop:replica bound}) only pertains to the case $k = 2, s = 1, t = 0$, the majority of our argument holds generally for any $O(1)$ value of the replica indices. In particular, we have Eqs.~(\ref{eq:prob bad}, \ref{eq:fO bad good}) for all $k,s,t \geq 0$. The final step involved the FKG inequalities in the form of Eqs.~(\ref{eq:FKG1}, \ref{eq:FKG2}), which is specific to the Ising model. Nevertheless, these inequalities can be viewed as a reflection of the fact that the spins in the Ising model interact ferromagnetically, leading to positive correlations between events that fix various spins to the particular state $I$. Since the spin model \eqref{eq:moment energy} also interact ferromagnetically, we speculate that these inequalities should generalize to arbitrary $k, s, t \geq 0$. This would suffice to show that the deviation of $\Xi^{(t)}_{k,s}$ from unity scales as \eqref{eq:error ising}.
	
	To generalize our result to non-diagonal observables, we need to explicitly consider the case $g = f_O$, where $f_O(\pi_A) = \tr[\text{Wg}(\pi_A) O]$, with $O \geq 0$ a positive semi-definite observable in the symmetric subspace. Since any such $O$ can be decomposed in terms of projectors onto irreps of the symmetric group $S_k$ on each site, the space of functions $f_O$ over all such choices of $O$ is the positive cone generated by the irrep characters of the group $S_k^A$, namely $\chi_{\mu_1, \ldots, \mu_A}(\pi_A) = \prod_{v \in A} \chi_{\mu_v}(\pi_v)$, with $\chi_{\mu_v} : S_k \rightarrow \mathbb{R}$ the character for a chosen irrep $\mu_v \in \text{Irr}(S_k)$. For example, in the case $k = 2$, we have $f_O(\pi_A) = \sum_{X \subseteq A} c_X \sigma_X$, where the coefficients $c_X \geq 0$, and $\sigma_X \coloneqq \prod_{v \in X} \pi_v \in \{\pm 1\}$.
	
	For $k = 2, s = 1, t = 0$ we can bound $r_{\rm good}[f_O]$ using another family of inequalities for the Ising model, namely the Griffiths-Kelly-Sherman (GKS) inequalities $\mathbb{E}[\sigma_X \sigma_Y] \geq \mathbb{E}[\sigma_X]\mathbb{E}[\sigma_Y]$, which hold for all positive couplings $J_{v_1, v_2} \geq 0$ and fields $h_v \geq 0$. Since the indicator function $\mathbbm{1}(\pi_X = I)$ can be written as $2^{-|X|}\sum_{Y \subseteq X} \sigma_Y$, which is in the positive cone generated by the characters $\sigma_Y$, one finds that $r_{\rm good}[f_O] \geq 1$.
	
	Unfortunately, in this instance the second term in \eqref{eq:fO bad good} does not have a definite sign, so we cannot convert this to a rigorous bound on $r[f_O]$ without deriving further bounds on the contributions from bad contour configurations. However, we believe it unlikely that the highly suppressing factor of $\mathbb{P}[\pi \text{ is bad}]$ can be overcome, and thus we conjecture that the same bounds apply to off-diagonal observables too. In future work, it would be interesting to explore whether correlation inequalities of FKG- and GKS-type can be derived in the more general spin models for higher replica indices, which would be useful in making these generalized arguments rigorous.
	
	\section{The stabilizer Scrooge ensemble and shallow 2D Clifford-random quantum circuits}
	
	\subsection{The stabilizer Scrooge ensemble}
	\label{sec:stabilizer_scrooge_ensemble}
	Looking at \Cref{def:scrooge}, we can define analogues of the Scrooge ensembles for $n$-qubit stabilizer states, which we refer to as stabilizer Scrooge ensembles. A stabilizer Scrooge ensemble is a discrete probability distribution over pure stabilizer states that is parametrized by a mixed stabilizer state $\rho$. We define this ensemble in a way that naturally generalizes the fourth characterization in \Cref{def:scrooge}, where Haar-random unitaries are replaced by uniformly random Clifford unitaries.
	\begin{definition}[Stabilizer Scrooge Distribution]
		\label{def:stabilizer_scrooge}
		For a mixed stabilizer state $\rho$, let $\ket{\Psi^\rho_{AB}}$ be any stabilizer purification, i.e., a stabilizer state such that $\tr_B[\Psi^\rho_{AB}] = \rho_A$. The stabilizer Scrooge ensemble $\mathcal{D}_{\rho}$ is the distribution of pure states obtained by applying a uniformly random Clifford unitary $U_B$ to $B$, followed by a projective measurement on $B$ in the computational basis, with outcome $x_B$, over the joint randomness induced by $U_B$ and $x_B$.
	\end{definition}
	
	We first provide a useful structural characterization of stabilizer Scrooge distributions.
	\begin{lemma}
		\label{lem:scrooge_plant}
		Let $\rho$ be an $n$-qubit mixed stabilizer state with independent stabilizer generators $\{g_1,\ldots,g_k\}$. Let $U$ be any $n$-qubit Clifford unitary such that $\{UZ_j U^{\dagger}:j\in\{1,\ldots,k\}\}=\{g_1,\ldots,g_k\}$. Then the stabilizer Scrooge distribution defined by $\rho$ is the uniform distribution over the set $\{U(\ket{0^k}\otimes \ket{\phi}):\ket{\phi}\text{ is an $(n-k)$-qubit stabilizer state}\}$.
	\end{lemma}
	\begin{proof}
		By the definition of $U$, we have
		\begin{equation}
			\rho=\frac{1}{2^{n-k}}\prod_{j=1}^k\frac{I^{\otimes n}+g_j}{2}=\frac{1}{2^{n-k}}\prod_{j=1}^k\frac{I^{\otimes n}+UZ_jU^\dagger}{2}=\frac{1}{2^{n-k}}U(\ket{0^{k}}\bra{0^{k}}\otimes I^{\otimes n-k})U^\dagger.
		\end{equation}
		We define $A_1$ to be the first $k$ qubits of $\rho$ and $A_2$ to be the last $n-k$ qubits of $\rho$. Notice that $\ket{\Psi}=\ket{0^k}_{A_1}\otimes\ket{\Phi}_{A_2B}$ is a purification of $\frac{1}{2^{n-k}}\ket{0^k}\bra{0^k}\otimes I^{\otimes n-k}$ where $\ket{\Phi}$ is the $2(n-k)$-qubit maximally entangled state and the ancilla register $B$ contains $n-k$ qubits. Then $(U_{A_1A_2}\otimes I_B)\ket{\Psi}$ is a purification of $\rho$. Thus, by \Cref{def:stabilizer_scrooge} and the transpose trick applied with $\ket{\Phi}$, the claim follows. 
	\end{proof}
	
	\noindent Given $\rho$ with stabilizer generators $\{g_1,\ldots,g_k\}$, one way to find a $U$ that satisfies $\{UZ_jU^\dagger:j\in\{1,\ldots,k\}\}=\{g_1,\ldots,g_k\}$ is to extend $\{g_1,\ldots,g_k\}$ to a complete set $\{g_1,\ldots,g_k,g_{k+1},\ldots,g_n\}$ of independent stabilizer generators for some $n$-qubit pure stabilizer state $\ket{\psi}$ and take a $U$ that satisfies $\ket{\psi}=U\ket{0^n}$. 
	
	From \Cref{lem:scrooge_plant}, we see that on one extreme, the stabilizer Scrooge distribution reduces to the uniform distribution over $n$-qubit pure stabilizer states if $\rho=\frac{I^{\otimes n}}{2^n}$ is the $n$-qubit maximally mixed state, while on the other extreme, the stabilizer Scrooge distribution always produces the same pure state $\ket{\psi}$ if $\rho=\ket{\psi}\bra{\psi}$ is pure. More generally, when $\rho$ has $2^k$ stabilizers for some $k\in\{0,1,\ldots,n\}$, we can think of $\rho$ as representing a stabilizer code that encodes $n-k$ logical qubits. Under this interpretation, the stabilizer Scrooge distribution is the uniform distribution over the set of all codewords of the stabilizer code represented by $\rho$.
	
	Before we proceed, we describe another way to sample from a stabilizer Scrooge ensemble.
	
	\begin{definition}[Alternative stabilizer Scrooge sampler]
		\label{def:alternative_scrooge_sampler}
		Let $\rho$ be an $n$-qubit mixed stabilizer state.
		\begin{enumerate}
			\item Draw a uniformly random $n$-qubit stabilizer state $\ket{\varphi}$.
			
			\item If $\rho\ket{\varphi}=0$, go back to step $1$.
			
			\item Output
			\begin{equation}
				\frac{\rho\ket{\varphi}}{\lVert\rho\ket{\varphi}\rVert}.
			\end{equation}
		\end{enumerate}
	\end{definition}
	
	\noindent We will implement the algorithm in \Cref{def:alternative_scrooge_sampler} to sample from stabilizer Scrooge ensembles in our numerical experiments in \Cref{sec:scrooge_numerics}. The next lemma shows that the alternative protocol indeed samples from the correct distribution. 
	
	\begin{lemma}
		On input $\rho$, the algorithm given in \Cref{def:alternative_scrooge_sampler} samples from the stabilizer Scrooge distribution defined by $\rho$.
	\end{lemma}
	\begin{proof}
		It suffices to show that under the conditions of \Cref{lem:scrooge_plant}, the algorithm given in \Cref{def:alternative_scrooge_sampler} also samples from the uniform distribution over the set $\{U(\ket{0^k}\otimes \ket{\phi}):\ket{\phi}\text{ is an $(n-k)$-qubit stabilizer state}\}$.
		
		First consider the case where $\rho=\frac{1}{2^{n-k}}(\ket{0^k}\bra{0^k}\otimes I)$. In this case, clearly, the algorithm only ever outputs states of the form $\ket{0^k}\otimes \ket{\phi}$ for some $(n-k)$-qubit stabilizer state $\ket{\phi}$. Moreover, the distribution is invariant under unitary transformations of the form $I\otimes V$ for every $(n-k)$-qubit Clifford unitary $V$. Therefore, the procedure samples from the uniform distribution over
		\begin{equation}
			\{\ket{0^k}\otimes\ket{\phi}:\ket{\phi}\text{ is an $(n-k)$-qubit stabilizer state}\}.
		\end{equation}
		By the definition of $\rho$, we have $\{g_1,\ldots,g_k\}=\{Z_1,\ldots,Z_k\}$, so it is not hard to see that $U$ must commute with the projector $|0^k\rangle\langle 0^k|\otimes I$. The claim then follows since the sampled distribution is invariant under $\ket{0^k}\otimes\ket{\phi}\rightarrow U(\ket{0^k}\otimes\ket{\phi})$.
		
		Next consider an arbitrary $n$-qubit stabilizer state $\rho$ with stabilizer group  $\langle g_1,g_2\ldots,g_k\rangle $. Recall that for a Clifford unitary $U$ satisfying $\{U Z_j U^{\dagger}:j\in\{1,\ldots,k\}\}=\{g_1,\ldots,g_k\}$, we can write
		\begin{equation}
			\rho=\frac{1}{2^{n-k}}U(\ket{0^{k}}\bra{0^{k}}\otimes I)U^\dagger.
		\end{equation}
		Therefore, we can write
		\begin{equation}
			\rho\ket{\varphi}=\frac{1}{2^{n-k}}U(\ket{0^{k}}\bra{0^{k}}\otimes I)U^\dagger\ket{\varphi}.
		\end{equation}
		By the unitary invariance property of $\ket{\varphi}\rightarrow U^\dagger\ket{\varphi}$, the algorithm in \Cref{def:alternative_scrooge_sampler} is equivalent to the following procedure.
		\begin{enumerate}
			\item Draw a uniformly random $n$-qubit stabilizer state $\ket{\varphi}$.
			\item If $(\ket{0^k}\bra{0^k}\otimes I)\ket{\varphi}=0$, go back to step $1$.
			\item Define
			\begin{equation}
				\ket{\alpha}=\frac{(\ket{0^k}\bra{0^k}\otimes I)\ket{\varphi}}{\lVert(\ket{0^k}\bra{0^k}\otimes I)\ket{\varphi}\rVert}.
			\end{equation}
			\item Output
			\begin{equation}
				U\ket{\alpha}.
			\end{equation}
		\end{enumerate}
		Notice that by the first case, $\ket{\alpha}$ is a uniformly random state drawn from the set 
		\begin{equation}
			\{\ket{0^k}\otimes\ket{\phi}:\ket{\phi}\text{ is an $(n-k)$-qubit stabilizer state}\}.    
		\end{equation}
		Therefore, $U\ket{\alpha}$ is a uniformly random state drawn from the set 
		\begin{equation}
			\{U(\ket{0^k}\otimes\ket{\phi}):\ket{\phi}\text{ is an $(n-k)$-qubit stabilizer state}\}.
		\end{equation}
	\end{proof}
	
	Next, we establish some properties about an illuminating special case of stabilizer Scrooge distributions where $\rho=\rho_A\otimes\rho_B$ is a product state. In particular, we will focus on understanding the distribution of the number of local and non-local $Z$-type stabilizers in a Scrooge-random state defined by product-state $\rho$. We first look at local $Z$-type stabilizers. In the following, let $[n]=A\cup B$ be a bipartition of the $n$ qubits with $|A|=n_A$ and $|B|=n_B$ so that $n_A+n_B=n$. Let $\rho_A$ be an $n_A$-qubit mixed stabilizer state with $2^{k_A}$ stabilizers, and let $\rho_B$ be an $n_B$-qubit mixed stabilizer state with $2^{k_B}$ stabilizers. Define $\rho=\rho_A\otimes \rho_B$ and let $\mathcal{D}_\rho$ denote the stabilizer Scrooge distribution defined by $\rho$. Let $S^Z_{\rho_A}$ denote the $Z$-type stabilizer subgroup of $\rho$ local on $A$, which is in bijection with the $Z$-type stabilizer subgroup of $\rho_A$. We use $\ket{\psi}$ to denote a Scrooge-random state drawn from $\mathcal{D}_\rho$, and we use $S^Z_{\psi_A}$ to denote the $Z$-type stabilizer subgroup of $\ket{\psi}$ local on $A$. In particular, this consists of all $Z$-type stabilizers of $\ket{\psi}$ that act nontrivially only on $A$ and as the identity on $B$. We first give a lower bound on the probability that $S_{\psi_A}^Z$ remains unchanged from $S_{\rho_A}^Z$.
	
	\begin{lemma}
		Let $\rho=\rho_A\otimes \rho_B$ be an $n$-qubit product stabilizer state, with subsystems $A \subseteq [n]$, $B = [n] \setminus A$. It holds that \begin{equation}
			\Pr_{\ket{\psi}\sim\mathcal{D}_\rho}(S^Z_{\psi_A}= S^Z_{\rho_A})\geq 1-\frac{1}{2^{n_B-k_B}},
		\end{equation}
		where $n_B=|B|$ and $\rho_B$ has $2^{k_B}$ stabilizers.
        
        \label{lem:scrooge_no_new_local_Z_type}
	\end{lemma}
	
	\begin{proof}
		By applying \Cref{lem:scrooge_plant} to $\rho_A\otimes \rho_B$, we can choose $n_A$- and $n_B$-qubit Clifford unitaries $U$ and $V$ such that $\mathcal{D}_\rho$ is the uniform distribution over the set
		\begin{equation}
			\{(U\otimes V)(\ket{0^{k_A}}\otimes\ket{0^{k_B}}\otimes\ket{\phi}):\ket{\phi}\text{ is an $(n-k_A-k_B)$-qubit stabilizer state}\}.
		\end{equation}
		Define 
		\[
		R^{*}=\pm U^{\dagger} \{I,Z\}^{\otimes {n_A}} U,
		\]
		i.e., $R^{*}$ is the set of Pauli operators $P$ on $A$ such that $UP U^{\dagger}$ is $Z$-type.
		
		For every $m\geq 1$, define $\mathcal{P}^m=\{I,X,Y,Z\}^{\otimes m}$. Define
		\begin{equation}
			R'=\{Q\in\mathcal{P}^{n_A-k_A}:Z(s)\otimes Q\in R^*\text{ for some $s\in\{0,1\}^{k_A}$}\}.
		\end{equation}
		Notice that $R'$ is a commuting set of Pauli operators supported on $n_A-k_A$ qubits with $+$ sign. Therefore, by maximality, $|R'|\leq 2^{n_A-k_A}$. Note that $I^{\otimes n_A-k_A}\in R'$ and $R'$ is determined entirely by $U$ which is independent of any randomness involving $\ket{\phi}$.
		
		Recall that we have $\ket{\psi}\sim\mathcal{D}_\rho$, so $\ket{\psi}=(U\otimes V)(\ket{0^{k_A}}\otimes\ket{0^{k_B}}\otimes\ket{\phi})$ where $\ket{\phi}$ is a uniformly random $(n-k_A-k_B)$-qubit stabilizer state. Let $S_{\phi}$ denote the stabilizer group of $\ket{\phi}$. Define
		\begin{equation}
			C=\{Q\in R':\pm Q\otimes I^{\otimes n_B-k_B}\in S_\phi\}.
		\end{equation}
		For every $Q\in C$, fix a representative $s_Q\in\{0,1\}^{k_A}$ that satisfies $Z(s_Q)\otimes Q\in R^*$, and let $c_Q\in\{-1,1\}$ be the sign that makes $c_Q\cdot Q\otimes I^{\otimes n_B-k_B}\in S_\phi$. Define
		\begin{equation}
			C'=\{c_Q\cdot U(Z(s_Q)\otimes Q)U^\dagger:Q\in C\}.
		\end{equation}
		Clearly, $|C'|=|C|$. Notice that by construction, $C'\subseteq S_{\psi_A}^Z$. 
		
		We next show that $S_{\psi_A}^Z\subseteq S_{\rho_A}^Z\times C'$, which implies $|S_{\psi_A}^Z|\leq |S_{\rho_A}^Z|\cdot |C|$. Let $P\in S_{\psi_A}^Z$. Then by the definition of $S_{\psi_A}^Z$, $P=\pm U(Z(s)\otimes Q)U^\dagger$ for some $s\in\{0,1\}^{k_A}$ and $Q\in\mathcal{P}^{n_A-k_A}$ satisfying $Z(s)\otimes Q\in R^*$ and $\pm Q\otimes I^{\otimes n_B-k_B}\in S_\phi$. Observe that the condition on $Q$ implies $Q\in C$, so $P'=c_Q\cdot U(Z(s_Q)\otimes Q)U^\dagger\in C'$. Define $Z(r)=Z(s)Z(s_Q)$. Then $P\in S_{\psi_A}^Z$ and $P'\in S_{\psi_A}^Z$ imply
		\begin{equation}
			PP'=U(Z(r)\otimes I^{\otimes n_A-k_A})U^\dagger\in S_{\psi_A}^Z.
		\end{equation}
		Notice that by the definition of $S_{\rho_A}^Z$, $PP'\in S_{\rho_A}^Z$. Therefore, $P=(PP')P'$ with $PP'\in S_{\rho_A}^Z$ and $P'\in C'$. Observe that
		\begin{equation}
			\Exp[|C|]=\sum_{Q\in R'}\Pr(\pm Q\otimes I^{\otimes n_B-k_B}\in S_\phi)\leq 1+\frac{2^{n_A-k_A}-1}{2^{n-k_A-k_B}+1}\leq 1+\frac{1}{2^{n_B-k_B}}.
		\end{equation}
		Therefore,
		\begin{equation}
			\Exp_{\ket{\psi}\sim\mathcal{D}_\rho}\left[\frac{|S_{\psi_A}^Z|}{|S_{\rho_A}^Z|}\right]\leq \Exp[|C|]\leq 1+\frac{1}{2^{n_B-k_B}}.
		\end{equation}
		By the law of total expectation,
		\begin{align}
			\Exp_{\ket{\psi}\sim\mathcal{D}_\rho}\left[\frac{|S_{\psi_A}^Z|}{|S_{\rho_A}^Z|}\right]&=1\cdot \Pr_{\ket{\psi}\sim\mathcal{D}_\rho}(S_{\psi_A}^Z=S_{\rho_A}^Z)+\Exp_{\ket{\psi}\sim\mathcal{D}_\rho}\left[\frac{|S_{\psi_A}^Z|}{|S_{\rho_A}^Z|}\middle|S_{\psi_A}^Z\neq S_{\rho_A}^Z\right]\left(1-\Pr_{\ket{\psi}\sim\mathcal{D}_\rho}(S_{\psi_A}^Z=S_{\rho_A}^Z)\right)\\
			&\geq \Pr_{\ket{\psi}\sim\mathcal{D}_\rho}(S_{\psi_A}^Z=S_{\rho_A}^Z)+2\left(1-\Pr_{\ket{\psi}\sim\mathcal{D}_\rho}(S_{\psi_A}^Z=S_{\rho_A}^Z)\right).
		\end{align}
		Therefore,
		\begin{equation}
			\Pr_{\ket{\psi}\sim\mathcal{D}_\rho}(S_{\psi_A}^Z=S_{\rho_A}^Z)\geq 2-\Exp_{\ket{\psi}\sim\mathcal{D}_\rho}\left[\frac{|S_{\psi_A}^Z|}{|S_{\rho_A}^Z|}\right]\geq 1-\frac{1}{2^{n_B-k_B}}.
		\end{equation}
	\end{proof}
	
	\Cref{lem:scrooge_no_new_local_Z_type} implies that as long as $n_B-k_B\geq \Omega(\log n)$, then with probability at least $1-1/\text{poly}(n)$, the $Z$-type stabilizer subgroup of $\ket{\psi}$ local on $A$ will be the same as that of $\rho_A$; in other words, applying Scrooge to $\rho_A\otimes\rho_B$ does not introduce to $\ket{\psi}$ new $Z$-type stabilizers local on $A$ with high probability. Since $Z$-type stabilizers dictate the output distribution of $\ket{\psi}$ when all qubits are measured in the computational basis, $S_{\psi_A}^Z=S_{\rho_A}^Z$ also implies that the marginal output distribution of $\ket{\psi}$ on $A$ will be the same as the output distribution of $\rho_A$. Also note that since $\rho_B$ is a mixed stabilizer state, $n_B-k_B$ equals the order-$\alpha$ Renyi entropy of $\rho_B$ for every $\alpha$. 
	
	In the following, we extend \Cref{lem:scrooge_no_new_local_Z_type} to the more general case where $\ket{\psi}$ is allowed to be drawn from not necessarily the exact stabilizer Scrooge distribution defined by $\rho_A\otimes\rho_B$ but instead from a relative error $\varepsilon$-approximate Scrooge $2$-design. We first define relative error $\varepsilon$-approximate Scrooge $2$-designs.
	
	\begin{definition}
		Let $\varepsilon\geq 0$. Let $\rho$ be an $n$-qubit mixed stabilizer state, and let $\mathcal{D}_\rho$ denote the stabilizer Scrooge distribution defined by $\rho$. We say a distribution $\mathcal{D}'_\rho$ of $n$-qubit pure states forms a relative error $\varepsilon$-approximate Scrooge $2$-design w.r.t. $\mathcal{D}_\rho$ if for every $2^{2n}$-by-$2^{2n}$ hermitian positive-semidefinite matrix $O$, it holds that
		\begin{equation}
			(1-\varepsilon)\tr\left(O\Exp_{\ket{\psi}\sim\mathcal{D}_\rho}\left[\ket{\psi}\bra{\psi}^{\otimes 2}\right]\right)\leq \tr\left(O\Exp_{\ket{\psi'}\sim\mathcal{D}'_\rho}\left[\ket{\psi'}\bra{\psi'}^{\otimes 2}\right]\right)\leq (1+\varepsilon)\tr\left(O\Exp_{\ket{\psi}\sim\mathcal{D}_\rho}\left[\ket{\psi}\bra{\psi}^{\otimes 2}\right]\right).
			\label{eq:scrooge_2_design_condition}
		\end{equation}
		\label{def:scrooge_2_design_def}
	\end{definition}
	
	\begin{corollary}
		\label{cor:scrooge_2_design_no_new_local_Z_type}
		Let $\mathcal{D}'_\rho$ be any distribution of $n$-qubit pure states which satisfies Eq.~\eqref{eq:scrooge_2_design_condition} with the specific observable
		\begin{equation}
			O=2^{n_A}\sum_{x\in\{0,1\}^{n_A}}\ket{x}\bra{x}\otimes I^{\otimes n_B}\otimes\ket{x}\bra{x}\otimes I^{\otimes n_B},
			\label{eq:the_diagonal_observable}
		\end{equation}
		and where $\rho=\rho_A\otimes\rho_B$ is a product state. For a state $\ket{\psi'}\sim\mathcal{D}'_\rho$, let $S^Z_{\psi'_A}$ denote the $Z$-type stabilizer subgroup of $\ket{\psi'}$ local on $A$. Then it holds that
		\begin{equation}
			\Pr_{\ket{\psi'}\sim\mathcal{D}'_\rho}(S^Z_{\psi'_A}= S^Z_{\rho_A})\geq 1-\varepsilon-\frac{1+\varepsilon}{2^{n_B-k_B}}.
			\label{eq:scrooge_2_design_no_new_local_Z_type_condition}
		\end{equation}
		
	\end{corollary}
	\begin{proof}
		In the proof of \Cref{lem:scrooge_no_new_local_Z_type}, we have established that for $\ket{\psi}$ drawn from the stabilizer Scrooge distribution defined by $\rho_A\otimes\rho_B$,
		\begin{equation}
			\Exp_{\ket{\psi}\sim\mathcal{D}_\rho}\left[\frac{|S_{\psi_A}^Z|}{|S_{\rho_A}^Z|}\right]\leq 1+\frac{1}{2^{n_B-k_B}}.
		\end{equation}
		Through standard manipulations, we can write
		\begin{equation}
		\Exp_{\ket{\psi}\sim\mathcal{D}_\rho}\left[|S_{\psi_A}^Z|\right]=\Tr\bigg(O\Exp_{\ket{\psi}\sim\mathcal{D}_\rho}\left[\ket{\psi}\bra{\psi}^{\otimes 2}\right]\bigg),
		\end{equation}
		where
		\begin{equation}
			O=2^{n_A}\sum_{x\in\{0,1\}^{n_A}}\ket{x}\bra{x}\otimes I^{\otimes n_B}\otimes\ket{x}\bra{x}\otimes I^{\otimes n_B}.
		\end{equation}
		Using Eq.~\eqref{eq:scrooge_2_design_condition} with the observable $O$ gives
		\begin{equation}
			\Exp_{\ket{\psi'}\sim\mathcal{D}'_\rho}\left[\frac{|S_{\psi'_A}^Z|}{|S_{\rho_A}^Z|}\right]\leq (1+\varepsilon)\Exp_{\ket{\psi}\sim\mathcal{D}_\rho}\left[\frac{|S_{\psi_A}^Z|}{|S_{\rho_A}^Z|}\right]\leq (1+\varepsilon)\left(1+\frac{1}{2^{n_B-k_B}}\right).
		\end{equation}
		Therefore,
		\begin{equation}
			\Pr_{\ket{\psi'}\sim\mathcal{D}'_\rho}(S_{\psi'_A}^Z=S_{\rho_A}^Z)\geq 2-\Exp_{\ket{\psi'}\sim\mathcal{D}'_\rho}\left[\frac{|S_{\psi'_A}^Z|}{|S_{\rho_A}^Z|}\right]\geq 1-\varepsilon-\frac{1+\varepsilon}{2^{n_B-k_B}}.
		\end{equation}
	\end{proof}
	
	Note that just like in \Cref{lem:scrooge_no_new_local_Z_type}, if we have $n_B-k_B\geq\Omega(\log n)$ and $\varepsilon\leq 1/\text{poly}(n)$ in \Cref{cor:scrooge_2_design_no_new_local_Z_type}, then we obtain $S_{\psi'_A}^Z=S_{\rho_A}^Z$ with probability $1-1/\text{poly}(n)$.  
	
	Lastly, we work out the distribution of the mutual information $I(A;B)$ for the computational-basis output distribution of $\ket{\psi}\sim\mathcal{D}_\rho$ where $\rho=\rho_A\otimes\rho_B$. Note that $I(A;B)$ is a random variable with randomness induced by $\ket{\psi}\sim\mathcal{D}_\rho$. Analogous to $S_{\psi_A}^Z$, let $S_{\psi_B}^Z$ denote the $Z$-type stabilizer subgroup of $\ket{\psi}$ local on $B$. We also let $S_{\psi}^Z$ denote the whole $Z$-type stabilizer subgroup of $\ket{\psi}$. We say $P\in S_{\psi}^Z$ is non-local if $P$ cannot be decomposed as a product $P=P_AP_B$ for some $P_A\in S_{\psi_A}^Z$ and $P_B\in S_{\psi_B}^Z$. To find a non-local $Z$-type stabilizer of $\ket{\psi}$, we can first find a set of independent generators $G_A^Z$ and $G_B^Z$ for $S_{\psi_A}^Z$ and $S_{\psi_B}^Z$ respectively, and then try to extend $G_A^Z\cup G_B^Z$ to a full set of independent generators $G^Z$ for $S_\psi^Z$. With $G^Z$ constructed this way, $\ket{\psi}$ has a non-local $Z$-type stabilizer if and only if $G^Z\neq G_A^Z\cup G_B^Z$ and every $P\in G^Z\setminus(G_A^Z\cup G_B^Z)$ is a non-local $Z$-type stabilizer of $\ket{\psi}$ (recall that $G^Z$ is a set of generators so $I\notin G^Z$). More generally, for a non-local $Z$-type stabilizer $P$ of $\ket{\psi}$, the product $PQ$ remains non-local for every $Q\in S_{\psi_A}^Z\cup S_{\psi_B}^Z$, i.e., non-local times local is non-local. We wish to characterize the distribution of the mutual information
	\begin{equation}
		I(A;B)=H(A)+H(B)-H(AB)=(n_A-|G_A^Z|)+(n_B-|G_B^Z|)-(n-|G^Z|)=|G^Z|-|G_A^Z|-|G_B^Z|
	\end{equation}
	which can also be interpreted as the number of independent non-local $Z$-type stabilizer generators of $\ket{\psi}$.  Recall that we have $\rho=\rho_A\otimes\rho_B$ where $\rho_A$ and $\rho_B$ have $2^{k_A}$ and $2^{k_B}$ stabilizers respectively. Let $\mathcal{D}$ denote the uniform distribution over the set of all $(n-k_A-k_B)$-qubit stabilizer states. We will view each $\ket{\phi}\sim\mathcal{D}$ as a bipartite state between the first $n_A-k_A$ and last $n_B-k_B$ qubits and consider the mutual information of the computational-basis output distribution of $\ket{\phi}$.
	
	\begin{theorem}
		It holds that for every integer $r\geq 0$,
		\begin{equation}
			\Pr_{\ket{\psi}\sim\mathcal{D}_\rho}\left[I(A;B)=r\right]=\Pr_{\ket{\phi}\sim\mathcal{D}}\left[I(A;B)=r\right].
		\end{equation}
		\label{thm:scrooge_nonlocal_Z_distribution}
	\end{theorem}
	
	\begin{proof}
		By applying \Cref{lem:scrooge_plant} to $\rho_A\otimes \rho_B$, we can choose $n_A$- and $n_B$-qubit Clifford unitaries $U$ and $V$ such that $\mathcal{D}_\rho$ is the uniform distribution over the set
		\begin{equation}
			\{(U\otimes V)(\ket{0^{k_A}}\otimes\ket{0^{k_B}}\otimes\ket{\phi}):\ket{\phi}\text{ is an $(n-k_A-k_B)$-qubit stabilizer state}\}.
		\end{equation}
		
		Let $\ket{\psi}=(U\otimes V)(\ket{0^{k_A}}\otimes\ket{0^{k_B}}\otimes\ket{\phi})\sim\mathcal{D}_\rho$. Let $A_1$ denote the first $k_A$ qubits of $A$ and let $A_2$ denote the last $n_A-k_A$ qubits of $A$. Similarly, let $B_1$ denote the first $k_B$ qubits of $B$ and let $B_2$ denote the last $n_B-k_B$ qubits of $B$. Note that under this further subdivision, the support of $\ket{\phi}$ is precisely $A_2\cup B_2$.
		
		For every set $S$ consisting of Pauli operators, we adopt the notation $\pm S=S\cup\{-P:P\in S\}$. Define
		\begin{align}
			R&=\pm(U^\dagger\otimes V^\dagger)\{I,Z\}^{\otimes n}(U\otimes V)=\{\pm(U^\dagger\otimes V^\dagger)Z(s)(U\otimes V):s\in\{0,1\}^n\}\\
			&=\pm (U^\dagger\{I,Z\}^{\otimes n_A}U)\otimes (V^\dagger\{I,Z\}^{\otimes n_B}V). \label{eq:R_product}
		\end{align}
		By construction, $(U\otimes V)R(U^\dagger\otimes V^\dagger)$ is the set of all $n$-qubit $Z$-type Pauli operators with $\pm$ signs. Note that $R$ is a maximal set of $n$-qubit commuting Pauli operators with $\pm$ signs, and for every Pauli operator $P\in R$, we call $P$ an $R$-type Pauli operator.
		
		Let us also define
		\begin{equation}
			T=\{P\in\pm\{I,X,Y,Z\}^{\otimes n-k_A-k_B}:Z(w)\otimes P\in R\text{ for some }w\in\{0,1\}^{k_A+k_B}\}.
		\end{equation}
		It is also the case that $T$ is a maximal set of $(n-k_A-k_B)$-qubit commuting Pauli operators with $\pm$ signs, i.e.,
		\begin{equation}
			|T|=2\cdot 2^{n-k_A-k_B}.
			\label{eq:sizeT}
		\end{equation}
		To establish Eq.~\eqref{eq:sizeT}, consider starting from the $n$-qubit state $(U^{\dagger}\otimes V^{\dagger})|0^n\rangle$, which has a stabilizer group $S'\subseteq R$, and then measuring all qubits in $A_1\cup B_1$ in the computational basis. The resulting state (conditioned on a particular measurement outcome) is an $(n-k_A-k_B)$-qubit pure state with a stabilizer group $S''$ such that $\pm S''=T$. Since the post-measurement state is pure, we have $|S''|=2^{n-k_A-k_B}=|T|/2$. Note that by Eq.~\eqref{eq:R_product}, the set $T$ factorizes across the $A_2B_2$ bipartition.
		
		We call every $P\in T$ a $T$-type Pauli operator. Note that the set $T$ is a fixed set fully determined by $U\otimes V$. 
		
		Next we show that the number of independent non-local $R$-type stabilizer generators of $\ket{0^{k_A}}\otimes \ket{0^{k_B}}\otimes \ket{\phi}$ equals the number of independent non-local $T$-type stabilizer generators of $\ket{\phi}$. More precisely, let
		\[
		|\Phi\rangle=|0^{k_A}\rangle\otimes |0^{k_B}\rangle\otimes |\phi\rangle
		\]
		and define $S(\Phi)$ to be the stabilizer group of $\Phi$. Let $S_A(\Phi),S_B(\Phi)$ be the subgroups of $S(\Phi)$ that are local on subsystems $A,B$ respectively. Finally, let
		\[
		S^{R}(\Phi)=S(\Phi)\cap R, \qquad S^{R}_A(\Phi)=S_A(\Phi)\cap R, \qquad S^{R}_B(\Phi)=S_B(\Phi)\cap R
		\]
		be the corresponding $R$-type subgroups.
		
		Similarly, define the group of stabilizers $S(\phi)$ of $|\phi\rangle$, its local subgroups $S_{A_2}(\phi),S_{B_2}(\phi)$, and their $T$-type subgroups:
		\[
		S^{T}(\phi)=S(\phi)\cap T, \qquad S^{T}_{A_2}(\phi)=S_{A_2}(\phi)\cap T, \qquad S^{T}_{B_2}(\phi)=S_{B_2}(\phi)\cap T.
		\]
		We are interested in the quotient groups
		\[
		H(\Phi)\equiv S^R(\Phi)/(S^{R}_A(\Phi)S^{R}_B(\Phi)), \qquad H(\phi)\equiv S^T(\phi)/(S^{T}_{A_2}(\phi)S^{T}_{B_2}(\phi)).
		\]
		The mutual information $I(A;B)$ of the output distribution of $|\psi\rangle$ is equal to $\log_2(|H(\Phi)|)$. Since $\ket{\phi}$ is a uniformly random stabilizer state, the distribution of $H(\phi)$ is invariant if we replace $|\phi\rangle\rightarrow W|\phi\rangle$ where $W=W_{A_2}\otimes W_{B_2}$ is an $(n-k_A-k_B)$-qubit Clifford that maps the set of $T$-type Pauli operators to the set of $Z$-type Pauli operators.  Therefore the distribution of $\log_2(|H(\phi)|)$ is the same as the distribution of the mutual information $I(A_2;B_2)$ of the output distribution of $\ket{\phi}$. The following claim then completes the proof.
		\begin{claim}
			$|H(\Phi)|=|H(\phi)|$
		\end{claim}
		\begin{proof}
			Let us define a function $f:H(\Phi)\rightarrow H(\phi)$ as follows. Let $P\in\pm\{I,X,Y,Z\}^{\otimes n}$ such that the equivalence class
			\[
			[P]\in H(\Phi).
			\]
			Then we can write
			\[
			P=Z(s)_{A_1B_1}\otimes Q_{A_2B_2}
			\]
			for some $s\in\{0,1\}^{k_A+k_B}$ and $Q\in\pm\{I,X,Y,Z\}^{\otimes n-k_A-k_B}$. Note that since $P$ is non-local, $Q$ is non-local. We define
			\[
			f([P])=[Q]\in H(\phi).
			\]
			Note that this function is well-defined; if we choose another representative $P'$ such that $[P']=[P]\in  H(\Phi)$ then $P'=PL$ where $L\in S^{R}_A(\Phi)S^{R}_B(\Phi)$ is a local $R$-type Pauli operator, i.e.
			\[
			P'= (Z(s)_{A_1B_1}\otimes Q_{A_2B_2})(Z(t)_{A_1B_1}\otimes K_{A_2B_2})(Z(u)_{A_1B_1}\otimes J_{A_2B_2})
			\]
			for some bit strings $t,u$ and Pauli operators $K$ supported on $A_2$ and $J$ supported on $B_2$. From the above we see that $K\in S^{T}_{A_2}(\phi)$ and $J\in S^{T}_{B_2}(\phi)$ and therefore $f([P])=[Q]=[Q\cdot K\cdot J]=f([P'])$, establishing that $f$ is well defined.
			
			To complete the proof of the claim, we show that $f$ is invertible, and therefore a bijection. Suppose that $[D]\in H(\phi)$ for some $(n-k_A-k_B)$-qubit Pauli operator $D$. Then by the definition of $T$ we have
			\[
			Z(r)\otimes D\in R
			\]
			for some $r\in\{0,1\}^{k_A+k_B}$. Then the inverse function is
			\[
			f^{-1}([D])=[Z(r)\otimes D]
			\]
			We can confirm that this function is well-defined by checking that (a) if we start with some  other representative $D'$ such that $[D']=[D]$ in $H(\phi)$ then $f^{-1}([D])=f^{-1}([D'])$, and (b) if there is another string $r'\in \{0,1\}^{k_A+k_B}$ such that $Z(r')\otimes D\in R$ then $[Z(r')\otimes D]=[Z(r)\otimes D]$ in $H(\Phi)$. Finally, we can confirm it is a valid inverse, i.e.,  $f^{-1}(f([P]))=[P]$ for all $[P]\in H(\Phi)$ and $f(f^{-1}([D]))=[D]$ for all $[D]\in H(\phi)$.
		\end{proof}
        
	\end{proof}
	
	Notice that for a uniformly random $(n-k_A-k_B)$-qubit stabilizer state $\ket{\phi}\sim\mathcal{D}$, when viewed as a bipartite state between the first $n_A-k_A$ and the last $n_B-k_B$ qubits, the probabilities for $\ket{\phi}$ to have a non-identity $Z$-type stabilizer local on $A$ or $B$ can be upper bounded by $\frac{1}{2^{n_B-k_B}}$ or $\frac{1}{2^{n_A-k_A}}$ respectively. Thus, when both $n_A-k_A$ and $n_B-k_B$ are moderately large, it becomes overwhelmingly unlikely to draw a $\ket{\phi}$ that has a local $Z$-type stabilizer. Therefore, when both $n_A-k_A$ and $n_B-k_B$ are large, $I(A;B)$ effectively becomes the number of independent $Z$-type stabilizer generators of $\ket{\phi}$. Note that the same is true for $I(A;C|B)$ for a tripartition $[m]=A\cup B\cup C$ where $|A|=n_A-k_A$, $|C|=n_B-k_B$, and $|B|=m-|A|-|C|$ since measuring all qubits in $B$ of an $m$-qubit uniformly random stabilizer state in the computational basis produces a post-measurement state that is a uniformly random $(n-k_A-k_B)$-qubit stabilizer state on $A\cup C$, i.e., a state $\ket{\phi}\sim\mathcal{D}$. 
	
	A closed-form formula for the number of independent $Z$-type stabilizer generators of a uniformly random stabilizer state is known.
	
	\begin{lemma}[\cite{bravyi2016improved}]
		For every integer $r\in\{0,1,\ldots,n\}$, the probability that a uniformly random $n$-qubit stabilizer state has $r$ independent $Z$-type stabilizer generators is
		\begin{equation}
			\frac{\eta(r)}{\sum_{k=0}^n\eta(k)}
		\end{equation}
		where $\eta(0)=1$ and 
		\begin{equation}
			\eta(r)=2^{-r(r+1)/2}\prod_{a=1}^r \frac{1-2^{r-n-a}}{1-2^{-a}}.
		\end{equation}
		\label{lem:uniform_formula}
	\end{lemma}
	
	As $n$ increases, the probabilities in \Cref{lem:uniform_formula} quickly converge to
	\begin{align}
		P(0)=P(1)&=\prod_{k=0}^{\infty}\left(1-\frac{1}{2^{2k+1}}\right)\approx 0.41942244,\\
		P(2)&=\frac{1}{3}P(1)\approx 0.13980748,\\
		P(3)&=\frac{1}{7}P(2)\approx 0.01997250,\\
		P(4)&=\frac{1}{15}P(3)\approx 0.00133150,\\
		P(5)&=\frac{1}{31}P(4)\approx 0.00004295.
	\end{align}
	\Cref{thm:scrooge_nonlocal_Z_distribution} together with \Cref{lem:uniform_formula} predict that as long as $n_A-k_A$ and $n_B-k_B$ are both reasonably large, $\Pr(I(A;B)=0)\approx 0.42$, $\Pr(I(A;B)=1)\approx 0.42$, $\Pr(I(A;B)=2)\approx 0.14$, and so on for $\ket{\psi}\sim\mathcal{D}_\rho$ with $\rho=\rho_A\otimes\rho_B$.
	
	We now have sufficient tools to explain the claim made in the main text, namely that the agreement between the distribution of CMI in the output of SRQCs above the critical depth is a hallmark of the emergence of Scrooge-randomness. To understand why, consider a different tripartition $\bar{A}\bar{B}\bar{C}$, where (referring to Fig.~\ref{fig:distribution_of_r} (a)) we define $\bar{A}=A\cup C$, choose $\bar{C}$ to consist of the $\sqrt{n}/3$ columns of qubits in the center of the grid, and $\bar{B}$ is all the other qubits (which shield $\bar{A}$ from $\bar{C}$). Since $U$ has constant-depth and the distance between $A$ and $C$ is increasing with $n$, the reduced state $\rho_{\bar{A}}=\rho_A\otimes\rho_C$ is a product state. Note that the stabilizer Scrooge hypothesis asserts that the ensemble $\mathcal{E}_{\bar{A}}(U_{\bar{A}\bar{B}},x_{\bar{B}})$ is approximated by the stabilizer Scrooge distribution with respect to $\rho_{\bar{A}}$. Since $\bar{A}$ and $\bar{B}$ have boundaries of size $\sqrt{n}$, the Scrooge hypothesis asserts that these states should have $\infty$-entropy $S_\infty(\rho_{\bar{A}}) = \Theta(\sqrt{n})$. Therefore, the distribution observed in the middle column of Fig.~\ref{fig:distribution_of_r}(b) serves as a signature of the Scrooge ensemble.
	
	\subsection{Numerical investigation of the output distribution of shallow 2D random Clifford circuits} \label{sec:Z_type_stablizer_numerics}
	
\begin{figure}[t]
		\centering
		\includegraphics[width=0.9\linewidth]{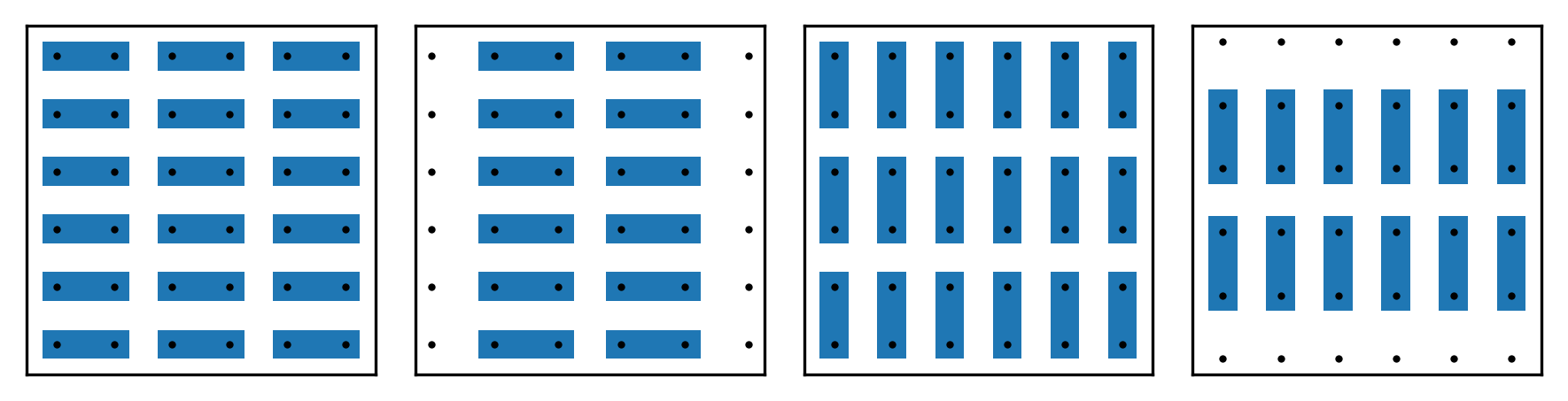}
		\caption{Gate patterns of the 2D brickwork architecture on a $6\times 6$ grid. Each black dot represents a qubit, and each blue rectangle represents a $2$-qubit gate acting on the qubits covered by the rectangle. The four panels from left to right depict the placement of $2$-qubit gates in layers one through four. Circuit layer five cycles back to use the pattern of the first layer (left most panel), and so on.}
		\label{fig:canonical_2D_brickwork_circuit}
	\end{figure}

	In this appendix, we investigate numerically a key question of this work: what does the output distribution of a typical constant-depth 2D brickwork random Clifford circuit look like? We consider a canonical 2D brickwork architecture where the gate patterns repeat in periods of four, e.g., the placement of the gates in the fifth layer cycles back to be the same as that of the first layer; see \Cref{fig:canonical_2D_brickwork_circuit} for an example on a $6\times 6$ grid of qubits. In this work, all numerical simulations of Clifford circuits are performed using Stim \cite{gidney2021stim}.
	
	We will focus our attention on depth-$7$ 2D brickwork random Clifford circuits because it has been observed in previous numerical investigations that w.r.t. this specific architecture, depth-$7$ is right after the conjectured complexity phase transition has taken place \cite{bene2025quantum}; see also \cite{mcginley2025measurement}.
	
	We first investigate the output distributions of typical 2D brickwork random Clifford circuits when all the qubits are measured in the computational basis. In the Clifford setting, this amounts to looking at the $Z$-type stabilizer subgroup of the output state since the output distribution of a stabilizer state is always the uniform distribution over an affine subspace of $\{0,1\}^n$ determined by its $Z$-type stabilizers. More concretely, let $\ket{\psi}$ be an $n$-qubit stabilizer state with $Z$-type stabilizer subgroup $S^Z_\psi$ and consider its computational-basis output distribution $q(x)=|\langle x|\psi\rangle|^2, x\in\{0,1\}^n$. Then $q$ will always be the uniform distribution over the set
	\begin{equation}
		\{x\in\{0,1\}^n:P\ket{x}=\ket{x}\text{ for every $P\in S^Z_\psi$}\}.
	\end{equation}
	Each $P\in S^Z_\psi$ imposes a linear constraint on a subset of the bits that a possible outcome string must satisfy. Equivalently, each $P\in S^Z_\psi$ (of weight at least $2$) enforces some correlation among the bits of an outcome string. 
	
\begin{figure}[t]
		\centering
		\includegraphics[width=0.9\linewidth]{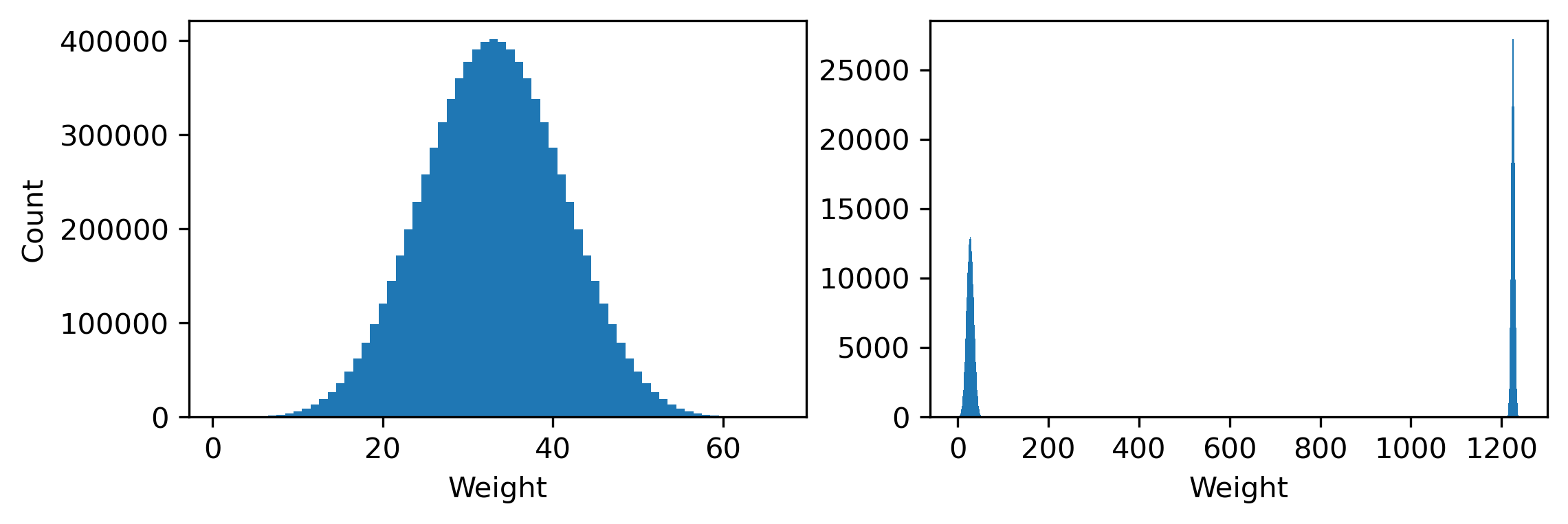}
		\caption{Examples of typical weight distributions of the $Z$-type stabilizers of the output states of depth-$7$ 2D brickwork random Clifford circuits acting on a $50\times 50$ grid of qubits. The example on the left panel shows a unimodal histogram over a total of $2^{23}-1$ non-identity $Z$-type stabilizers centered around a weight of $33$. The right panel shows a sharply different bimodal histogram over a total of $2^{19}-1$ non-identity $Z$-type stabilizers, with a low-weight peak centered around $28$ and a far-apart high-weight peak centered around $1225$. In both cases, the total number of non-identity $Z$-type stabilizers, $2^{23}-1$ and $2^{19}-1$ respectively, far exceed the average value of roughly $1$ expected in the anti-concentration regime.}
		\label{fig:all_Z_type_histogram}
	\end{figure}

	What do the $Z$-type stabilizers of the output state of a typical depth-$7$ 2D brickwork random Clifford circuit look like? We know that on average, an output state should have at least $\exp(\Omega(n))$ $Z$-type stabilizers in total \cite{dalzell2022random} where the constant in $\Omega(\cdot)$ scales inverse exponentially with the circuit depth. In contrast, when the circuit depth reaches the anti-concentration regime, which is believed to occur at $\Omega(\log n)$ depth for our specific architecture, the expected total number of non-identity $Z$-type stabilizers converges to approximately $1$ \cite{dalzell2022random}. Recall that on average, an $n$-qubit uniformly random stabilizer state has roughly one non-identity $Z$-type stabilizer, whose weight is concentrated around $\frac{n}{2}$. We numerically simulated depth-$7$ 2D brickwork random Clifford circuits acting on a $50\times 50$ grid of qubits and solved for the $Z$-type stabilizer subgroups of the output states. We found that the most common $Z$-type stabilizer weight distributions/histograms follow the shapes exemplified in \Cref{fig:all_Z_type_histogram}. 
	
	In the case depicted in the right panel of \Cref{fig:all_Z_type_histogram}, it is possible to find a set of independent $Z$-type stabilizer generators for the $Z$-type stabilizer subgroup consisting of $18$ low-weight generators, whose supports are confined to small-sized local regions, and a single high-weight generator, whose support spans the entire grid. In this particular case, due to the clear gap between the two peaks in the weight histogram, we can be certain that the sole high-weight generator cannot be obtained as a product involving only low-weight $Z$-type stabilizers. We say that this high-weight generator is indecomposable, in the sense that it cannot be decomposed into a product of a large number of $Z$-type stabilizers, each supported on a local region. An indecomposable $Z$-type stabilizer embeds an indecomposable long-range correlation into the output distribution, serving as a signal of global randomness in the output state. Note that a $Z$-type stabilizer being high-weight does not mean that it must be indecomposable; for example, $Z^{\otimes n}$ is a maximal weight $Z$-type stabilizer of the state $\ket{0^n}$, but $Z^{\otimes n}$ is clearly decomposable as $Z^{\otimes n}=\prod_{j=1}^n Z_j$. A high-weight $Z$-type stabilizer being present in a particular generating set also does not mean that it is indecomposable, as we could extend $\{Z^{\otimes n}\}$ to a generating set for the stabilizer group of $\ket{0^n}$, which will include the high-weight but decomposable $Z^{\otimes n}$ as a generator by construction.
	
	Motivated by our findings so far, we move on to investigate the distribution of the number of indecomposable $Z$-type stabilizers of the output state $\ket{\psi}$ of a constant-depth 2D brickwork random Clifford circuit. To make this question well-defined, we consider a tripartition of the $\sqrt{n}\times\sqrt{n}$ grid where we take $A$ to be the top row of qubits, $C$ to be the bottom row of qubits, $B$ to be all the other qubits, and consider a rather stringent notion of decomposability. Let $S_\psi^Z$ denote the $Z$-type stabilizer subgroup of $\ket{\psi}$, and for every $D\subseteq[n]$, let $S^Z_{\psi_{D}}$ denote the subgroup of $S_\psi^Z$ local on $D$. We call $P\in S^Z_\psi$ indecomposable if $P\neq QR$ for every $Q\in S^Z_{\psi_{AB}}$ and $R\in S^Z_{\psi_{BC}}$ and decomposable otherwise. This operational notion of decomposability is captured precisely by a quotient group structure: every indecomposable $P\in S^Z_\psi$ belongs to a non-identity element of the quotient group $S^Z_\psi/(S^Z_{\psi_{AB}}S^Z_{\psi_{BC}})$. As a result, the number of indecomposable $Z$-type stabilizer generators is given by
	\begin{align}
		\log_2\left(\frac{|S^Z_\psi|}{|S^Z_{\psi_{AB}}S^Z_{\psi_{BC}}|}\right)&=\log_2(|S^Z_\psi|)-\log_2(|S^Z_{\psi_{AB}}|)-\log_2(|S^Z_{\psi_{BC}}|)+\log_2(|S^Z_{\psi_{B}}|)\\
		&=(|AB|-\log_2(|S^Z_{\psi_{AB}}|))+(|BC|-\log_2(|S^Z_{\psi_{BC}}|))-(n-\log_2(|S^Z_\psi|))-(|B|-\log_2(|S^Z_{\psi_{B}}|))\\
		&=H(AB)+H(BC)-H(ABC)-H(B)\\
		&=H(A|B)+H(C|B)-H(AC|B)\\
		&=I(A;C|B)
	\end{align}
	where $H$ denotes the Shannon entropy, and $I(A;C|B)$ is the conditional mutual information between $A$ and $C$ given $B$ in the computational-basis output distribution of $\ket{\psi}$. 
	
	In the following, we numerically simulate depth-$7$ 2D brickwork random Clifford circuits acting on a $100\times 100$ grid of qubits and compute the value of $I(A;C|B)$ for each circuit instance. Since
	\begin{align}
		I(A;C|B)&=\log_2(|S^Z_\psi|)-\log_2(|S^Z_{\psi_{AB}}|)-\log_2(|S^Z_{\psi_{BC}}|)+\log_2(|S^Z_{\psi_{B}}|)\\
		&=\log_2(|S^Z_\psi/S^Z_{\psi_B}|)-\log_2(|S^Z_{\psi_{AB}}/S^Z_{\psi_B}|)-\log_2(|S^Z_{\psi_{BC}}/S^Z_{\psi_B}|),
	\end{align}
	we can compute $I(A;C|B)$ using the following procedure. For each circuit instance $U$, we first compute its output state $\ket{\psi}=U\ket{0^n}$ and then measure all the qubits of $\ket{\psi}$ in $B$ in the computational basis to get a pure post-measurement state $\ket{\psi'}$ on $A$ and $C$. Next, we solve for a set of independent generators $G^Z$ for the $Z$-type stabilizer subgroup of $\ket{\psi'}$ and sets of independent generators $G_A^Z$ and $G_C^Z$ for the $Z$-type stabilizer subgroups of $\ket{\psi'}$ local on $A$ and $C$, respectively. Then we have $I(A;C|B)=|G^Z|-|G_A^Z|-|G_C^Z|$, which can also be interpreted as the number of independent non-local $Z$-type stabilizer generators of $\ket{\psi'}$. \Cref{fig:distribution_of_r} in the main text shows the results from numerically estimating the distribution of the integer $r=I(A;C|B)$ using $10^5$ random circuit samples for depth-$5$ and depth-$7$ 2D brickwork random Clifford circuits. The full data including the data for depth-$6$ are recorded in \Cref{tab:distribution_of_r_full}.
	
	\begin{table}[t]
		\centering
		\begin{tabular}{ |c|c|c|c|c| }
			\hline
			$r$ & $\mathrm{Pr}[I(A;C|B)=r]$ & $\mathrm{Pr}[I(A;C|B)=r]$ & $\mathrm{Pr}[I(A;C|B)=r]$ & $\mathrm{Pr}[I(A;C|B)=r]$\\ 
			&   2D  Clifford & 2D Clifford & 2D Clifford & random stab. state\\
			& depth-$5$ & depth-$6$ & depth-$7$ & as $n\rightarrow \infty$\\
			\hline
			0 & 0.99999 & 0.46207 &  0.41815 & 0.41942 \\ 
			\hline
			1 & 0.00001 & 0.41336 & 0.42077 & 0.41942 \\
			\hline
			2 & 0 & 0.11267 & 0.13952 & 0.13981 \\
			\hline
			3 & 0 & 0.01141 & 0.02022 & 0.01997 \\
			\hline
			4 & 0 & 0.00049 & 0.00131 & 0.00133 \\
			\hline
			5 & 0 & 0 & 0.00003 & 0.00004 \\
			\hline
		\end{tabular}
		\caption{Numerical estimation of the distribution of the integer $r=I(A;C|B)$ for depth-$d\in\{5,6,7\}$ 2D brickwork random Clifford circuits acting on a $100\times 100$ grid of qubits using $10^5$ random circuit samples. The columns for $d\in\{5,7\}$ and uniformly random stabilizer states are reproduced from the main text for ease of comparison. From the data for depth-$6$, we see that while $I(A;C|B)>0$ with probability over $0.5$, the estimated distribution is visibly different from that of depth-$7$ and uniformly random stabilizer states.}
		\label{tab:distribution_of_r_full}
	\end{table}
	
	Our findings at $d=6$ are inconclusive. It is possible that there are strong finite-size effects near criticality, and the distribution of $I(A;C|B)$ for $d=6$ will eventually converge to either the distribution for $d=5$ or $d=7$ as $n\rightarrow\infty$. However, given the data, it is also possible that as $n\rightarrow\infty$, the distributions of $I(A;C|B)$ are qualitatively distinct for $d=6$ and $d=7$, extensive long-range MIE is present at $d=6$, yet the Scrooge hypothesis is false for $d=6$. In contrast and perhaps surprisingly, at depth-$7$, the estimated distribution of $I(A;C|B)$ turns out to be statistically indistinguishable from the large-$n$ limiting distribution of the number of independent $Z$-type stabilizer generators of a uniformly random stabilizer state. The results in \Cref{tab:distribution_of_r_full} predict that the unimodal and bimodal weight histograms shown in \Cref{fig:all_Z_type_histogram} each occur with probability approximately $0.419$, corresponding to the cases $r=0$ and $r=1$. Our numerical findings strongly suggest that some form of global randomness resides within the output distribution of depth-$7$ 2D brickwork random Clifford circuits.
	
	To better understand the characteristics of the global randomness in the output distribution of depth-$7$ 2D brickwork random Clifford circuits predicted by our previous numerics, we next look at the marginal output distributions of those circuits. Recall that for an $n$-qubit uniformly random stabilizer state $\ket{\phi}$, the computational-basis marginal output distribution of $\ket{\phi}$ on a subset $A$ of $m=n-\Omega(\log n)$ qubits will be the uniform distribution over $\{0,1\}^m$ with high probability. We can again understand this intuitively through the $Z$-type stabilizers $S^Z_\phi$ of $\ket{\phi}$. The marginal output distribution is dictated by the $Z$-type stabilizer subgroup of the reduced state $\Tr_{[n]\setminus A}(\ket{\phi}\bra{\phi})$, which is in bijection with $S^Z_{\phi_A}$, the subgroup of $S^Z_\phi$ local on $A$. Recall that $\ket{\phi}$ typically has at most a few high-weight $Z$-type stabilizers, so that with high probability, all have non-trivial support on the $\Omega(\log n)$-sized subset $[n]\setminus A$. Equivalently, with high probability, $S^Z_{\phi_A}=\emptyset$, so the marginal output distribution will be uniform. 
	
	\begin{figure}[t]
		\centering
		\includegraphics[width=0.9\linewidth]{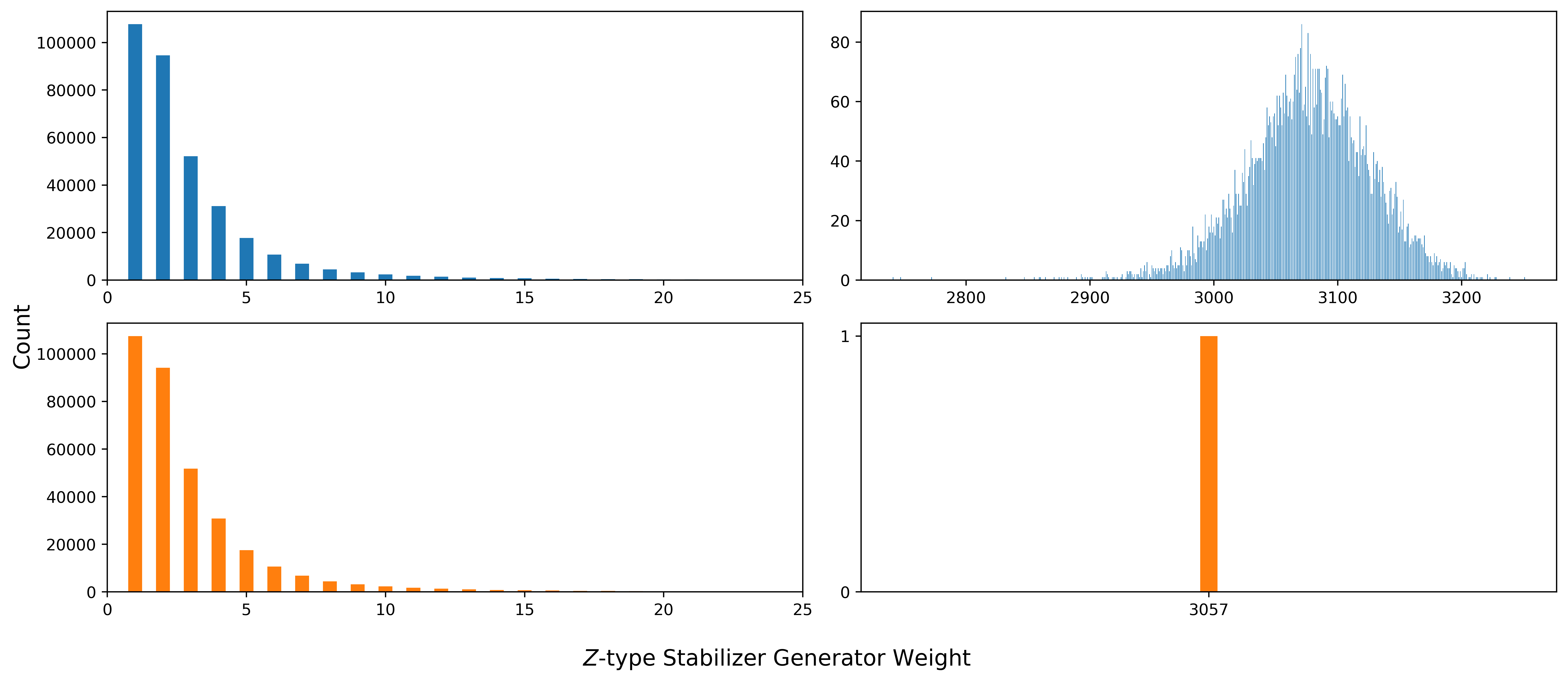}
		\caption{Weight histograms of the $Z$-type stabilizer generators of the full output states (top left and top right panels) versus reduced output states (bottom left and bottom right panels) of $(80\times 80)$-qubit depth-$7$ 2D brickwork random Clifford circuits aggregated over $10^4$ circuit instances. The top left and right panels show that a full output state tends to admit a $Z$-type stabilizer subgroup generating set containing many low-weight (at most $166$) generators and perhaps a few high-weight (around $3000$) generators. There are $1603$ generators with weights between $26$ and $166$ not shown in the top left panel. In our experiment, after tracing out $15$ random qubits, the reduced output state admits a $Z$-type stabilizer subgroup generating set with only low-weight (at most $166$) generators in all but one out of the $10^4$ circuit instances for which the solved generating set contained an element of weight $3057$. There are $1485$ generators with weights between 26 and 166 not shown in the bottom left panel.}
		\label{fig:marginal_Z_stabilizer}
	\end{figure}
	
	We do not expect the marginal output distribution of a depth-$7$ 2D brickwork random Clifford circuit to be uniform on a large region $A$ of size $n-\Omega(\log n)$ because the output state will have many low-weight $Z$-type stabilizers with support contained in $A$. However, what could happen is that the reduced state $\Tr_{[n]\setminus A}(\ket{\psi}\bra{\psi})$ on $A$, with $\ket{\psi}$ being the output state of the circuit, may no longer have any high-weight indecomposable $Z$-type stabilizers, for the same reason that all the indecomposable $Z$-type stabilizers are inevitably supported non-trivially on $[n]\setminus A$. Note that this can be certified by the existence of a set of independent generators for $S^Z_{\psi_A}$ for which every generator has low-weight and is locally supported.
	
	In this experiment, we numerically simulate $(80\times 80)$-qubit depth-$7$ 2D brickwork random Clifford circuits with $10^4$ circuit samples. For each circuit instance $U$ with output state $\ket{\psi}=U\ket{0^n}$, we first solve for a set of independent generators for $S^Z_\psi$, then choose a uniformly random subset $B$ of $15$ qubits, and lastly solve for a set of independent generators for $S^Z_{\psi_A}$ with $A=[n]\setminus B$. \Cref{fig:marginal_Z_stabilizer} plots the aggregated weight histograms of all the generators encountered throughout the $10^4$ circuit samples. As we have explained earlier, while the presence of a high-weight $Z$-type stabilizer in a generating set does not imply that it is necessarily indecomposable, we do suspect that all the high-weight generators encountered in this experiment are actually indecomposable due to the significant weight difference between the low-weight and high-weight generators. In contrast, in $9999$ out of the $10^4$ instances, after tracing out $|B|=15$ random qubits, $S^Z_{\psi_A}$ admits an independent generating set consisting of only low-weight generators, and this fact does imply that every $Z$-type stabilizer of the reduced state is decomposable. Similar to uniformly random stabilizer states, our numerics suggest that any global correlation present in the output distribution of $\ket{\psi}$ ceases to exist in its marginal output distributions even after tracing out just a small subset of qubits.
	
	In summary, the following intuitive picture emerges from our numerical investigation of the output distribution, or equivalently the $Z$-type stabilizer subgroup, of depth-$7$ 2D brickwork random Clifford circuits. From \Cref{fig:all_Z_type_histogram}, we infer that the $Z$-type stabilizer subgroup $S^Z_\psi$ of the output state of a typical circuit instance $\ket{\psi}=U\ket{0^n}$ admits an independent generating set consisting of a large number of low-weight local $Z$-type generators and perhaps a few high-weight non-local generators. From \Cref{tab:distribution_of_r_full}, we infer that the distribution of the number of indecomposable $Z$-type stabilizer generators closely matches the distribution of the number of $Z$-stabilizer generators of a uniformly random stabilizer state. From \Cref{fig:marginal_Z_stabilizer}, we infer that just like for uniformly random stabilizer states, any global correlation present in the output distribution of $\ket{\psi}$ is fragile against depolarizing noise; the $Z$-type stabilizer subgroup of the reduced state of $\ket{\psi}$ admits a generating set composed solely of low-weight local $Z$-type stabilizers after tracing out only a small number of qubits. 
	
	\FloatBarrier
	
	\subsection{Structure of the output distribution of shallow 2D random Clifford circuits}
	\label{sec:structure_of_clifford}
	
	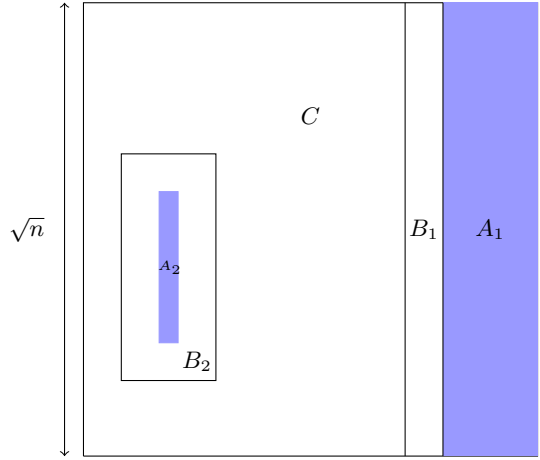
\begin{figure}[ht]
		\centering
		\begin{tikzpicture}
			\draw (0,0)--(6,0)--(6,6)--(0,6)--(0,0);
			\draw[<->] (-0.25,0)--(-0.25,6);
			\draw node at (-0.75,3) {$\sqrt{n}$};
			
			\filldraw[blue!40!white] (1,1.5) rectangle (1.25,3.5);
			\draw node at (1.14,2.5) {\tiny{$A_2$}};
			
			\draw (0.5,1) rectangle (1.75,4);
			\draw node at (1.5,1.25)  {$B_2$};
			
			\fill[blue!40!white] (4.75,0) rectangle (6,6);
			\draw (4.75,0)--(4.75,6);
			\draw node at (5.375,3)  {$A_1$};
			
			\draw (4.25,0)--(4.25,6);
			\draw node at (4.5,3)  {$B_1$};
			
			\draw node at (3,4.5)  {$C$};
		\end{tikzpicture}
		\caption{An example of an $ABC$ tripartition in the set $\mathcal{T}$. In this example, the shaded $A_1$ and $A_2$ regions come from $A_1\in \mathcal{K}_R$ and $A_2\in\mathcal{H}_L$. We have $|B_1|=\sqrt{n}\times b\log_2 n$, $|A_2|=a\log_2 n\times 1$, and $|B_2|=O(\log n)\times O(\log n)$ for some constants $a$ and $b$. Notice that $A_1$ and $A_2$ are always lightcone separated.}
		\label{fig:structral_lemma_regions}
	\end{figure}
	
	In this appendix, we show how we can use our main Scrooge Hypothesis to deduce detailed structural properties of the output distribution of shallow 2D random Clifford circuits that are fully consistent with what we observed numerically in \Cref{sec:Z_type_stablizer_numerics}. As such, the Scrooge Hypothesis provides an explanation for why we observed what we did in \Cref{sec:Z_type_stablizer_numerics}. The results in this appendix, together with \Cref{sec:Z_type_stablizer_numerics}, could be viewed as further evidence for the Scrooge Hypothesis in the stabilizer setting. For every integer $n\geq 1$, we define $[n]=\{1,\ldots,n\}$. 
	
	We consider depth-$d$ 2D brickwork random Clifford circuits acting on a $\sqrt{n}\times\sqrt{n}$ grid of qubits with $d>d^*$, and we assume the Scrooge Hypothesis holds. We will label the columns of the grid $1$ through $\sqrt{n}$ from left to right. In the following, whenever we refer to a rectangular region of $y\times x$ qubits, $y$ denotes the vertical side length, and $x$ denotes the horizontal side length. Let $L$ be the set of qubits in columns $1$ through $\sqrt{n}/2$, and let $R$ be the set of qubits in columns $\sqrt{n}/2+1$ through $\sqrt{n}$, so $L$ and $R$ partition the grid into left and right halves. Let $a\geq 1$ be a constant to be chosen later. Let $\mathcal{H}_L$ and $\mathcal{H}_R$ denote the set of all rectangular regions of $a\log_2 n\times 1$ qubits in $L$ and $R$ respectively, so each region is a thin rectangular strip of height $a\log_2 n$ and width $1$. Note that $|\mathcal{H}_L|=|\mathcal{H}_R|\leq n/2$. Then, for every $j=1,\ldots,\sqrt{n}/2$, let $K_j$ denote the set of all qubits in columns $1$ through $j$, and define $\mathcal{K}_L=\{K_j:j\in[\sqrt{n}/2]\}$. Similarly, for every $j=\sqrt{n}/2+1,\ldots,\sqrt{n}$, let $K_j$ denote the set of all qubits in columns $j$ through $\sqrt{n}$, and define $\mathcal{K}_R=\{K_j:j\in\{\sqrt{n}/2+1,\ldots,\sqrt{n}\}\}$. Note that $|\mathcal{K}_L|=|\mathcal{K}_R|=\sqrt{n}/2$. Next, we define a set of $ABC$ tripartitions of the $\sqrt{n}\times\sqrt{n}$ grid, with each tripartition being specified by an element in the set
	\begin{equation}
		\mathcal{T}=\{(A_1,A_2)\in\mathcal{K}_L\times\mathcal{H}_R:\text{dist}(A_1,A_2)\geq \sqrt{n}/3\}\cup\{(A_1,A_2)\in\mathcal{K}_R\times\mathcal{H}_L:\text{dist}(A_1,A_2)\geq \sqrt{n}/3\}.
	\end{equation}
	Let $b\geq 1$ be another constant to be chosen later. Each $(A_1,A_2)\in\mathcal{T}$ specifies an $ABC$ tripartition where $A=A_1\cup A_2$ and $B=B_1\cup B_2$ is the disjoint union of two shielding regions separating $A$ from $C$, the rest of the qubits, so that $\text{dist}(A,C)=b\log_2 n+1$; see \Cref{fig:structral_lemma_regions} for an example with $(A_1,A_2)\in\mathcal{K}_R\times\mathcal{H}_L$. Note that for every $ABC$ tripartition in $\mathcal{T}$, the $C$ region contains a square region of qubits of size at least $\sqrt{n}/4\times\sqrt{n}/4$, and that $|\mathcal{T}|\leq n^{1.5}$. 
	
Let $\mathcal{U}$ denote the distribution of depth-$d$ 2D brickwork random Clifford circuits with $d>d^*$. Next, we invoke the Scrooge hypothesis with $t=2$ for a fixed $U\sim\mathcal{U}$ on every $ABC$ tripartition in $\mathcal{T}$ twice in the following way. Let $c\geq 2$ be a constant. Let $ABC$ with $A=A_1\cup A_2$ and $B=B_1\cup B_2$ be a tripartition specified by an element in $\mathcal{T}$. We choose the constant $b$ to be large enough so that the Scrooge hypothesis holds for the $ABC$ tripartition with $\varepsilon\leq 1/n^{c}$. If we were to apply the circuit, measure all qubits in $B$, and trace out all qubits in $C$, we would obtain a background state $\rho=\rho_{A_1}\otimes\rho_{A_2}$ since $A_1$ and $A_2$ are lightcone separated. Depending on the measurement outcome $x_B$, $\rho$ belongs to an equivalence class $[\rho]$ with $\rho$ being a representative where all $\rho'\in[\rho]$ have the same stabilizers as $\rho$ modulo the $\pm$ signs. Next, we invoke \Cref{cor:scrooge_2_design_no_new_local_Z_type} on $[\rho_{A_1}\otimes\rho_{A_2}]$ with the single observable, so the event in \eqref{eq:scrooge_2_design_no_new_local_Z_type_condition} occurs with probability $1-\varepsilon-(1+\varepsilon)/2^{n_{A_2}-k_{A_2}}$ for all $\rho'\in[\rho_{A_1}\otimes\rho_{A_2}]$. To lower bound $n_{A_2}-k_{A_2}$, we choose the constant $a$ to be large enough so that the Scrooge hypothesis holds for another tripartition $\bar{A}=A_2$, $\bar{B}=B_2$, and $\bar{C}=A_1\cup B_1\cup C$ with $\varepsilon\leq 1/n^{c}$ and $S_{\infty}(\rho'_{A_2})=n_{A_2}-k_{A_2}\geq c\log_2 n$ for all $\rho'_{A_2}\in[\rho_{A_2}]$. With this, we get
\begin{equation}
1-\varepsilon-\frac{1+\varepsilon}{2^{n_{A_2}-k_{A_2}}}\geq1-\frac{3}{n^c}.
\end{equation}
By a union bound, all of the above events occur simultaneously with probability at least $1-5n^{1.5}/n^{c}$, which is $1-1/\mathrm{poly}(n)$. We call a $U\sim\mathcal{U}$ Good if all the above events occur simultaneously. We have established that we draw a Good circuit with probability at least $1-1/\text{poly(n)}$.
	
	In the following, let $\ket{\psi}$ be the output state of a Good constant-depth 2D random Clifford circuit. Let $P$ be a non-identity $Z$-type stabilizer of $\ket{\psi}$. Let $j$ and $k$ be the indices of the leftmost and rightmost $\sqrt{n}\times 1$ columns of qubits on which $P$ has non-trivial support. We define the column width of $P$ to be $cw(P)=k-j+1$. We say $P$ is column-wise decomposable if $P$ can be written as a product
	\begin{equation}
		P=\prod_{j=1}^r Q_j
	\end{equation}
	for some $r\geq 1$ and every $Q_j$ is a $Z$-type stabilizer of $\ket{\psi}$ that has column width at most $b\log_2 n+1$. For every $P=P_1\otimes\cdots\otimes P_n$ and $R\subseteq[n]$, we define the truncation of $P$ to $R$ to be the $|R|$-qubit Pauli operator $P_R=\bigotimes_{j\in R}P_j$. We say $P$ has a hole $H\subseteq[n]$ if $H$ has dimension $a\log_2(n)\times 1$ and $P$ is all identity on $H$, or equivalently, $P_H=I^{\otimes |H|}$.
	
	\begin{theorem}
		Let $\ket{\psi}$ be the output state of a Good constant-depth 2D random Clifford circuit $U$. For every non-identity $Z$-type stabilizer $P$ of $\ket{\psi}$, if $P$ has a hole, then $P$ is column-wise decomposable. That is, $P$ can be written as a product
		\begin{equation}
			P=\prod_{j=1}^r Q_j
		\end{equation}
		where $r\leq cw(P)$ and every $Q_j$ is a $Z$-type stabilizer of $\ket{\psi}$ with column width at most $b\log_2 n+1$. 
		\label{thm:column-wise_decomposable}
	\end{theorem}
	\begin{proof}
Let $S^Z$ be the set of all non-identity $Z$-type stabilizers of $\ket{\psi}$. We proceed by induction on the column width. The base case includes all $P\in S^Z$ that have $cw(P)\leq b\log_2 n+1$ for which the claim holds trivially. For the inductive step, let $P\in S^Z$ such that $P$ has a hole and $cw(P)>b\log_2 n+1$. Let $j$ and $k$ be the indices of the leftmost and rightmost columns of qubits that $P$ has non-trivial support on. Let $H_k$ and $H_j$ be any holes of $P$ that have the largest column-wise distance to $j$ and $k$ respectively. Without loss of generality, assume $H_j$ and $k$ are the more distant pair. Note that the column-wise distance between $H_j$ and $k$ is necessarily at least $\sqrt{n}/2-2$. Define $A_1$ to be the union of the columns $k$ through $\sqrt{n}$. Define $A_2=H_j$. Let $R$ denote the $k$-th column of qubits. First, notice that since $P$ is a $Z$-type stabilizer of $\ket{\psi}$ and $P_{A_2}=I^{\otimes|A_2|}$, if we were to measure all qubits of $\ket{\psi}$ in $[n]\setminus(A_1\cup A_2)$ in the computational basis to get a pure post-measurement state $\ket{\psi'}$, $P_R$ or $-P_R$ will be a $Z$-type stabilizer of $\ket{\psi'}$ that is local on $A_1$. Our choice of $(A_1,A_2)$ corresponds to an $ABC$ tripartition contained in $\mathcal{T}$ with a corresponding region $B=B_1\cup B_2$. Let $W$ denote the backward lightcone circuit of $A_1\cup B_1$, and let $[\rho_{A_1}\otimes\rho_{A_2}]$ denote the equivalence class of background states sampled when invoking the Scrooge hypothesis on this $ABC$ tripartition. Since $U$ is Good, $P_R$ or $-P_R$ being a $Z$-type stabilizer of $\ket{\psi'}$ local on $A_1$ implies that $P_R$ or $-P_R$ is a $Z$-type stabilizer of $\rho_{A_1}$. By lightcone cancellation, this implies that $(W\otimes I)\ket{0^n}$ has a $Z$-type stabilizer $Q$ supported on $R\cup B_1$ such that $Q_R=P_R$, and $Q$ is also a $Z$-type stabilizer of $\ket{\psi}$. Define $Q'=PQ$. Then we have $P=QQ'$ satisfying $cw(Q)\leq b\log_2 n+1$, $Q'\in S^Z$, $Q'$ having a hole, and $cw(Q')\leq cw(P)-1$. Notice that $cw(P)>b\log_2 n+1$ implies $P\neq Q$, which implies $Q'\neq I^{\otimes n}$. Thus, the claim follows by applying the inductive hypothesis to $Q'$. 
	\end{proof}
	
	As a sanity check, note that the converse of \Cref{thm:column-wise_decomposable} is not true. Clearly, $P$ could still be column-wise decomposable even if $P$ does not have a hole. The contrapositive of \Cref{thm:column-wise_decomposable} implies that every $Z$-type stabilizer of $\ket{\psi}$ that is not column-wise decomposable cannot have any holes, so the weight of a column-wise indecomposable $Z$-type stabilizer needs to be at least $\frac{n}{a\log_2(n)}$.
	
	\begin{corollary}
		Let $\ket{\psi}$ be the output state of a Good constant-depth 2D random Clifford circuit. The $Z$-type stabilizer subgroup of $\ket{\psi}$ admits a set of independent generators $G$ such that for every $P\in G$, either $cw(P)\leq b\log_2 n+1$ or $P$ is not column-wise decomposable and has no hole.
	\end{corollary}
	\begin{proof}
		Let $S^Z$ be the set of all non-identity $Z$-type stabilizers of $\ket{\psi}$. Consider constructing a set of independent generators for $S^Z$ using the following procedure. We sort every element in $S^Z$ in increasing order of column width with arbitrary tie breaking, so that $S^Z=\{P_1,P_2,\ldots,P_{|S^Z|}\}$ with $cw(P_1)\leq cw(P_2)\leq\cdots\leq cw(P_{|S^Z|})$. Define $G_0=\emptyset$. We then scan through every element in $S^Z$. For every $j\in[|S^Z|]$, for the $j$-th iteration , we set $G_j=G_{j-1}$ if $P_j\in\langle G_{j-1}\rangle$ and $G_{j}=G_{j-1}\cup \{P_j\}$ otherwise. By construction, $G_{|S^Z|}$ is a set of independent generators for $S^Z$. Let $k$ be the cutoff such that $cw(P_j)\leq b\log_2 n+1$ if and only if $j\leq k$. Suppose for some $j>k$, $P_j$ is added to $G_{j-1}$ in the $j$-th iteration. This implies $P_j\notin\langle G_{j-1}\rangle$, which implies $P_j$ is not column-wise decomposable. If $P_j$ has a hole, by \Cref{thm:column-wise_decomposable}, $P_j$ would be column-wise decomposable. Therefore, $P_j$ cannot have a hole. 
	\end{proof}
	
	At this point, we would like to apply the Scrooge Hypothesis three more times with $A$ being the leftmost column, the rightmost column, and the union of the leftmost and rightmost columns together with \Cref{thm:scrooge_nonlocal_Z_distribution} to argue that the distribution of the number of column-wise indecomposable $Z$-type stabilizer generators converges to the distribution characterized in \Cref{lem:uniform_formula} as $n\rightarrow\infty$. Unfortunately, we do not know how to prove a generalization of \Cref{thm:scrooge_nonlocal_Z_distribution} for $\varepsilon$-approximate Scrooge $k$-designs, and we do not get exact Scrooge from the Scrooge Hypothesis. Intuitively speaking, since the distribution in \Cref{lem:uniform_formula} effectively has a sample space of size $6$, and every discrete probability distribution over a sample space of size $k$ is completely characterized by its first $k-1$ moments, a generalization requiring only the guarantees of an $\varepsilon$-approximate Scrooge $k$-design seems plausible.
	
	If we trace out from $\ket{\psi}$ any subset of qubits $H\subseteq[n]$ that contains a hole of dimension $a\log_2(n)\times 1$, then every $Z$-type stabilizer of $\ket{\psi}$ that does not have any hole is deleted, which includes every $Z$-type stabilizer of $\ket{\psi}$ that is not column-wise decomposable. Therefore, every $Z$-type stabilizer of $\Tr_{H}(\ket{\psi}\bra{\psi})$ is column-wise decomposable in terms of the $Z$-type stabilizers of $\ket{\psi}$. However, it may not be generically true that a $Z$-type stabilizer of $\Tr_{H}(\ket{\psi}\bra{\psi})$ is column-wise decomposable using only the $Z$-type stabilizers of $\Tr_{H}(\ket{\psi}\bra{\psi})$. To illustrate this subtlety, consider a $3$-qubit example where $S^Z_\psi=\langle ZZI,IZZ\rangle$. We have $ZIZ\in S^Z_\psi$ and $ZIZ$ is decomposable as $ZIZ=(ZZI)(IZZ)$. However, if we trace out the middle qubit, $ZIZ$ is the sole non-identity element in $S^Z_\psi$ that is not deleted, so it becomes indecomposable. Nevertheless, we show below that in our case, all the $Z$-type stabilizers of the reduced state are column-wise decomposable in terms of the $Z$-type stabilizers of the reduced state in a slightly relaxed sense.
	
	\begin{corollary}
		Let $\ket{\psi}$ be the output state of a Good constant-depth 2D random Clifford circuit. Let $H\subseteq[n]$ be a hole of dimension $a\log_2n\times 1$, and define $\rho=\tr_H(\ket{\psi}\bra{\psi})$. It holds that every non-identity $Z$-type stabilizer $P'$ of $\rho$ can be decomposed as 
		\begin{equation}
			P'=\prod_{j=1}^{r'}Q_j'
		\end{equation}
		where $r'\leq cw(P')$ and each $Q_j'$ is a $Z$-type stabilizer of $\rho$ with column width at most $2b\log_2n+1$.
		\label{corol:reduced_state_decomposable}
	\end{corollary}
	\begin{proof}
Let $P'$ be a non-identity $Z$-type stabilizer of $\rho$. Then $P=P'\otimes I^{\otimes |H|}$ is a $Z$-type stabilizer of $\ket{\psi}$ with a hole. By \Cref{thm:column-wise_decomposable}, let $P=\prod_{j=1}^r Q_j$ be a column-wise decomposition of $P$. Note that there could be $j\in[r]$ such that $Q_j$ is supported non-trivially on $H$. Define $B=\{j\in[r]:(Q_j)_H\neq I^{\otimes |H|}\}$ and consider $S=\prod_{j\in B}Q_j$ and $T=\prod_{j\in [r]\setminus B} Q_j$. Since $P=ST$ and $P_H=T_H=I^{\otimes |H|}$, we have $S_H=I^{\otimes |H|}$. Since $cw(Q_j)\leq b\log_2n+1$ for every $j\in B$, we deduce that $cw(S)\leq 2b\log_2n+1$. Therefore,
\begin{equation}
P'=\tr_H(S)\prod_{j\in[r]\setminus B}\tr_H(Q_j)
\end{equation}
is a decomposition of $P'$ satisfying all the requirements.
	\end{proof}
	
	\FloatBarrier
	
	\subsection{Numerical evidence for the Scrooge hypothesis in shallow 2D random Clifford circuits} \label{sec:scrooge_numerics}
	
	\begin{figure}[t]
		\centering
		\subfloat[Row-by-row]{
			\includegraphics[width=0.25\textwidth]{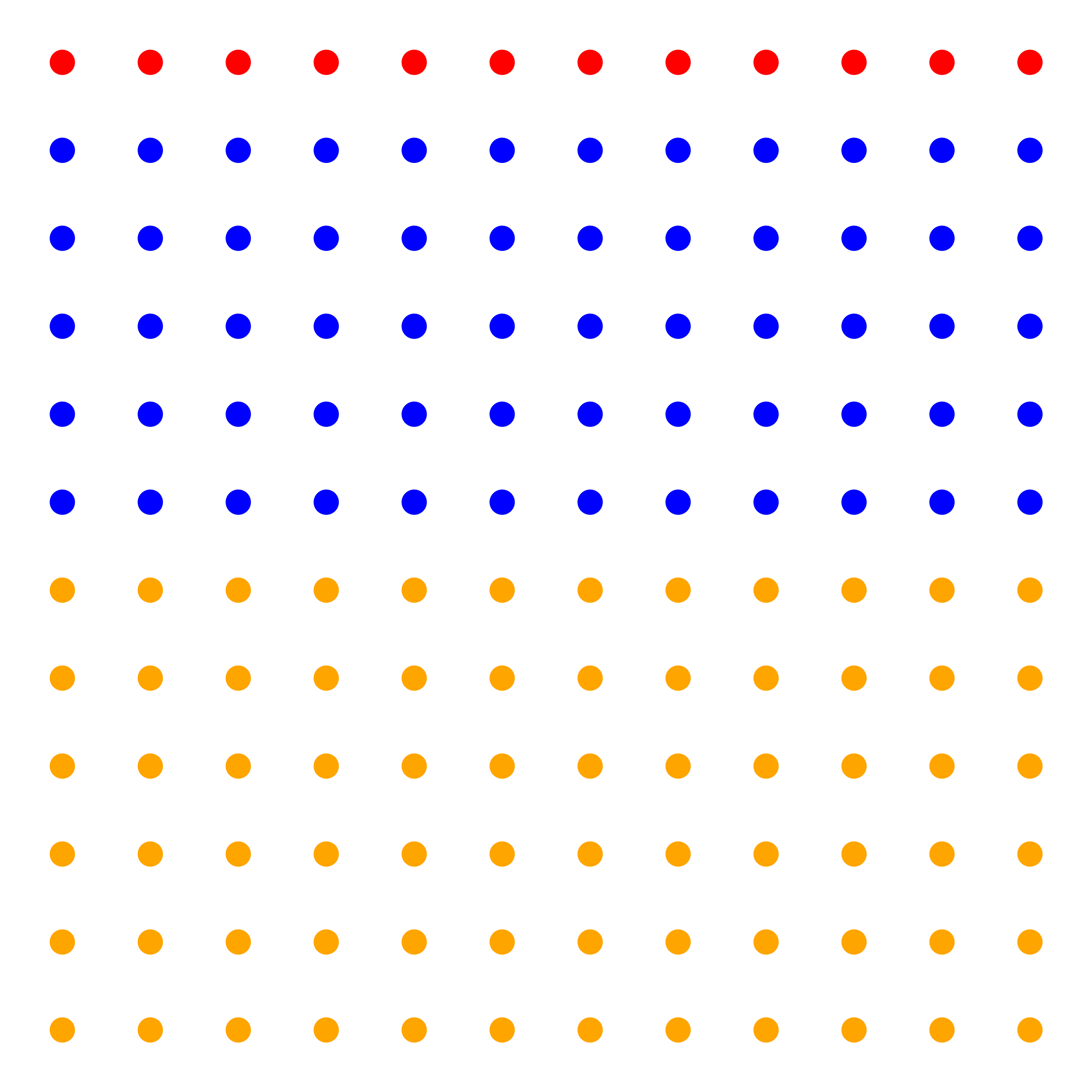}
		}
		\hspace{2cm}
		\subfloat[Row-by-row with two corners]{
		  \includegraphics[width=0.25\textwidth]{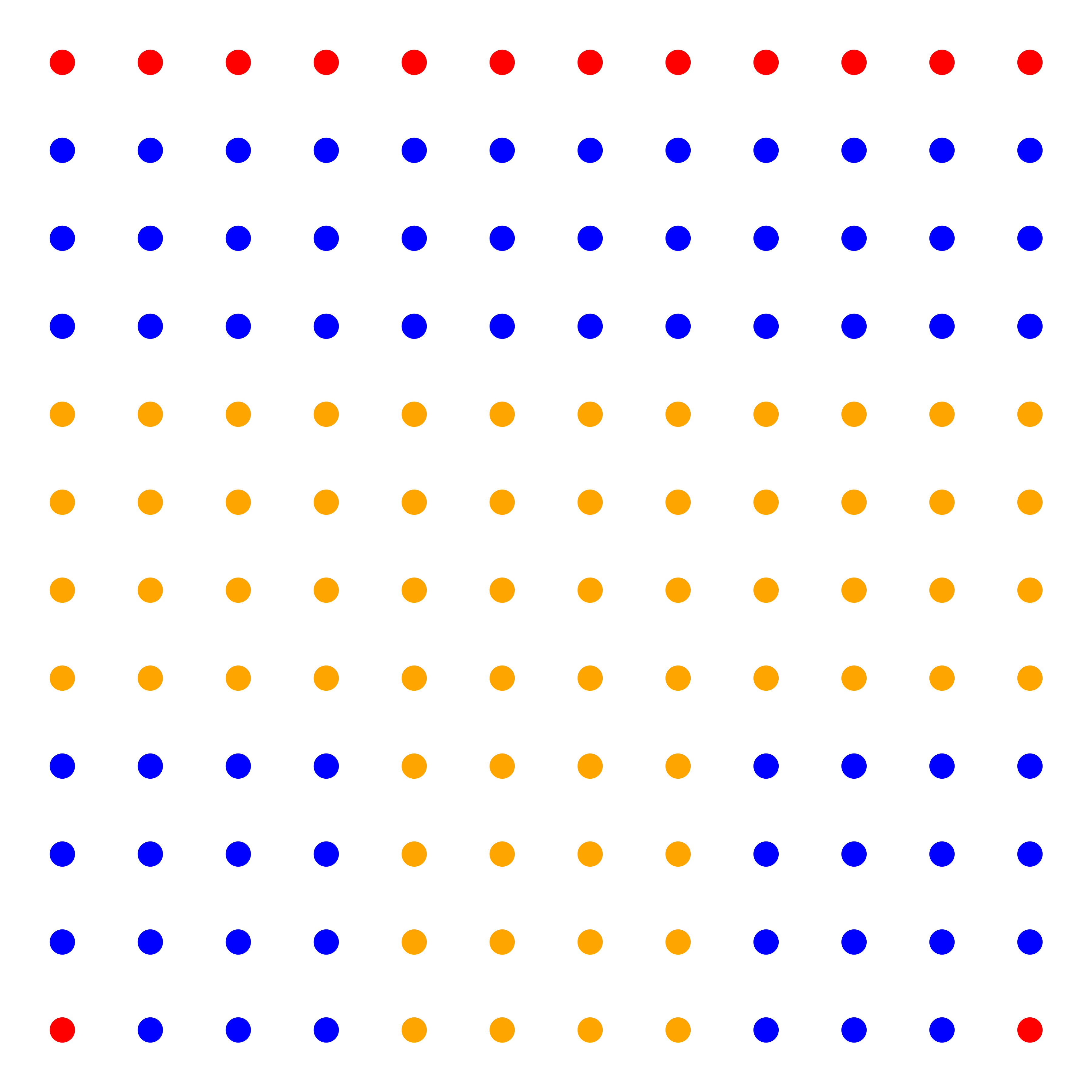}
		}
		\caption{$12\times 12$ examples of the $ABC$ tripartitions considered in our numerical experiments. For both cases, the red, blue, and orange grid points depict the qubits in the regions $A$, $B$, and $C$ respectively. For (a) the row-by-row tripartition, we always have $|A|=L_x$, and for (b) with the two corners, we always have $|A|=L_x+2$. In the example (a), $\text{width}(B)=5$ while in the example (b), $\text{width}(B)=3$. We will always simulate a square grid ($L_y=L_x$) of qubits for the row-by-row with two corners tripartition, but we will consider the row-by-row tripartition also on rectangular grids with $L_y\neq L_x$.}
		\label{fig:ABC_tripartition_numerics}
	\end{figure}
	
	In this appendix, we numerically investigate the Scrooge Hypothesis, which is the central conjecture of this work, in the stabilizer Scrooge setting for diagonal observables like \eqref{eq:the_diagonal_observable}. All the following numerical experiments share a common setup. We take an $ABC$ tripartition of an $L_x\times L_y$ grid of qubits, and we define $\text{width}(B)=\text{dist}(A,C)-1$ where $\dist(A,C)=\min_{x\in A,y\in C}\dist(x,y)$ denotes the minimum L1 distance between $x\in A$ and $y\in C$ on the 2D grid. The two types of $ABC$ tripartitions we consider are depicted in \Cref{fig:ABC_tripartition_numerics}, where in both figures, each dot represents a qubit, and the red, blue, and orange qubits form the regions $A$, $B$, and $C$ respectively. We write a depth-$7$ 2D brickwork circuit $U$ as $U=U_CU_{AB}$ where $U_{AB}$ is a backward lightcone circuit acting non-trivially on $A\cup B$ and some boundary qubits $C'\subseteq C$, and $U_C$ is a forward lightcone circuit acting non-trivially on $C$. To generate a random $\rho_A$ sample, we first sample random Clifford gates for $U_{AB}$, then apply $U_{AB}$ to $\ket{0^n}$, and finally measure all qubits in $B$ in the computational basis to get an outcome $x_B$ and a  pure post-measurement state $\ket{\varphi}_{AC}=\ket{\psi}_{AC'}\otimes\ket{0^{|C\setminus C'|}}$ on $A\cup C$. With this, we obtain $\rho_A=\tr_C\left(\ket{\varphi}\bra{\varphi}\right)$. Conditioned on each $\rho_A$ generated in this way, we consider two distributions of pure stabilizer states supported on $A$. The first is $\mathcal{D}_\rho$, the stabilizer Scrooge distribution defined by $\rho_A$. The second, $\mathcal{E}_A=\mathcal{E}_A(U_{AB},x_B)$, is the distribution sampled by the following procedure: first draw independent fresh random gates for $U_C$, then apply $U_C$ to $\ket{\varphi}_{AC}$, and finally measure all qubits in $C$ in the computational basis to get a pure post-measurement state on $A$.
	
	For constant $t\geq 1$ and diagonal PSD $t$-th moment observable $O$, the Scrooge Hypothesis pertains to the scaling of the relative error
	\begin{equation}
		\left|\frac{\Exp_{\ket{\psi}\sim\mathcal{E}_A}\left[\bra{\psi}^{\otimes t} O \ket{\psi}^{\otimes t}\right]-\Exp_{\ket{\psi}\sim\mathcal{D}_\rho}\left[\bra{\psi}^{\otimes t} O \ket{\psi}^{\otimes t}\right]}{\Exp_{\ket{\psi}\sim\mathcal{D}_\rho}\left[\bra{\psi}^{\otimes t} O \ket{\psi}^{\otimes t}\right]}\right|
        \label{eq:relative error O app}
	\end{equation}
	as a function of $\text{width}(B)$ and $|\partial A|$ when the ABC tripartition satisfies some regularity conditions. While the above depends on a particular choice of observable $O$, we can quantify the convergence of moments for many observables simultaneously using the following construction.
	
	Let $\mathcal{M}$ denote the completely dephasing channel, or equivalently the computational-basis measurement channel, which sets all off diagonal entries of an input density matrix to zero. We call the expression
	\begin{equation}
		\Exp_{\ket{\psi}\sim\mathcal{E}_A}\left[\mathcal{M}\left(\ket{\psi}\bra{\psi}^{\otimes t}\right)\right]
        \label{eq:diagonal moment operator app}
	\end{equation}
	the $t$-th diagonal moment operator of $\mathcal{E}_A$. Notice that 
	\begin{equation}
		\Exp_{\ket{\psi}\sim\mathcal{E}_A}\left[\bra{\psi}^{\otimes t} O \ket{\psi}^{\otimes t}\right]=\Exp_{\ket{\psi}\sim\mathcal{D}_\rho}\left[\bra{\psi}^{\otimes t} O \ket{\psi}^{\otimes t}\right]
	\end{equation}
	for every diagonal $t$-th moment observable $O$ if and only if
	\begin{equation}
		\Exp_{\ket{\psi}\sim\mathcal{E}_A}\left[\mathcal{M}\left(\ket{\psi}\bra{\psi}^{\otimes t}\right)\right]=\Exp_{\ket{\psi}\sim\mathcal{D}_\rho}\left[\mathcal{M}\left(\ket{\psi}\bra{\psi}^{\otimes t}\right)\right].
		\label{eq:diagonal_moment_operator}
	\end{equation}
    The $t$-th diagonal frame potential is equal to the purity of the diagonal moment operator \eqref{eq:diagonal moment operator app}. Specifically,
    \begin{equation}
			\label{eq:DFP def appendix}
			\text{DFP}^{(t)}(\mathcal{E}_A) \coloneqq \tr\left[\left(\Exp_{\ket{\psi}\sim\mathcal{E}_A}\left[\mathcal{M}\left(\ket{\psi}\bra{\psi}^{\otimes t}\right)\right] \right)^2 \right] = \Exp_{\ket{\psi}, \ket{\phi}\sim\mathcal{E}_A}  \left( \sum_x |\braket{x|\psi}|^2 |\braket{x|\phi}|^2   \right)^t.
		\end{equation}
    This single scalar is relatively straightforward to estimate in our numerics, and for $t \leq 3$ can be related to the relative error \eqref{eq:relative error O app} for typical diagonal observables in the following sense: if the relative deviation between the DFP for the post-measurement and Scrooge ensembles satisfies
    \begin{align}
        \left| \frac{\text{DFP}^{(t)} (\mathcal{E}_A) }{\text{DFP}^{(t)} (\mathcal{D}_\rho)} - 1 \right| \leq \varepsilon
        \label{eq:DFP deviation}
    \end{align}
    for some $0 \leq \varepsilon < 1$, then, for a randomly chosen diagonal observable $O = \dyad{x_1, \ldots, x_t}$, with each bitstring $x_i : i = 1, \ldots, t$ sampled independently from the uniform distribution over all bitstrings in the support of $\rho_A$, the relative error \eqref{eq:relative error O app} will be at most $\varepsilon'$ with probability at most $1 - \varepsilon'$, where $\varepsilon' \leq ((t!)^2 \varepsilon)^{1/3} + 2^{-\Omega(S(\rho_A))}$. The proof of the above statement is provided at the end of this section.

    We focus on diagonal observables here for two reasons. First, these are the observables relevant to our classical simulability results, in particular \cref{thm:classical-algorithm}. Second, if we were to study the convergence of the moments for completely arbitrary observables, then it is possible that we encounter particular adversarial choices of $O$ for which the relative error \eqref{eq:relative error O app} may be large. Indeed, we emphasize that the Scrooge Hypothesis captures the behavior of moments where the observable $O$ is chosen independently of the randomness $(U_{AB}, x_B)$. It is conceivable that, for typical instances of $(U_{AB}, x_B)$, we could adaptively choose $O$ such that the corresponding moments in $\mathcal{E}_A(U_{AB}, x_B)$ and $\mathcal{D}_\rho$ differ significantly; however, this possibility lies beyond the scope of our conjecture. By restricting to diagonal observables, we limit the possible effects of these adversarial instances.\\
	
	We estimate the diagonal frame potential using the following procedure. After having sampled a list of $|A|$-qubit states $\ket{\psi_1},\ldots,\ket{\psi_T}$ from $\mathcal{E}_A$ or $\mathcal{D}_\rho$, we can estimate the corresponding $t$-th diagonal frame potential using the empirical average
	\begin{equation}
		\frac{1}{\binom{T}{2}}\sum_{i=1}^T\sum_{j=i+1}^T\bigg(\sum_{x\in\{0,1\}^{|A|}}|\langle x|\psi_i\rangle|^2|\langle x|\psi_j\rangle|^2\bigg)^t.
		\label{eq:diagonal_FP_average}
	\end{equation}
	For a pair of stabilizer states $\ket{\psi_i}$ and $\ket{\psi_j}$, the inner-most sum satisfies
	\begin{equation}
		\sum_{x\in\{0,1\}^{|A|}}|\braket{x|\psi_i}|^2\cdot|\braket{x|\psi_j}|^2=\frac{|\text{Supp}(\ket{\psi_i})\cap\text{Supp}(\ket{\psi_j})|}{|\text{Supp}(\ket{\psi_i})|\cdot |\text{Supp}(\ket{\psi_j})|}
		\label{eq:diagonal_FP_inner_sum}
	\end{equation}
	where the RHS can be computed using linear algebra in $O(|A|^3)$ time.
	
\begin{figure}[t]
		\centering
		\includegraphics[width=0.9\linewidth]{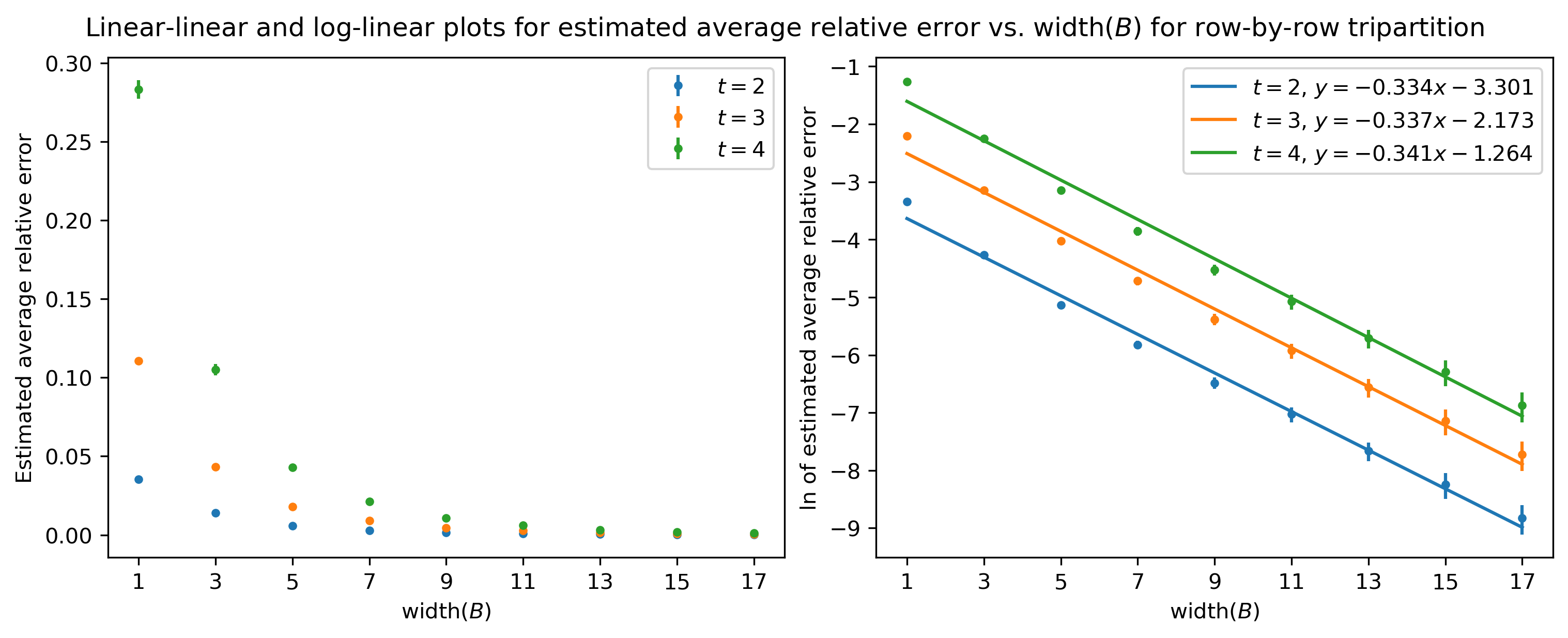}
		\caption{In this experiment, we investigate the dependence of the average relative error on $\text{width}(B)$ for the row-by-row tripartition. The other architectural parameters are held constant at $L_x=100$, $|A|=100$, and $|C|=600$. Each data point is estimated using $5000$ independent $\rho_A$ samples, and for each $\rho_A$, the relative error is computed from $t$-th diagonal frame potentials of $\mathcal{E}_A$ and $\mathcal{D}_\rho$ estimated using $10^4$ independent state samples. From the right panel, we see that as $\text{width}(B)$ increases from $1$ to $17$, the data points show good agreement with the linear least squares best fits, and that the $y$-axis spans a good range from, for example, $e^{-3}$ to $e^{-9}$ for $t=2$. Therefore, the data predict that the average relative error scales inverse exponentially w.r.t. $\text{width}(B)$. The error bars show $2\times$ standard error.}
		\label{fig:row_by_row_B}
	\end{figure}

\begin{figure}[t]
		\centering
		\includegraphics[width=0.9\linewidth]{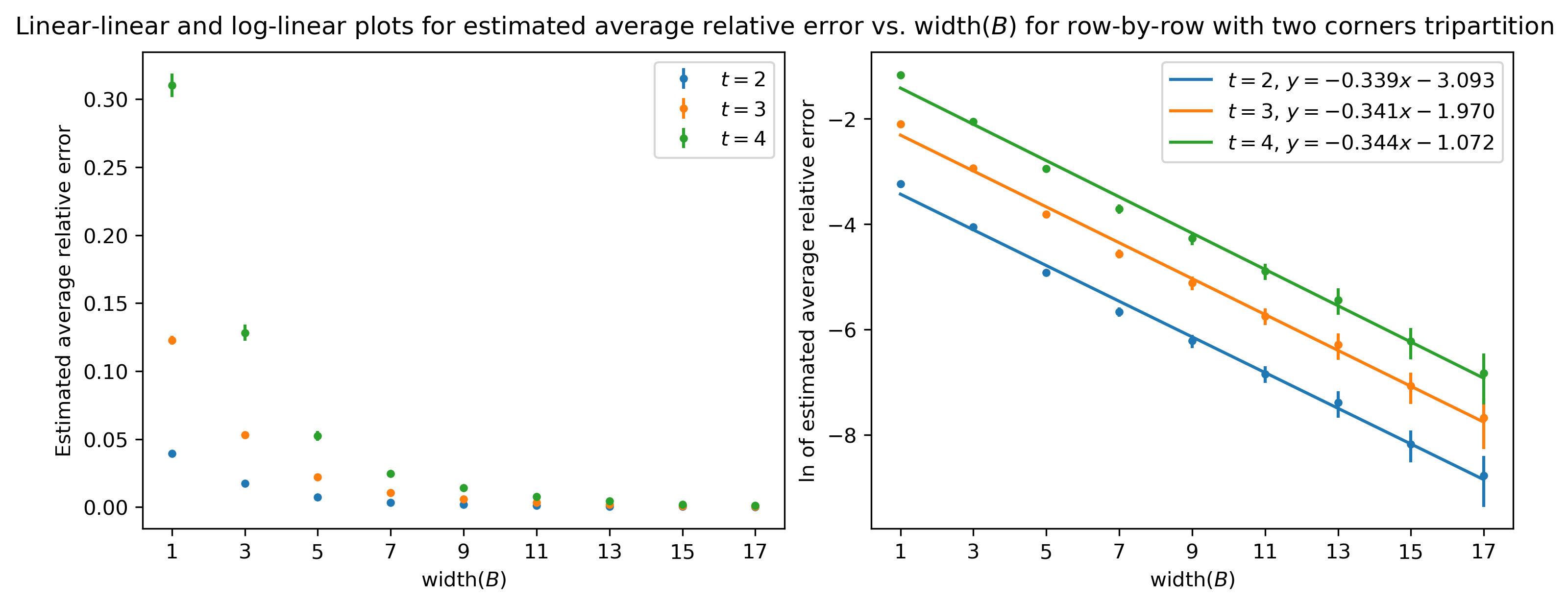}
		\caption{This experiment investigates the dependence of the average relative error on $\text{width}(B)$ for the row-by-row with two corners tripartition. we fix $L_x=L_y=60$ so $|A|=62$. Note that as $|B|$ increases, $|C|$ decreases accordingly as the total number of qubits is always fixed to be $60\times 60$. All the other experimental parameters are set identically to those used for \Cref{fig:row_by_row_B}, and the same overall qualitative scaling behaviour is observed.}
		\label{fig:row_by_row_two_corners_B}
	\end{figure}

	In our first experiment, we investigate the scaling of the average relative error between the $t$-th diagonal frame potentials of $\mathcal{E}_A$ and $\mathcal{D}_\rho$ as a function of $\text{width}(B)$ for the two types of ABC tripartitions depicted in \Cref{fig:ABC_tripartition_numerics}. For the row-by-row tripartition, we fix $L_x=100$ and $A$ to be the top row of qubits, so $|A|=L_x=100$. We let $\text{width}(B)$ vary from $1$ to $17$ and fix $C$ to contain $6$ rows of qubits below $B$, so $|C|=600$. To estimate each data point, we generate $5000$ independent $\rho_A$ (strictly speaking $\ket{\varphi}_{AC}$) samples using the procedure described earlier. Conditioned on each $\rho_A$, we draw $10^4$ independent state samples from $\mathcal{E}_A$ and $\mathcal{D}_\rho$ to estimate the $t$-th diagonal frame potentials of $\mathcal{E}_A$ and $\mathcal{D}_\rho$ using \eqref{eq:diagonal_FP_average} and \eqref{eq:diagonal_FP_inner_sum}. In \Cref{fig:row_by_row_B}, we plot the estimated average relative error against $\text{width}(B)$ for $t\in\{2,3,4\}$, as well as in log-linear scale with the best linear fits. The results shown in \Cref{fig:row_by_row_B} predict an inverse exponential relationship between the average relative error and $\text{width}(B)$. Also note that we were not bottlenecked by the dimension of $C$. We repeat the same experiment for the row-by-row with two corners tripartition on a $60$-by-$60$ grid, and the results are shown in \Cref{fig:row_by_row_two_corners_B}. Again, the data are consistent with a scaling that is inverse exponential in $\text{width}(B)$.
	
	\begin{figure}[t]
		\centering
		\includegraphics[width=0.8\linewidth]{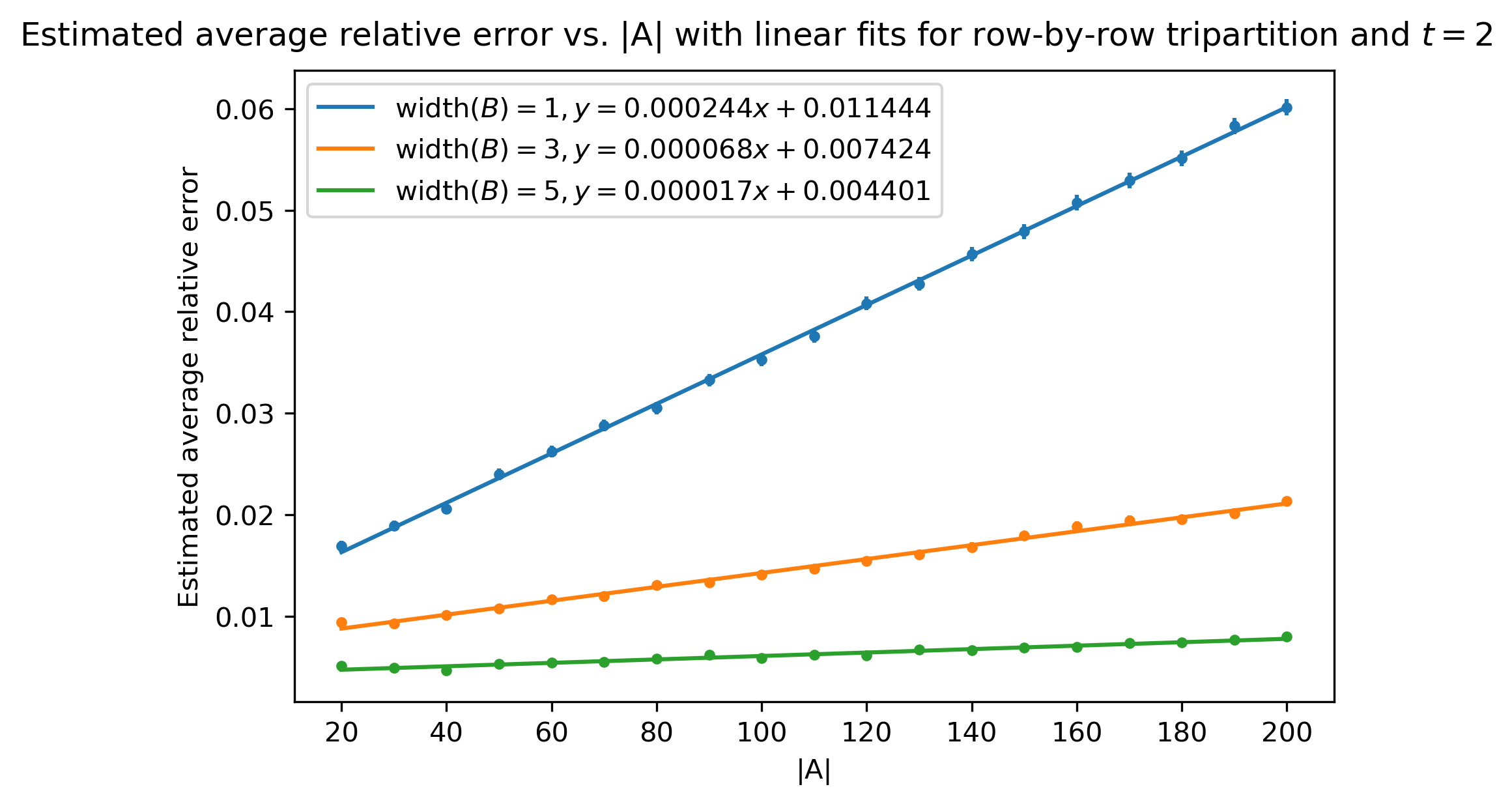}
		\caption{In this experiment, we probe the dependence of the average relative error on $|\partial A|=|A|$ for the row-by-row tripartition and $t=2$. We fix $C$ to contain $6$ rows of qubits below $B$, but note that $|C|$ increases with $|A|$. In all three cases $\text{width}(B)\in\{1,3,5\}$, the data points are well-explained by a linear model with slope decreasing in $\text{width}(B)$.}
		\label{fig:row_by_row_A}
	\end{figure}
	
	In our second experiment, we probe the scaling of the average relative error between the $t$-th diagonal frame potentials of $\mathcal{E}_A$ and $\mathcal{D}_\rho$ as a function of $|\partial A|$. We focus on $t=2$ and the row-by-row tripartition with which $|\partial A|=|A|$. For this experiment, we set $\text{width}(B)\in\{1,3,5\}$ and let $|A|$ range from $20$ to $200$. We still set $C$ to always contain $6$ rows of qubits below $B$, but note that $|C|$ becomes larger as $|A|$ increases. Again, we estimate each data point using $5000$ independent $\rho_A$ samples, and conditioned on each $\rho_A$, $10^4$ independent state samples are used to estimate the $t$-th diagonal frame potentials of $\mathcal{E}_A$ and $\mathcal{D}_\rho$. The results are shown in \Cref{fig:row_by_row_A}. As the values on the $y$-axis only spanned a narrow range, all we could conclude is that for each $\text{width}(B)\in\{1,3,5\}$, the estimated average relative error is well-modeled by the linear least squares fit. Nonetheless, note that as $\text{width}(B)$ increases from $1$ to $3$ and from $3$ to $5$, the slopes of the best linear fits decay by a factor of $3.5$ to $4$, which is consistent with the inverse exponential in $\text{width}(B)$ scaling predicted by the previous experiment shown in \Cref{fig:row_by_row_B}.
	
	In summary, all the scaling behaviours we observed numerically in \Cref{fig:row_by_row_B,fig:row_by_row_two_corners_B,fig:row_by_row_A} show strong agreement with those predicted by our analytical calculations.
	
	\textit{Relating convergence of the diagonal frame potential to relative error of moments.---}Here we prove the claim made earlier in this section, which we now state formally. Let $\mathcal{E}$ be an ensemble of pure $n$-qubit stabilizer states, with average state $\mathbb{E}_{\psi \sim \mathcal{E}} \dyad{\psi} = \rho$, and let $\mathcal{D}_\rho$ be the stabilizer Scrooge ensemble with the same average state $\rho$. Then, if the relative deviation between the diagonal frame potential of the ensembles $\mathcal{E}$ and $\mathcal{D}_\rho$ is bounded as
    \begin{align}
        \left|\frac{\text{DFP}^{(t)}(\mathcal{E})}{\text{DFP}^{(t)}(\mathcal{D}_\rho)} - 1 \right| \leq \varepsilon,
    \end{align}
    then the following holds. For any $t \leq 3$, choose a random $t$-copy diagonal observable $O = \dyad{x_1, \ldots, x_t}$, with each $x_i$ sampled independently from the uniform distribution over all bitstrings for which $\braket{x_i|\rho|x_i} \neq 0$. Then the corresponding moments $\chi_{\mathcal{E}}^{(t)}(O) \coloneqq \mathbb{E}_{\phi \sim \mathcal{E}} \tr[O \dyad{\phi}^{\otimes t}]$ and $\chi_{\mathcal{D}_\rho}^{(t)}(O)\coloneqq \mathbb{E}_{\psi \sim \mathcal{D}_\rho} \tr[O \dyad{\psi}^{\otimes t}]$ will be $\varepsilon'$-close in relative error (in the sense of Eq.~\eqref{eq:observable_relative_error_def}) with probability $1 - \varepsilon'$, where
    \begin{align}
        \varepsilon' \leq ((t!)^2\varepsilon)^{1/3} + 2^{-\Omega(S(\rho))},
    \end{align}
    where $S(\rho) = -\tr[\rho \log \rho]$ is the von Neumann entropy of $\rho$. Note that $\chi_{\mathcal{E}}(O) = \chi_{\mathcal{D}_\rho}(O) = 0$ trivially if one of the bitstrings $x_i$ is outside the support of $\rho$.
    
    Let the number of bitstrings $x_i$ for which $\braket{x_i|\rho|x_i} \neq 0$ be $2^{m}$, where $m$ is a non-negative integer due to $\rho$ being a stabilizer state. Then,
    \begin{align}
    \Exp_O[\chi^{(t)}_{\mathcal{E}}(O)] &= \Exp_O[\chi^{(t)}_{\mathcal{D}_\rho}(O)] = \frac{1}{2^{tm}} \label{eq:diag moment mean O} \\ 
        \Exp_{O} [\chi^{(t)}_{\mathcal{E}}(O) ^2] &= \frac{1}{2^{tm}} \text{DFP}^{(t)}(\mathcal{E}) & \Exp_{O} [\chi^{(t)}_{\mathcal{D}_\rho}(O) ^2] &= \frac{1}{2^{tm}} \text{DFP}^{(t)}(\mathcal{D}_\rho)
    \end{align}
    Now, by \cref{lem:scrooge_plant}, the distributions $\ket{\phi} \sim \mathcal{E}$ and $\ket{\psi} \sim \mathcal{D}_\rho$, both of which have the same average state $\rho$, can be more easily characterized by identifying a Clifford unitary $V$ such that $V^\dagger \rho V = 2^{-(n-k)} (\dyad{0^k} \otimes I_{n-k})$ for some integer $k$. In which case we find
	\begin{align}
		\ket{\phi} \sim \mathcal{E} \; &\Rightarrow \; \ket{\phi} = V(\ket{0^k} \otimes \ket{\phi'}) & \ket{\phi'} &\sim \mathcal{E}' \\
		\ket{\psi} \sim \mathcal{D}_\rho \; &\Rightarrow \;\ket{\psi} = V(\ket{0^k} \otimes \ket{\psi'}) & \ket{\psi'} &\sim \mathcal{D}' \label{eq:Scrooge dist rotate}
	\end{align}
	where $\mathcal{E}'$ is some distribution of $(n-k)$-qubit stabilizer states with average state $\mathbb{E}_{\ket{\phi'} \sim \mathcal{E}'} \dyad{\phi'} = 2^{-{(n-k)}}I_{n-k}$, and $\mathcal{D}'$ is the uniform distribution over $(n-k)$-qubit stabilizer states. For convenience, we will define $d \coloneqq 2^{n-k}$, and we note that $S(\rho) = \log d = n-k$.

    For each computational basis state $\ket{x}$ in the support of $\rho$, we can define an unnormalized state $\ket{\alpha_x} = (\bra{0^k}\otimes I_{n-k}){V^\dagger}\ket{x}$ such that $\braket{x|\psi} = \braket{\alpha_x|\psi'}$. In particular, every such state $\ket{\alpha_x}$ has the same norm, which we denote $\sqrt{\beta}$. This is because $\braket{\alpha_x|\alpha_x} \propto \braket{x|\rho|x}$ and the output distribution of $\rho$ is flat by virtue of $\rho$ being a stabilizer state.
    
    We now write $\omega_{\mathcal{D}'}^{(t)} \coloneqq \mathbb{E}_{\ket{\psi'} \sim \mathcal{D}'} \dyad{\psi'}^{\otimes t}$ for the matrix of all $t$-th moments of a random $(n - k)$-qubit stabilizer state $\ket{\psi'} \sim \mathcal{D}'$. In particular, since random stabilizer states form a 3-design \cite{kueng2015qubitstabilizerstatescomplex}, for $t \leq 3$ we have $\omega_{\mathcal{D}'}^{(t)} = \frac{1}{d^{(t)}}\sum_{\pi \in S_t}\pi$, where $\pi$ are operators that permute the $t$ copies of Hilbert space, and $d^{(t)} \coloneqq d(d+1) \cdots (d+t-1)$. Thanks to Eq.~\eqref{eq:Scrooge dist rotate}, for any fixed valid choice of $O = \dyad{x_1, \ldots, x_t}$ and $t \leq 3$, we have
    \begin{align}
        \chi^{(t)}_{\mathcal{D}_\rho}(O) = \braket{\alpha_{x_1} \cdots \alpha_{x_t}| \omega_{\mathcal{D}'}^{(t)}| \alpha_{x_1} \cdots \alpha_{x_t}} = \frac{1}{d^{(t)}} \sum_{\pi \in S_t} \braket{\alpha_{x_1} \cdots \alpha_{x_t}| {\pi}| \alpha_{x_1} \cdots \alpha_{x_t}} \geq  \frac{1}{d^{(t)}} \prod_{i=1}^t \braket{\alpha_{x_t}|\alpha_{x_t}} = \frac{\beta^t}{d^{(t)}}.
    \end{align}
    The inequality in the above is a consequence of Marcus' permanent inequality, which states that for any $t \times t$  positive semi-definite matrix $X$, we have $\text{per(X)} \geq \prod_{a=1}^t X_{aa}$, where $\text{per}(X) = \sum_{\pi \in S_t} \prod_{a=1}^t X_{a, \pi(a)}$ is the permanent \cite{marcus1963permanent}. We apply this to the Gram matrix $X$, with matrix elements $X_{ab}= \braket{\alpha_{x_a}|\alpha_{x_b}}$, which is indeed positive semi-definite. This implies that the sum of all the terms for which $\pi \neq I$ is always non-negative.
    
    The corresponding mean square is at most
    \begin{align}
        \Exp_O[\chi^{(t)}_{\mathcal{D}_\rho}(O)^2] &= \frac{1}{2^{tm}} \sum_{x_1, \ldots, x_t} \braket{\alpha_{x_1}\cdots \alpha_{x_t}|\omega_{\mathcal{D}'}^{(t)}|\alpha_{x_1}\cdots \alpha_{x_t} }^2 \nonumber\\ &\leq \frac{1}{2^{tm}} \sum_{x_1, \ldots, x_t} \|\omega_{\mathcal{D}'}^{(t)}\|^2_\infty \prod_{j=1}^t \braket{\alpha_{x_j}|\alpha_{x_j}}^2 = \|\omega_{\mathcal{D}'}^{(t)}\|_\infty^2 \beta^{2t} = \left(\frac{t!}{d^{(t)}} \beta^t \right)^2
    \end{align}
    where the final equality is due to the fact that $\omega_{\mathcal{D}'}^{(t)}$ is the maximally mixed state on the symmetric subspace of $t$ copies of a Hilbert space of dimension $d = 2^{n-k}$, and accordingly this subspace has dimension $\binom{d+t-1}{t} = \frac{d^{(t)}}{t!}$. Therefore, $\chi^{(t)}_{\mathcal{D}_\rho}(O) \geq \frac{1}{t!} \sqrt{\mathbb{E}_{O'}[\chi^{(t)}_{\mathcal{D}_\rho}(O')^2]}$ with certainty over the randomness in $O$. Accordingly, we obtain
    \begin{align}
        \Pr_O\left(\left| \frac{\chi^{(t)}_{\mathcal{E}}(O)}{\chi^{(t)}_{\mathcal{D}_\rho}(O)} - 1 \right| \geq \eta \right) \leq \Pr_O\left(t! \frac{|\chi^{(t)}_{\mathcal{E}}(O) - \chi^{(t)}_{\mathcal{D}_\rho}(O)|}{\sqrt{\mathbb{E}_{O'}[\chi^{(t)}_{\mathcal{D}_\rho}(O')^2]}  }   \geq \eta \right) \leq \frac{(t!)^2}{\eta^2} \frac{\mathbb{E}_O[(\chi^{(t)}_{\mathcal{E}}(O) - \chi^{(t)}_{\mathcal{D}_\rho}(O))^2]}{\mathbb{E}_O[\chi^{(t)}_{\mathcal{D}_\rho}(O)^2]}.
    \end{align}
    Now we claim that
    \begin{align}
        \Exp_O[ \chi^{(t)}_{\mathcal{E}}(O)\chi^{(t)}_{\mathcal{D}_\rho}(O)] \geq (1 - \delta_\rho)\Exp_O[\chi^{(t)}_{\mathcal{D}_\rho}(O)^2]
        \label{eq:diag cross ineq}
    \end{align}
    for any ensemble of stabilizer states $\mathcal{E}$ whose mean state is $\rho$, where $\delta_\rho =  2^{-\Omega(S(\rho))}$. This then implies
    \begin{align}
        \Pr_O\left(\left| \frac{\chi^{(t)}_{\mathcal{E}}(O)}{\chi^{(t)}_{\mathcal{D}_\rho}(O)} - 1 \right| \geq \eta \right) \leq \frac{(t!)^2}{\eta^2} \left(\frac{\mathbb{E}_O[\chi^{(t)}_{\mathcal{E}}(O)^2]}{\mathbb{E}_O[\chi^{(t)}_{\mathcal{D}_\rho}(O)^2]} - 1 + 2\delta_\rho \right) \leq \frac{(t!)^2}{\eta^2} (\varepsilon + 2 \delta_\rho).
    \end{align}
    Setting $\eta = [(t!)^2(\varepsilon + 2\delta_\rho)]^{1/3}$ then proves the desired result.

    We now prove \eqref{eq:diag cross ineq}. If we define $\omega_{\mathcal{E}'}^{(t)} \coloneqq \mathbb{E}_{\ket{\phi'} \sim \mathcal{E}'} \dyad{\phi'}^{\otimes t}$ by analogy to $\omega_{\mathcal{D}'}^{(t)}$, then this is equivalent to
    \begin{align}
        &\sum_{x_1, \ldots, x_t} \sum_{\pi \in S_t} \braket{\alpha_{x_1} \cdots \alpha_{x_t}|\pi| \alpha_{x_1} \cdots \alpha_{x_t}}  \braket{\alpha_{x_1} \cdots \alpha_{x_t}| \omega_{\mathcal{E}'}^{(t)} | \alpha_{x_1} \cdots \alpha_{x_t}}\\
        \geq& (1 - \delta_\rho)\sum_{x_1, \ldots, x_t} \sum_{\pi \in S_t} \braket{\alpha_{x_1} \cdots \alpha_{x_t}|\pi| \alpha_{x_1} \cdots \alpha_{x_t}}  \braket{\alpha_{x_1} \cdots \alpha_{x_t}| \omega_{\mathcal{D}'}^{(t)} | \alpha_{x_1} \cdots \alpha_{x_t}}.
        \label{eq:DFP target ineq}
    \end{align}
    Recall that $\braket{\alpha_{x}|\alpha_x} = \beta$ independent of $x$. From their definition, the moment matrices $\omega_{\mathcal{E}'}^{(t)}$, $\omega_{\mathcal{D}'}^{(t)}$ both have unit trace, and we also have $\sum_x \dyad{\alpha_x} = I_{n-k}$. As a result, the $\pi = I$ terms on both sides of \eqref{eq:diag cross ineq} are equal to $\beta^t$. Again using Marcus' permanent inequality, one can also show that the sum of all the $\pi \neq I$ terms on the left hand side is non-negative.
    
    Finally, we can bound the relative size of the $\pi \neq I$ terms on the right hand side as
    \begin{align}
		\frac{1}{d^{(t)}}\sum_{\pi \neq I} \sum_{\pi'} \sum_{x_1, \ldots, x_t} \braket{\alpha_{x_1}\cdots \alpha_{x_t}|\pi|\alpha_{x_1}\cdots \alpha_{x_t}} \braket{\alpha_{x_1}\cdots \alpha_{x_t}|\pi'|\alpha_{x_1}\cdots \alpha_{x_t}}
        \leq \frac{t! \beta^t}{d^{(t)}} \sum_{\pi \neq I}\sum_{x_1, \ldots, x_t} \left|\prod_{a=1}^t \braket{\alpha_{x_a}|\alpha_{x_{\pi(a)}}}\right| 
		\label{eq:dfp non-indentity}
	\end{align}
	where the second line follows from the fact that $|\braket{\alpha_{x_1}\cdots \alpha_{x_t}|\pi'|\alpha_{x_1}\cdots \alpha_{x_t}} | \leq \prod_{a=1}^t\| \ket{\alpha_{x_a}} \|^2 = \beta^t$. From the definition of the states $\ket{\alpha_x}$, we have $\braket{\alpha_x|\alpha_y} = \braket{x|\Pi|y} \eqqcolon \Pi_{x,y}$, where $\Pi = VV^\dagger = 2^{n-k}\rho$ is the stabilizer projector on which $\rho$ is supported. Let $S_Z$ be the $Z$-type stabilizers of $\Pi$, with corresponding projector $\Pi_Z \propto \sum_{g \in S_Z} g$, and $G_\perp$ a minimal set of independent generators that together with $S_Z$ generate all the stabilizers of $\Pi$. Let $S_\perp$ be the group generated by $G_\perp$, elements $g$ of which act on computational basis states as $g\ket{x} = \omega_{g,x}\ket{x \oplus u(g)}$, where $\omega_{g,x}$ is a phase and $u : S_\perp \rightarrow \mathbb{F}_2^n$ is a homomorphism. Note that $g \in \text{Ker}(u)$ implies $g$ is a $Z$-type stabilizer, so $u$ must have trivial kernel and is hence injective.
	
	Now define the relation $x \sim y \Leftrightarrow x \oplus y \in \text{Im}(u)$. We then have
	\begin{align}
		|\braket{\alpha_x|\alpha_y}| = \begin{dcases}
			\beta & x \sim y \\ 0 & \text{otherwise} 
		\end{dcases}
	\end{align}
	This follows since, if $x \sim y$ then there is exactly one $g \in S_\perp$ such that $x \oplus y = u(g)$, in which case $\braket{x|\Pi|y} = \kappa \omega_{x,y}$ for some real positive constant $\kappa$ independent of $x,y$ and phase $\omega_{x,y}$. We have $\kappa = \beta$ from the case $x = y$. Therefore, if $i, j \in [t]$ are two indices in the same cycle of $\pi$, we have
	\begin{align}
		\left|\prod_{a=1}^t \braket{\alpha_{x_a}|\alpha_{x_{\pi(a)}}}\right|  \leq |\braket{\alpha_{x_i}|\alpha_{x_j}}|^2 \prod_{\substack{a = 1\\ a \neq i, j}}^t \braket{\alpha_{x_{a}}|\alpha_{x_{a}}}.
	\end{align}
    If $\pi$ is any permutation other than $I$, then it must have a cycle of length at least two, and thus we can find such a pair of indices $i,j$. Thus, 
    \begin{align}
        \sum_{x_1, \ldots, x_t} \left|\prod_{a=1}^t \braket{\alpha_{x_a}|\alpha_{x_{\pi(a)}}}\right| &\leq \left(\sum_{x_i,x_j} |\braket{\alpha_{x_i}|\alpha_{x_j}}|^2\right) \prod_{\substack{a=1 \\ a \neq i,j}}^t\left(\sum_{x_a} \braket{\alpha_{x_a}|\alpha_{x_a}} \right) = d^{t-1} & \forall \pi \neq I.
    \end{align}
    In the last equality, we use $\sum_x \braket{\alpha_x|\alpha_x} = \tr[I_{n-k}] = d$, and
    \begin{align}
        \sum_{x,y} |\braket{\alpha_x|\alpha_y}|^2 = \sum_{x,y} \tr[ \dyad{x}V(\dyad{0}\otimes I_{n-k})V^\dagger \dyad{y} V(\dyad{0}\otimes I_{n-k})V^\dagger] = \tr[(\dyad{0}\otimes I_{n-k})^2] = d.
    \end{align}
    We substitute this into the right hand side of Eq.~\eqref{eq:dfp non-indentity} and sum over $\pi \neq I$, which gives an upper bound of $(t!)^2 \beta^t \times \frac{d^{t-1}}{d^{(t)}}$. Since we have already shown that the $\pi = I$ term evaluates to $\beta^t$, and we have  $d^{(t)} \geq d^t$, the desired inequality \eqref{eq:DFP target ineq} holds with $\delta_\rho = (t!)^2 d^{-1} = 2^{-\Omega(S(\rho))}$.
    
\end{document}